%% file: main_arxiv.tex
\documentclass[11pt]{article}
\usepackage[margin=1in]{geometry}

\usepackage{natbib}
\bibpunct[, ]{(}{)}{,}{a}{}{,}%

\usepackage{my_style_arxiv}
\usepackage{my_notation_arxiv}
\newif\ifarxiv
\arxivtrue

\title{Network Synthetic Interventions: A Causal Framework \\for Panel Data Under Network Interference}

\author[1]{Anish Agarwal}
\author[2]{Sarah H. Cen\thanks{Correspondence to Sarah H. Cen at shcen@mit.edu.}}
\author[2]{Devavrat Shah}
\author[3]{Christina Lee Yu}
\affil[1]{Industrial Engineering and Operations Research, Columbia University}
\affil[2]{Electrical Engineering and Computer Science, Massachusetts Institute of Technology}
\affil[3]{Operations Research and Information Engineering, Cornell University}
\date{\vspace{-5ex}}

\begin{document}
	\maketitle

\input{sections/abstract.tex}

\input{sections/intro.tex}

\input{sections/prob_statement.tex}

\input{sections/estimator.tex}

\input{sections/theoretical_results.tex}

\input{sections/validity_tests.tex}
\input{sections/experimental_procedure.tex}

\input{sections/simulations.tex}

\input{sections/discussion.tex}

\section*{Acknowledgments}
The authors gratefully acknowledge funding from the MIT-IBM project on Causal Representation, the
National Science Foundation (NSF) grant CNS-1955997, 
and the
Air Force Research Laboratory (AFOSR) grant FA9550-23-1-0301.

\bibliography{bib.bib}{}
\bibliographystyle{apalike}

\newpage 

\begin{center}
{  \LARGE \bf Appendix}
\end{center}

\appendix
	
	\section{Preliminaries} \label{app:preliminaries}

\input{sections/app_preliminaries.tex}

	\section{Helper lemmas}\label{app:helpers}

\input{sections/app_useful_lemmas}

	\section{Proofs for Section \ref{sec:results}} \label{app:proofs}
	\input{sections/app_proofs.tex}

	\section{Proofs for Section \ref{sec:validity_test}}\label{app:proofs_tests}

\input{sections/app_valid_test_proofs.tex}

	\section{Proofs for Section \ref{sec:experimental_proc}}\label{app:proofs_exp}
	\input{sections/app_proofs_exp_design.tex}

	\section{Simulation details} \label{app:simulations}
	\input{sections/app_simulations.tex}

\end{document}

%% file: sections/abstract.tex
\ifarxiv
\begin{abstract}
\else
\ABSTRACT{
\fi
We propose a generalization of the synthetic controls and synthetic interventions methodology to incorporate network interference. 
We consider the estimation of unit-specific potential outcomes from panel data in the presence of spillover across units and unobserved confounding. 
Key to our approach is a novel latent factor model that takes into account network interference and generalizes the factor models typically used in panel data settings.
We propose an estimator, \emph{Network Synthetic Interventions} (NSI), and show that it consistently estimates the mean outcomes for a unit under an arbitrary set of counterfactual treatments for the network.
We further establish that the estimator is asymptotically normal. 
We furnish two validity tests for whether the NSI estimator reliably generalizes to produce accurate counterfactual estimates. 
We provide a novel graph-based experiment design that guarantees the NSI estimator produces accurate counterfactual estimates, and also analyze the sample complexity of the proposed design.
We conclude with simulations that corroborate our theoretical findings.
\ifarxiv
\end{abstract}
\else
}
\fi

%% file: sections/intro.tex
\section{Introduction}\label{sec:intro}

There is growing interest in the identification and estimation of causal effects in the context of spillover on networks, in which the outcomes of a unit are affected by the treatments assigned to other units, known as the unit’s ``neighbors.''
Here, a unit could be an individual, customer cohort, or region, and correspondingly, treatments could be recommendations, discounts, or legislation.
For example, whether an individual gets COVID-19 is a function of not only the individual’s vaccination status but also the vaccination status of that individual’s social network. 
In the setting of e-commerce, the number of goods sold of a particular product is a function of not only whether that product gets a discount, but the discount level of other products that are substitutes or complements of it. 
That is, there is \emph{network interference}.

\begin{figure*}[t]
	\centering
	\includegraphics[width=\textwidth]{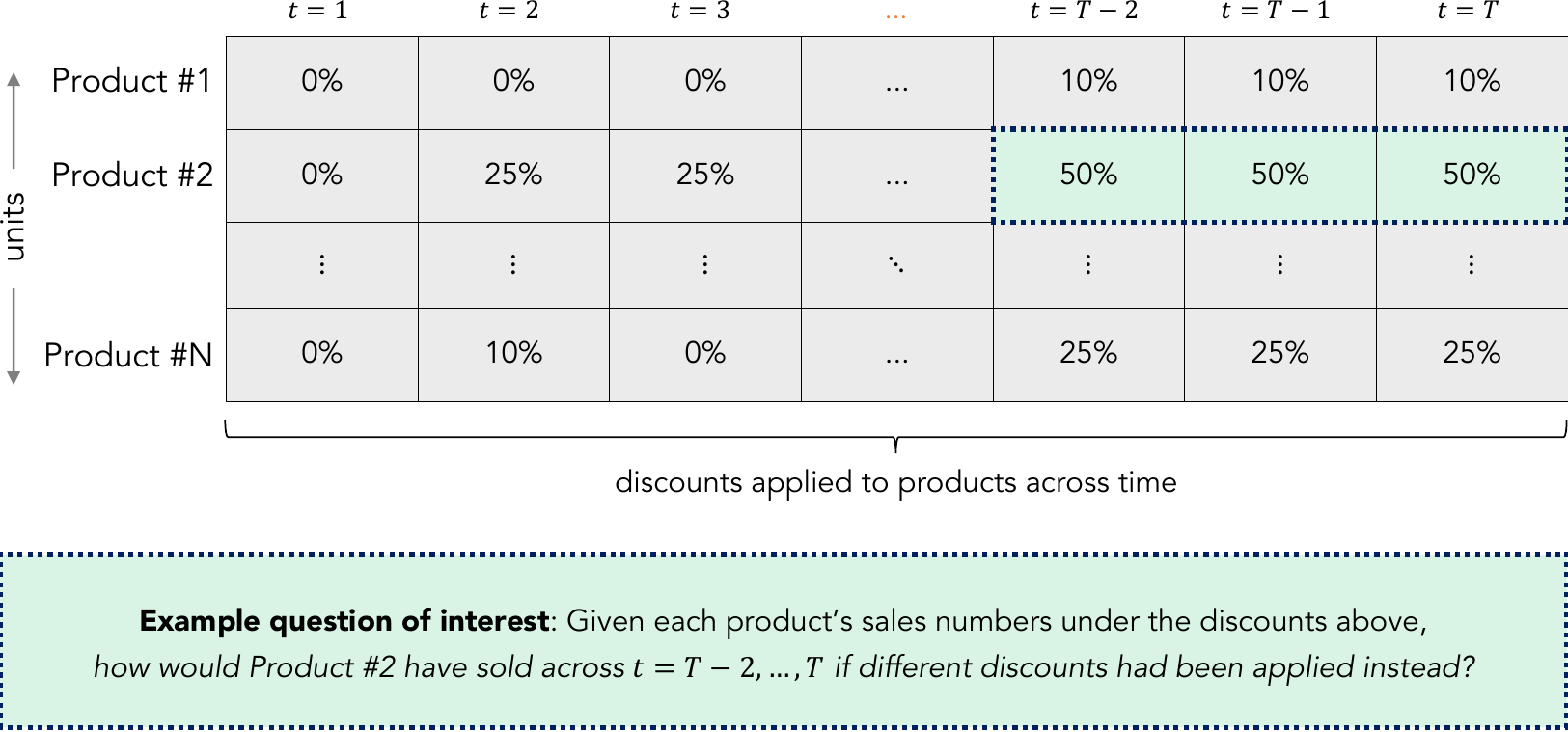}
	\caption{
		Panel data setting illustrated via an online retail example.  
		Each row corresponds to a product (or unit). 
		Each column corresponds to a week (or measurement). 
		A discount (the treatment) is applied to each product each week.
		In this work, we ask questions of the form: 
		Given every product's sales numbers across time and under various discounts,  
		what would the sales numbers (i.e., potential outcomes)
		have been for a specific unit, like Product \#2, under a different set of end-of-year discounts?}
	\label{fig:panel_data_intro}
\end{figure*}

In this work, we focus on network inference with {\em panel data}, a ubiquitous manner in which data is structured, where we collect multiple measurements of different units, and each unit can undergo a different sequence of treatments.
See Figure \ref{fig:panel_data_intro} for an example of panel data and the type of causal question we are interested in.
Causal inference with panel data has recently received significant attention, and a popular class of estimators in such settings are known as \emph{matching estimators}, where one represents the outcomes of one unit as some combination of other units to answer counterfactual questions.
Such estimators have been very popular in practice due to their flexibility and simplicity in addition to the fact that they provide valid causal estimates under unobserved confounding with appropriate assumptions.   
Some examples of matching estimators with panel data include Difference-in-Differences (DiD) \citep{bertrand2004much}, Synthetic Controls (SC) \citep{abadie2021using}, and variants thereof. 
However, such matching estimators rely on the Stable Unit Treatment Value Assumption (SUTVA), which implies that there is no spillover across units, i.e., the treatment applied to one unit does not affect the outcomes of other units. 
Failing to account for spillovers can lead to %
biased estimates.

We propose a novel latent factor model---which is a generalization of models studied in the panel data literature---that accounts for network interference. 
Given this model, we establish an identification result where the counterfactual potential outcome for a given unit and its neighbors can be written as a linear combination of the observed outcomes of a carefully selected set of other units.
This identification result leads to a natural estimator, which we call \emph{Network Synthetic Interventions (NSI)}, a simple two-step procedure, that estimates the mean counterfactual potential outcome for a given unit. 
We then show that, given our latent factor model, the NSI estimator is finite-sample consistent and asymptotically normal under suitable conditions. 
NSI and our analysis of it can be viewed as a generalization of the Synthetic Interventions \citep{agarwal2020synthetic}  and, in turn,  Synthetic Controls frameworks to account for network interference.

We furnish two validity tests that verify whether the treatment assignment pattern and the observed data have enough variation such that valid counterfactual estimates can be produced.
Motivated by these tests, we provide a novel graph-based experiment design.  

To explain the efficacy of the experiment design and the NSI estimator, we consider the setting of a regular network graph with degree $d \geq 2$. 
We show that the proposed experiment design requires only $O(d^3)$ training samples in order to guarantee that the training data has enough variation such that it is possible 
to generalize to a given target counterfactual treatment. 
Further, NSI obtains an estimate within error $\varepsilon$ with high probability under the proposed experiment design when $O\big({\sf poly}(d) / \varepsilon^4\big)$ training samples per unit are available. 
This is a significant improvement over the $O({\sf exp}(d) / \varepsilon^2)$ training samples that a naive procedure would require.

We conclude with simulations showing that NSI is robust to spillovers under which existing estimators are biased.

\subsection{Related work}\label{sec:related_work}

The literature on causal inference with network interference or spillover effect has mostly considered the setting of a single measurement per unit, whether in the setting of a randomized experiment or an observational study. 
Under fully arbitrary interference, it has been shown that it is impossible to estimate any desired causal estimands as the model is not identifiable \citep{Manski13, AronowSamii17, BasseAiroldi17, karwa2018systematic}. 
Subsequently, various models have been proposed in the literature that impose restrictions on the exposure functions \citep{Manski13, AronowSamii17, viviano2020experimental, auerbach2021local, li2021causal}, interference neighborhoods \citep{UganderKarrerBackstromKleinberg13, bargagli2020heterogeneous, SussmanAiroldi17, pmlr-v115-bhattacharya20a}, parametric structure \citep{ToulisKao13, BasseAiroldi15, cai2015social, GuiXuBhasinHan15,EcklesKarrerUgander17}, two-sided platforms \citep{johari2022experimental, bajari2021multiple} or a combination of these, each leading to a different solution concept.
A comprehensive review on network interference models is given by \cite{de2017econometrics}.
In this work, we focus on network interference that is additive across the neighbors, referred to in the literature as the joint assumptions of neighborhood interference, additivity of main effects, or additivity of interference effects \citep{SussmanAiroldi17, yu2022graph, cortez2022exploiting, cortez2022graph}. 

Distinct to our work is that we consider a {\em panel data} setting in which there are multiple measurements (e.g., a time series) for each unit.
The potential outcomes function is thus also dependent on both the unit and the measurement.
Additionally, we allow for the estimation of unit-specific counterfactuals under {\em multiple treatments}, whereas the existing literature has largely focused on binary treatments. 
Key to our approach is a novel latent factor model that takes into account network interference and is a generalization of the factor models typically used in panel data settings.
Previous work has focused on causal estimands that capture population-level effects, such as the average direct treatment effect (the average difference in outcomes if only one unit and none of its neighbors get treated \citep{BasseAiroldi15, JagadeesanPillaiVolfovsky17, SavjeAronowHudgens17, SussmanAiroldi17, leung2019causal, ma2021causal}) and the average total treatment effect (the average difference in outcomes if all units get treated versus if they do not \citep{UganderKarrerBackstromKleinberg13, EcklesKarrerUgander17, chin2019regression, yu2022graph, cortez2022exploiting, cortez2022graph}). 
Alternately there has been some literature that focuses on hypothesis testing for the presence of network interference \citep{Aronow12, BowersFredricksonPanagopoulos12,AtheyEcklesImbens17,PougetAbadieSaveskiSaintJacquesDuanXuGhoshAiroldi17,saveski2017detecting}; these results do not immediately extend to estimation as they are based on randomization inference with a fixed network size, and focus on testing the sharp null hypotheses.

While a majority of the literature focuses on randomized experiments, there is a growing interest in the literature to account for network interference when analyzing observational studies.
The existing literature generally assumes partial interference, where the network consists of many disconnected sub-communities \citep{TchetgenVanderWeele12, perez2014assessing, liu2016inverse, DiTraglia2020, vazquez-bare2022}. 
Without this strong clustering condition, other works impose strong parametric assumptions on the potential outcomes function, assuming that the potential outcomes only depend on a known statistic of the neighborhood treatment, e.g. the number or fraction of treated \citep{verbitsky2012causal, chin2019regression, ogburn2017causal}. 
This reduces estimation to a regression task under requirements of sufficient diversity in the treatments.
\cite{belloni2022neighborhood} also consider a setting in which the exposure mapping is known
but allow the ``radius'' of interference to vary across units, then learn this radius from data to devise a doubly robust estimator.  
\citet{forastiere2021identification} consider a general exposure mapping model alongside an inverse propensity weighted estimator, but the estimator has high variance when the exposure mapping is complex.
\cite{de2018recovering} and \cite{de2019identifying} derive identification conditions when the observational panel data contains no information about the social ties (i.e., network). 
Further, building on recent works in panel data \citep{agarwal2020synthetic, causalmatrixcompletion}, we allow for unobserved confounding in treatment assignment as long as there exist low-rank latent factors that mediate the unobserved confounding, i.e., there is ``selection on latent factors''.

%% file: sections/prob_statement.tex
\section{Setup \& Model}\label{sec:setup}

We begin with some notation. 
Let $[X] \defeq \{1,\dots, X\}$ for any positive integer $X$.
For vector $\ba \in [D]^N$ and set $S \subseteq [N]$, 
let $\ba_S \in [D]^{|S|}$ denote the vector containing the elements of $\ba$ indexed by $S$ and $a_i \in [D]$ denote the $i$-th element of $\ba$.
Let $\bbI_x$ denote the $x \times x$ identity matrix and $\otimes$ denote the Kronecker product. 
Let $\ind(\cdot)$ denote the indicator function. 
Let $\norm{\cdot}_{\psi_2}$ denote the Orlicz norm. 
Let $O_p$ denote a probabilistic version of big-$O$ notation and  $\tilde{\Omega}$ denote the variation on big-$\Omega$ notation that ignores logarithmic terms (see Appendix \ref{app:preliminaries} for precise definitions). 
For sets of indices $S_1 \subseteq [m_1]$ and $S_2 \subseteq [m_2]$ and a matrix $\Pi \in \bbR^{m_1 \times m_2}$, let $\Pi[S_1, S_2] \in \bbR^{|S_1| \times |S_2|}$ denote the submatrix corresponding to the rows indexed by $S_1$ and columns index by $S_2$. 
We use ``$\col$'' as a shorthand for all indices such that $\Pi[\col, S_2] \in \bbR^{m_1 \times |S_2|}$ and $\Pi[S_1, \col] \in \bbR^{|S_1| \times m_2}$.
Let $\cX^*$ denote the $*$-product space, where its length is not pre-determined. 
Let $\Pi^+$ denote the pseudo-inverse of $\Pi$.

\subsection{Setup}

\begin{figure*}[t]
\centering
\includegraphics[width=\textwidth]{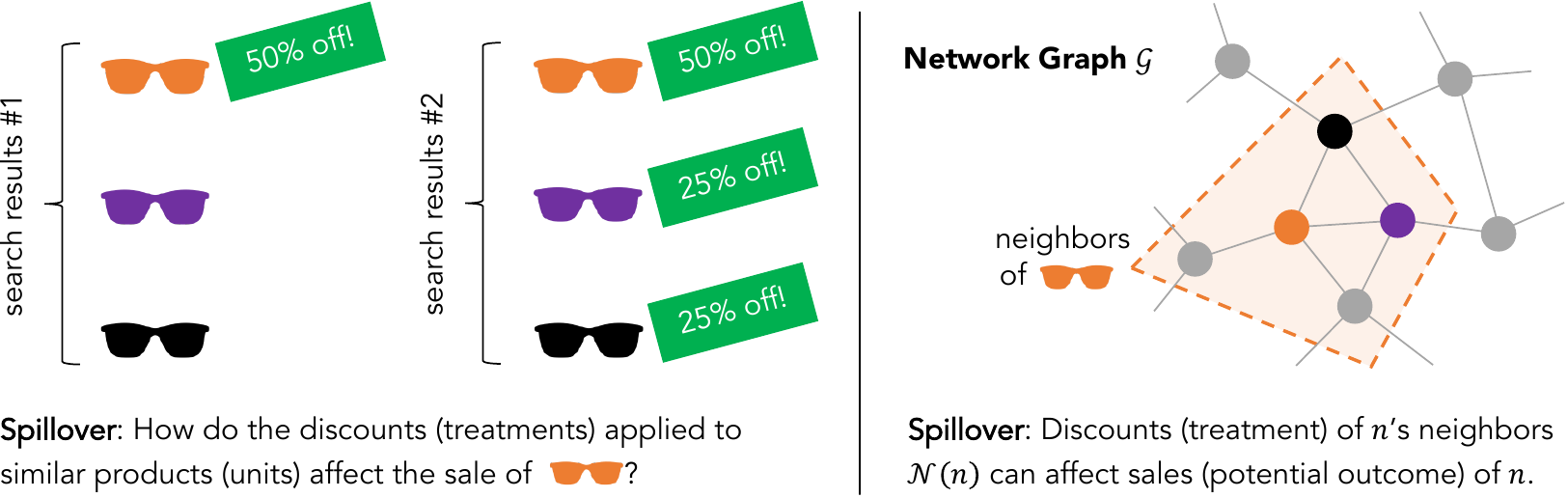}
\caption{Example of spillover effects and how they can be captured via a graph network. 
On the left, suppose that an online retailer presents similar products alongside one another. 
Then, the sale of one product (e.g., orange sunglasses) is affected by the discounts applied to similar products; in this case, other sunglasses.
On the right, spillover is often modeled via a network graph $\cG$, in which the treatments applied to the neighbors $\cN(n)$ of a unit $n$ may affect the potential outcomes of $n$. 
}
\label{fig:network_example}
\end{figure*}

Consider $N \geq 1$ units, $D \geq 1$ treatments, and $T \geq 1$ measurements of interest. 
We denote the potential outcome for a given unit $n$ and measurement $t$  by the real-valued random variable $Y^{(\bintv)}_{t,n}$,
where $\bintv \in [D]^{N}$ denotes the vector of treatments over \emph{all} $N$ units.
This definition allows for spillover effects because the potential outcome for a given unit is a function of the treatment assignment of all units.
To model spillover across units, we use a network graph.
Let $\cG = ([N] , \cE)$ denote a graph over the $N$ units, where $\cE \subseteq [N] \times [N]$ denotes the edges of the graph.
Throughout, we assume that $\cG$ is {\em fixed} and {\em known}.
Let $\cN(n)$ denote the neighbors of unit $n \in [N]$ with respect to $\cG$
such that $j \in \cN(n) \iff (j, n) \in \cE$.
For simplicity of notation, let self-edges be included, i.e.,  $(n, n) \in \cE$ for all $n \in [N]$.
We assume that the network graph $\cG$ captures spillover effects in the following way. 
\begin{assumption}[Stable Neighborhood Treatment Value Assumption (SNTVA)]\label{assumption:network_sutva}
The potential outcome of measurement $t \in [T]$ for unit $n \in [N]$ under treatments $\bintv \in [D]^{N}$ is given by%
\begin{align*}
Y^{(\bintv)}_{t, n} = Y^{(\bintv_{\cN(n)})}_{t, n},
\end{align*}
where $\bintv_{\cN(n)} \in [D]^{|\cN(n)|}$ denotes the treatments assigned to the units in $n$'s neighborhood $\cN(n)$ for measurement $t$. 
That is, the potential outcome of unit $n$ depends on its neighbors' treatments but does not depend the treatment of any other 
unit $j \in [N] \setminus \cN(n)$.
\end{assumption}
See Figure \ref{fig:network_example} for an example of spillover and its network representation.
Several prior works on network interference also assume SNTVA, e.g., as the \emph{Neighborhood Interference Assumption (NIA)}
\citep{sussman2017elements}. 
It can be viewed as a particular instantiation of exposure mappings, as defined by \cite{aronow2017estimating}, and effective treatment functions (e.g., under the constant treatment response assumption) \citep{Manski13}. 

\begin{remark}
SNTVA only captures first-order spillover effects, i.e., assumes that the potential outcome of unit $n$ is only affected by the treatments of its \emph{immediate} neighbors. 
One could capture higher-order spillover effects by adding edges to $\cG$.  
The trade-off is that, as the number of edges in $\cG$ increases, the estimation bounds for the NSI estimator in Section \ref{sec:results} get correspondingly weaker.  
\end{remark}

\begin{remark}
Although we assume $\cG$ is an undirected graph, our results can be  adapted for directed graphs
by changing the definition of $\cN(i)$. 
When $\cG$ is directed, $j \in \cN(i)$ if and only if $(j, i) \in \cE$. 
\end{remark}

\subsection{Network latent-factor model} \label{sec:lf_model}
In this section, we introduce the model that we use to develop our estimator and formal results. 
\begin{assumption}\label{asm:model}
Let the potential outcome of measurement $t \in [T]$ for unit $n \in [N]$ under graph $\cG$ and treatments $\ba \in [D]^N$ be given by:
\begin{equation}
    \begin{aligned}
        Y^{(\bintv_{\cN(n)})}_{t, n} 
        &= 
        \left< \bu_{n, n} , \bw_{t , a_n} \right> 
        +
        \sum_{j \in \cN(n) \setminus n} \left< \bu_{j, n} , \bw_{t , a_j} \right>  + \epsilon_{t, n}^{(\ba_{\cN(n)})} , 
    \end{aligned}
    \label{eq:full_model}
\end{equation}
where $\bu_{\cdot , \cdot} \in \bbR^r$ and $\bw_{\cdot , \cdot} \in \bbR^r$ represent latent (unobserved) factors; 
$\epsilon_{t,  n}^{(\ba_{\cN(n)})}$ represents additive, idiosyncratic shocks,
and $r$ is the ``rank'' or model complexity. 
Further, we assume that $\bbE \big[ \epsilon_{t , i}^{(\ba_{\cN(i)})} \, | \, \LF \big] = 0 $, where
$\LF \defeq \big\{ \bu_{j , i} , \bw_{t, a} :  i, j \in [N] \, ,  \, t \in [T], \text{ and } a \in [D] \big\}.$
\end{assumption}
We make several remarks.
First, we note that Assumption \ref{asm:model} automatically satisfies Assumption \ref{assumption:network_sutva}.
Second, the latent factor $\bu_{j, n}$ captures the effect in the potential outcome $Y^{(\bintv_{\cN(n)})}_{t, n} $ due to the interaction between node $n$ and its neighbour $j$;
analogously $\bw_{t , a_j}$ captures the effect due to the treatment that neighbor $j$ receives (i.e., $a_j$) for measurement $t$.
Specifically, their effect is captured through the inner product $\left< \bu_{j, n} , \bw_{t , a_j} \right>$.
In this sense, the spillover effect of different neighbors in \eqref{eq:full_model} is additive. 
Lastly, \eqref{eq:full_model} can be equivalently written as 
\begin{gather}
    Y_{t, n}^{(\ba)} = Y_{t , n}^{(\ba_{\cN(n)})} 
    = \left< \tilde{\bu}_{n, \cN(n)} ,  \tilde{\bw}_{t, {\bintv_{\cN(n)}}} \right> + \epsilon_{t , n}^{(\ba_{\cN(n)})} , 
    \label{eq:abbrev_model}
\end{gather}
where 
\ifarxiv
\begin{align*}
	 \tilde{\bu}_{n, \cN(n)} 
&\defeq [
\bu_{\cN_1(n), n}^\top \, , \,
\hdots \, ,  \,
\bu_{\cN_{|\cN(n)|}(n), i}^\top 
]^\top  ,
\\
 \tilde{\bw}_{t , \ba_{\cN(n)}} 
&\defeq [
\bw_{t, a_{\cN_1(n)}}^\top \, , \,
\hdots \, ,  \,
\bw_{t, a_{\cN_{|\cN(n)|}(n)}}^\top 
]^\top .
\end{align*}
\else
$ \tilde{\bu}_{n, \cN(n)} 
    \defeq [
        \bu_{\cN_1(n), n}^\top \, , \,
        \hdots \, ,  \,
        \bu_{\cN_{|\cN(n)|}(n), i}^\top 
    ]^\top $ and
$ \tilde{\bw}_{t , \ba_{\cN(n)}} 
    \defeq [
        \bw_{t, a_{\cN_1(n)}}^\top \, , \,
        \hdots \, ,  \,
        \bw_{t, a_{\cN_{|\cN(n)|}(n)}}^\top 
    ]^\top$.
    \fi
Here, $\cN_i(n)$ refers to $i$-th neighbor of $n$.
\eqref{eq:abbrev_model} is reminiscent of classical interactive fixed effects models studied in the literature.
Indeed, we can think of $\tilde{\bu}_{n, \cN(n)}  \in \mathbb{R}^{r |\cN(n)|}$ and $\tilde{\bw}_{t, a_{\cN(n)}} \in \mathbb{R}^{r |\cN(n)|}$ as the \emph{network-adjusted} latent factors
and $r |\cN(n)| \in \bbN_{> 0}$ as denoting the \emph{network-adjusted} ``rank'' (note that $r|\cN(n)|$ is actually an upper bound on the model's rank, but we will refer to it as the network-adjusted rank for convenience).

\subsection{Examples of latent-factor model}
We discuss how examples of latent factor models 
previously studied in the literature are captured by the model we propose in Assumption \ref{asm:model}.
Further, we discuss how additive non-linear latent factor models can be approximated by the linear additive model we propose.

\begin{example}\label{ex:f_no_spillover}
Consider a setting with \emph{no spillover effects}, i.e., $\cN(n) = \{ n \}$ for all $n \in [N]$.
Then, the latent factor model in \eqref{eq:full_model} reduces to 
\begin{align}\label{eq:no_int_model}
Y_{t, n}^{(\ba)}  = Y_{t, n}^{(a_n)} 
                &=  \langle \bu_{n, n} , \bw_{t , a_n} \rangle + \epsilon_{t, n}^{(a_n)}, 
\end{align}
This recovers the model considered in \citep{agarwal2020synthetic}. 
As explained in \citep{agarwal2020synthetic}, this also captures the models considered in \citep{abadie2021using} and \citep{arkhangelsky2019synthetic}.
\end{example}

\begin{example}\label{ex:linear_model}
There are several prior works that assume that network interference is additive.
For instance, consider the model proposed by \citet{yu2022graph} in which $D = 2$, i.e., the treatments are binary, denoted by $\{0, 1\}$, and
\begin{align}
    Y_{n}^{(\ba)} =
    Y_{n}^{(\ba_{\cN(n)})}
    &= u_{0,n} + u_{n,n} a_n + \sum_{j \in \cN(n) \setminus n} u_{j, n} a_j + \epsilon_{n}^{(\ba_{\cN(n)})}, \label{eq:yu_model}
\end{align}
where $u_{0,n}, u_{n, n}, u_{j, n}, \epsilon_{n}^{(\ba_{\cN(n)})} \in \Rb$. 
One can verify that \eqref{eq:yu_model} can be recovered from \eqref{eq:full_model} by taking $r = 1$; $T = 1$ (i.e., no index $t$); $w_{a} = a$; and there is an auxiliary node $0$ for which $a_0 = 1$ and $0 \in \cN(n)$ for all $n \in [N]$.

That is, both our model \eqref{eq:full_model}  and \eqref{eq:yu_model} assume that spillover is additive, 
and in order to exploit the structure across measurements $t$ that exists in panel data, we extend \eqref{eq:yu_model} by: (i) allowing for multiple measurements $t$, and (ii) assuming that $u_{j, n} a_j$ in \eqref{eq:yu_model} has the measurement-dependent latent-factor representation $\left< \bu_{j, n}, \bw_{t , a_j} \right>$.
These repeated measurements are exactly what lets us create personalized counterfactual trajectories per units and implicitly correct for unobserved confounding.
\end{example}

\begin{example}\label{ex:f_spillover}
Consider a setting where network interference is additive but the effect of the latent factors is {\em non-linear}. 
Precisely, consider the following variation of \eqref{eq:full_model}:
\begin{equation}
    \begin{aligned}
        Y_{t, n}^{(\ba)} 
        =
        Y_{t, n}^{(\ba_{\cN(n)})}
        &= 
        h(\bu_{n, n} , \bw_{t , a_n})
        +
        \sum_{j \in \cN(n) \setminus n} g(\bu_{j, n} , \bw_{t , a_j})  + \epsilon_{t, n}^{(\ba_{\cN(n)})} , 
    \end{aligned}
    \label{eq:nl_full_model}
\end{equation}
where $h, g: \bbR^r \times \bbR^r \to \bbR$ are potentially non-linear functions. 
If the latent factors take value in a bounded domain, say ${\cal C} \subset \bbR^r$, and $h, g$ are Lipschitz continuous (or more generally {\em smooth}), then it can be argued that (see Theorem 1 by \citet{shah2020sample} for example) for any given $\delta > 0$, there is some $r' = r'(\delta)$ large enough and choice of functions $\{ \phi_k, \psi_k, \phi'_k, \psi'_k : \bbR^r \to \bbR, ~k\leq r' \}$ such that
\ifarxiv
\begin{align*}
&\Big| h(\bu, \bw) - \sum_{k=1}^{r'} \phi_k(\bu)\psi_k(\bw) \Big|  \leq \delta ,
\\
&\Big| g(\bu, \bw) - \sum_{k=1}^{r'} \phi'_k(\bu)\psi'_k(\bw) \Big| \leq \delta ,
\end{align*}
\else
$
\Big| h(\bu, \bw) - \sum_{k=1}^{r'} \phi_k(\bu)\psi_k(\bw) \Big|  \leq \delta$
and
$\Big| g(\bu, \bw) - \sum_{k=1}^{r'} \phi'_k(\bu)\psi'_k(\bw) \Big| \leq \delta
$ 
\fi
for all $\bu, \bw \in {\cal C}$. Then, by setting
$\tilde{\bu} = [\phi_k(\bu): k \leq r']$, 
$\tilde{\bw} = [\psi_k(\bu): k \leq r']$, 
$\tilde{\bu}' = [\phi'_k(\bu): k \leq r']$, 
$\tilde{\bw}' = [\psi'_k(\bu): k \leq r']$, 
it follows that \eqref{eq:nl_full_model} is pointwise $\delta$-approximated as a linear latent factor model as given in \eqref{eq:full_model}, with $\bu, \bw$ appropriately replaced by
$\tilde{\bu}, \tilde{\bw}, \tilde{\bu}',$ and $\tilde{\bw}'$.
\end{example}

\subsection{Target causal estimand}

\begin{figure*}[t]
\centering
\includegraphics[width=1\textwidth]{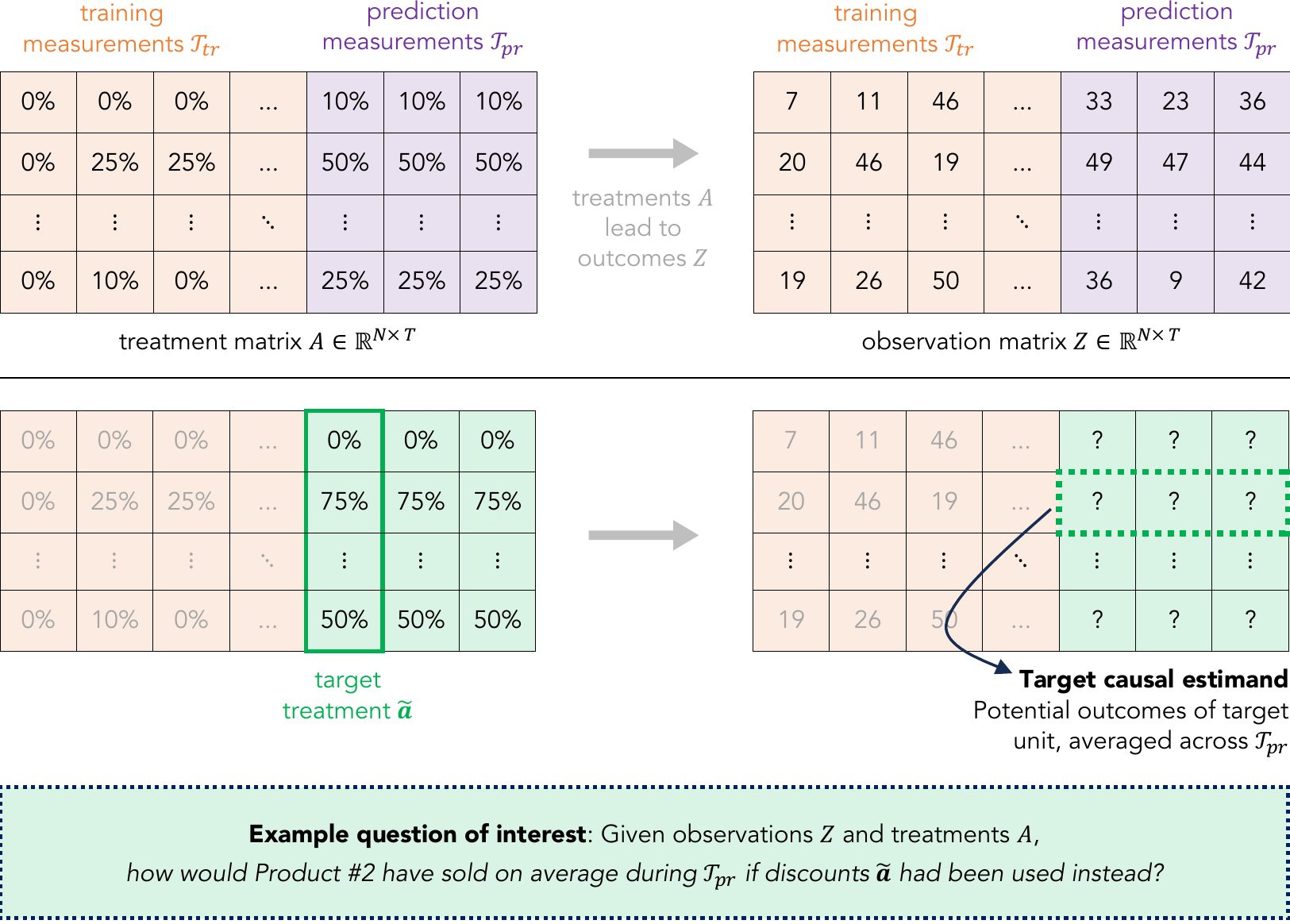}
\caption{
Illustration of our setup and target causal estimand. 
On the top-left is an example $N \times T$ treatment matrix $A$, 
split into the training and prediction measurements $\cT_\pre$ and $\cT_{\post}$. 
On the top-right is the corresponding  observation matrix $Z$,
i.e., the $(n, t)$-th element of $Z$ is the outcome of unit $n$ at measurement $t$ under treatments $\ba^t$. 
The bottom-left gives the counterfactual treatment matrix, i.e., the treatments during $\cT_\pre$ remain intact and the counterfactual prediction treatment is given by $\tilde{\ba}$. 
Lastly, on the bottom-right are the potential outcomes of interest under  $\tilde{\ba}$. 
The target causal estimand is the average of potential outcomes of a specific unit under $\tilde{\ba}$ across $\cT_\post$. 
}
\label{fig:panel_data}
\end{figure*}

Recall that we consider the panel data setting in which we observe $T$ measurements (e.g., a time series) for every unit. 
Let $a^t_n \in [D]$ denote the treatment assignment for unit $n$ at measurement $t$;
 $\ba^t \in [D]^N$ denote the vector of treatment assignments for all $N$ units at  $t$;
and $A \defeq [ \ba^{1}, \ba^{2}, \hdots, \ba^{T} ] \in [D]^{N \times T}$ denote the sequence of treatment assignments across all $N$ units and $T$ measurements. 
Note that the treatment assignments $A$ are observed, and the potential outcome $Y_{t,n}^{(\ba^t)}$ is observed for every unit $n \in [N]$ and measurement $t \in [T]$.
We denote the \emph{observed} outcomes for unit $n$ at measurement $t$ by $Z_{t,n} = Y^{(\ba^{t})}_{t,n}$ for all $t \in [T]$ and the matrix of observations by $Z \in \bbR^{T \times N}$. 

To define the target causal estimand of interest, let $\cT_{\post} \subseteq [T]$ refer to a subset of measurements for which we would like to make counterfactual predictions; let $T_{\post} = |\cT_{\post}| \leq T$.
To simplify notation, we assume without loss of generality that the treatment assignments are fixed across the measurements in $\cT_{\post}$, i.e., $A[:, t]  = \ba^\post \in [D]^N$ for all $t  \in \cT_\post$ (see Remark \ref{rem:const_pred_treatment}).

For any given unit $n$ and target treatment assignment $\tilde{\ba} \in [D]^N$, our goal is to estimate the individual potential outcome averaged over the prediction period:
\begin{align}
\IPO{n} & = \frac{1}{T_{\post}} \sum_{t \in \cT_{\post}} \Ex\Big[Y^{(\tilde{\ba}_{\cN(n)} )}_{t , n} \mid \LF \Big],\label{eq:estimand}
\end{align}
using observations $Z$, 
where we condition on the latent factors, $\LF$.
Under Assumption \ref{assumption:network_sutva}, the outcome of unit $n$ depends only on the treatments applied to $\cN(n)$, i.e., on $\tilde{\ba}_{\cN(n)}$ rather than the
entire $\tilde{\ba}$. 
See Figure \ref{fig:panel_data} for an illustration of our target causal estimand and observation pattern.

Note that our results are given for any $T_\post \geq 1$. 
That is, we show identifiability and finite-sample consistency even for point estimates (i.e., for $T_\post = 1$). 
\begin{remark}\label{rem:const_pred_treatment}
Our assumption $A[:, t]  = \ba^\post \in [D]^N$ for all $t  \in \cT_\post$ is without loss of generality. 
First, our work does not allow for spillover across measurements, i.e.,  $Y_{t,n}^{(\ba)}$ does not depend on treatments other than those assigned at $t$. 
We can therefore extract measurements in $\cT_{\post}$ that share the same treatment $\ba^\post$, i.e., we can redefine the prediction set as $\cT_\post'$ so that $\ba^t = \ba^\post$ for all $t \in \cT_\post'$.
We can repeat this for all unique prediction treatments and apply NSI separately to each. 
Further, we note that our consistency and normality results allow for $T_\post = 1$, and so our results go through even if we have a different target prediction treatment for every measurement in $\cT_{\post}$.
\end{remark}

%% file: sections/estimator.tex
\section{Network Synthetic Interventions (NSI) Estimator}\label{sec:estimator}
We now describe our estimator for the estimand of interest \eqref{eq:estimand}, which we term {\em Network Synthetic Intervention} (NSI).
It can be seen as a natural extension of the Synthetic
Interventions (SI) estimator \citep{agarwal2020synthetic}, which is itself a generalization of Synthetic Controls (SC) \citep{abadie2021using} estimator,  to settings in which there is \emph{network interference}. 
For the remainder of this work, we fix the unit $n$ and counterfactual treatment assignment $\cfA$ of interest.

\subsection{Donor set}
To define the NSI estimator, we introduce some necessary concepts.
First, let $\cT_{\pre} \subset [T]$ denote a subset of the measurements known as {\em training} measurements.
Without loss of generality, let $\cT_{\pre} \coloneqq \{1, 2, \hdots, T_{\pre}\}$, $\cT_{\post} \coloneqq \{T_{\pre} + 1, \dots, T\}$, $T_{\pre} \defeq |\cT_{\pre}|$, and $T_{\post} \defeq |\cT_{\post}|$. 
We note that $\cT_{\pre}$ does not need to be $[T] \backslash \cT_{\post}$ but we keep it as such to simplify the exposition.
Recall that $A \in [D]^{N \times T}$ denotes the treatment assignments to the various units over time. 
Let $A^\pre \defeq A[: , \cT_{\pre}]$ and $A^\pre_n \defeq A[\cN(n) , \cT_{\pre}]$.
Next, we introduce the notion of  a ``donor set."
\begin{definition}[Donors]\label{def:node_donors}
For a given unit $n \in [N]$ and counterfactual treatment assignment $\cfAn \in [D]^{|\cN(n)|}$, we consider $i \in [N] \setminus \{n\}$ a ``donor unit'' if the following conditions hold:
\begin{enumerate}
\item $|\hspace{1pt} \cN(i)| = | \hspace{1pt}  \cN(n) |$, i.e., donor unit $i$ has the same number of neighbors as unit $n$.
\item There exists a permutation $\pi_i : [\cN(i)] \to [\cN(i)]$ such that:
\begin{enumerate}
\item $A[\pi_i(\cN(i)),\cT_{\pre}] = A[\cN(n),\cT_{\pre}]$, i.e., the training treatment assignment of donor unit $i$ and its neighbors match that of unit $n$ and its neighbors, once permuted by $\pi_i$.
\item $\ba^\post_{\pi_i(\cN(i))} = \cfAn$, i.e., the prediction treatment assignment of donor unit $i$ and its neighbors matches the target counterfactual treatment assignment $\cfAn$, once permuted by $\pi_i$.
\end{enumerate}
\end{enumerate}
\end{definition}
For the remainder of this work, we fix the unit $n$ and counterfactual treatments $\cfAn$ of interest and let $\cI^{(n)} \subset [N] \setminus \{ n \}$ denote the corresponding set of donors. 
One can think of the donor set $\cI^{(n)}$ as units whose \emph{observed} outcomes can be used to estimate the unobserved \emph{potential} outcome of unit $n$ under the counterfactual treatments of interest.

\subsection{NSI Estimation procedure}\label{sec:estimation_proc}

Recall that $Z \in \bbR^{T \times N}$ denotes the matrix of observations. 
We define
$\bz_{\pre, n} \defeq  Z[\cT_{\pre}, n] \in \bbR^{T_{\pre}}$,
$Z_{\pre, \cI^{(n)}}  \defeq Z[\cT_{\pre}, \cI^{(n)}] \in \bbR^{T_{\pre} \times | \cI^{(n)} |} $, and
$Z_{\post, \cI^{(n)}}  \defeq
Z[\cT_{\post}, \cI^{(n)}]\in \bbR^{T_{\post} \times | \cI^{(n)} |}$.
NSI takes in one hyperparameter $\kappa \in [\min(T_{\pre}, |\cI^{(n)}|)]$ and proceeds in two steps, as follows. 
\ifarxiv 
\\
\else
\medskip
\fi

\noindent 
\emph{1. Point estimate}.
Let $\{(\hat{s}_\ell, \hat{\boldsymbol{\mu}}_\ell, \hat{\boldsymbol{\nu}}_\ell)\}_{\ell = 1}^{\min(T_\pre, |\cI^{(n)}|)}$ denote the set of singular values, left singular vectors, and right singular vectors for the observed matrix $Z_{\pre, \cI^{(n)}}$, where $\hat{s}_1 \geq \hat{s}_2 \geq \hdots \geq \hat{s}_{\min(T_\pre, |\cI^{(n)}|)} \geq 0$. 
The NSI estimator produced a point estimate as follows:
\begin{align}\label{eq:NSI_estimator_point}
\hIPO{n} &= \frac{1}{T_{\post}} {\bf 1}^T 
Z_{\post, \cI^{(n)}} ~\hat{\bbE}  [ Z_{\pre, \cI^{(n)}} | \LF , \cO]^+ \bz_{\pre, n}.
\end{align}
For the given hyperparameter $\kappa$,
$
\hat{\bbE}  [ Z_{\pre, \cI^{(n)}} | \LF , \cO]^+
 = \sum_{\ell = 1}^\kappa \frac{1}{\hat{s}_\ell}
     \hat{\boldsymbol{\nu}}_\ell
     \hat{\boldsymbol{\mu}}_\ell^\top,
$
can be viewed as an estimate of $\bbE [ Z_{\pre, \cI^{(n)}} | \LF , \cO]$  that is obtained via hard singular value thresholding, where only the top $\kappa$ components are preserved.
This estimator, which begins with singular value thresholding, has been shown to be equivalent to principal component regression \citep{PCR_1, PCR_2}.
\ifarxiv 
\\
\else
\medskip
\fi

\noindent 
\emph{2. Confidence interval}.
Let $\hat{\boldsymbol{\alpha}} = \hat{\bbE}  [ Z_{\pre, \cI^{(n)}} | \LF , \cO]^+  \bz_{\pre, n}$.
Then, the $\textsc{CI}$-percent confidence interval can be constructed as:
\begin{align*}
\IPO{n}
&\in 
\left[
    \hIPO{n}
    \pm
    \frac{\Phi^{-1}(\textsc{CI}/100) \hat{\sigma} \norm{\hat{\boldsymbol{\alpha}}}_2}
    {\sqrt{T_{\post}}}
\right],
\end{align*}
where $\Phi$ denotes the cumulative distribution function (CDF) of the standard normal distribution, $\Phi^{-1}$ is the inverse CDF, and
\ifarxiv
$$\hat{\sigma}^2 
= 
\frac{1}{T_{\pre}}
\norm{
	\bz_{\pre, n}  - Z_{\pre, \cI^{(n)}} \hat{\boldsymbol{\alpha}} 
}_2^2 ,$$
which can be interpreted as the in-sample prediction error of the NSI estimator.
\else
$\hat{\sigma}^2 
	= 
	\frac{1}{T_{\pre}}
	\norm{
		\bz_{\pre, n}  - Z_{\pre, \cI^{(n)}} \hat{\boldsymbol{\alpha}} 
	}_2^2$, which can be interpreted as the in-sample prediction error of the NSI estimator.
	\fi

\subsection{Discussion of NSI}\label{sec:discussion_NSI}

We briefly provide intuition for the NSI estimator, 
then compare it to the traditional Synthetic Control \citep{abadie2021using} and Synthetic Interventions \citep{agarwal2020synthetic} estimators. 
\ifarxiv 
\\
\else
\medskip
\fi

\noindent
{\bf NSI linearly combines donor outcomes.}\label{sec:linear_comb}
NSI begins by finding units, called ``donors,'' whose outcomes can be used to estimate the potential outcomes of unit $n$. 
In Section \ref{sec:results}, under suitable assumptions, we establish that the expected potential outcome of unit $n$ can be expressed as a linear combination of the expected outcomes of the donor units, i.e., 
\begin{align}
\Ex[Y_{t, n}^{(\tilde{\ba})}] &= \sum_{j \in \cI^{(n)}} \alpha_j \cdot 
\Ex[Y_{t, j}^{(\ba)}],
\label{eq:NSI_linear}
\end{align}
where $\alpha_j \in \bbR$ (note that $\alpha_j$ can be negative). 
NSI can be viewed as a method for estimating the coefficients $\{ \alpha_j \}$.
More precisely, recall that $\hat{\boldsymbol{\alpha}} = \hat{\bbE}  [ Z_{\pre, \cI^{(n)}} | \LF , \cO]^+  \bz_{\pre, n}$. 
Then, the NSI estimator \eqref{eq:NSI_estimator_point} can be rewritten as 
\ifarxiv
$$
\hIPO{n} = \frac{1}{T_{\post}} \sum_{t \in \cT_{\post}} \sum_{j \in \cI^{(n)}} \hat{\alpha}_j Z[t, j],
$$
\else
$
\hIPO{n} = \frac{1}{T_{\post}} \sum_{t \in \cT_{\post}} \sum_{j \in \cI^{(n)}} \hat{\alpha}_j Z[t, j],
$
\fi
which is precisely what would follow from expressing the target causal estimand \eqref{eq:estimand} using \eqref{eq:NSI_linear}.
\ifarxiv 
\\
\else
\medskip
\fi

\begin{figure*}[t]
\centering
\includegraphics[width=\textwidth]{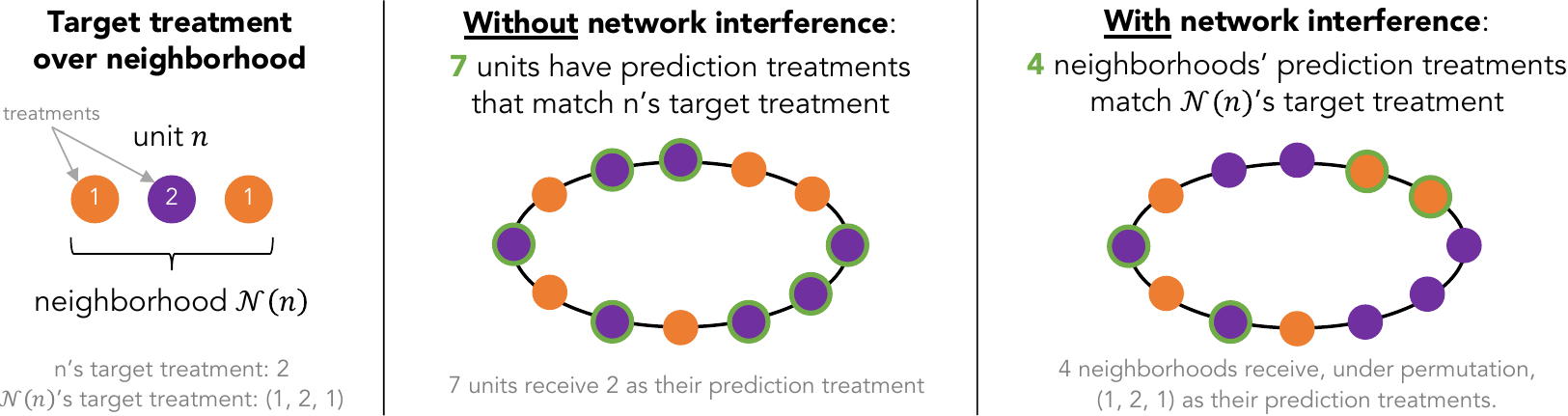}
\caption{One of the main differences between the NSI estimator and previous estimators (such as SC and SI) is the choice of donors. To illustrate this point, consider a unit $n$ whose target treatment is $2$, as given in the left panel. 
Suppose that $\cG$ is a ring graph and the prediction treatments are assigned as given in the middle and right panels. 
Then, under SC and SI, there are 7 units (with green borders) whose prediction treatments match unit $n$'s target treatment (middle panel).
Under NSI, however, the donor requirements are stricter. 
Specifically, NSI looks at the target treatment $(1, 2, 1)$ across all neighbors $\cN(n)$, as given in the left panel. 
The only units $j$ that could be considered as potential donors must have neighborhood $\cN(j)$ that receive prediction treatments $(1, 2, 1)$ subject to permutation (recall Definition \ref{def:node_donors}). 
As shown on the right, only 4 units (with green borders) meet this requirement. }
\label{fig:nsi_donors_ring}
\end{figure*}

\noindent
{\bf Comparing NSI to SI \& SC: Choosing donors appropriately.}
NSI is a generalization of SI and SC. 
The key difference between SC/SI and NSI is the choice of donor units.
In SC/SI, a valid donor unit only needs to undergo the same training and prediction treatments as the training and target prediction treatments of $n$. 
In NSI, there are more stringent requirements on donors, as given by Definition \ref{def:node_donors} and illustrated in Figure \ref{fig:nsi_donors_ring}. 
These more stringent requirements on how donors are chosen are how NSI removes the bias that SI and SC suffer from when there is spillover. 

Under the appropriate choice of donors, the way that the linear model \eqref{eq:NSI_linear} is learned can depend on modeling assumptions. 
NSI uses principal component regression (PCR), which is motivated by \cite{agarwal2020synthetic}). 
However, other estimators, such as convex regression \citep{abadie2021using} and variants thereof can also be used. 
In Section \ref{sec:results}, we detail the conditions under which PCR produces consistent and asymptotically normal estimates.

%% file: sections/theoretical_results.tex
\section{Formal results} \label{sec:results}

In this section, we provide formal results for the NSI estimator. 
We characterize conditions under which $\IPO{n}$ can be identified.
We then establish that NSI provides a consistent estimate of $\IPO{n}$ and its estimation error is asymptotically normal, justifying the confidence interval given in Step 3 of Section \ref{sec:estimation_proc}.
As before, we restrict our attention to a specific unit $n$ and target counterfactual treatment assignment $\cfAn$. 
All proofs are given in Appendices \ref{app:helpers}-\ref{app:proofs}. 

\subsection{Identification Result}\label{sec:id}
We now discuss the key assumptions we make about the intervention assignments $A$.
We begin with an assumption on the treatment assignment. 
\begin{assumption}[Conditional exogeneity]\label{asm:conditional_exo}
For all $n \in [N]$, $t \in [T]$, and $\ba \in [D]^N$, we have that $Y^{(\ba_{\cN(n)})}_{n, t} \perp \cO \, | \, \LF$.
\end{assumption}
Given Assumption \ref{asm:model}, this conditional independence is equivalent to assuming that $\epsilon_{t, n}^{(\ba_{\cN(n)})} \perp \cO \, | \, \LF $.
Similar conditions of ``selection on latent factors'' have been considered in the literature (see \citep{agarwal2020synthetic}
and discussion therein).
In this work, we analogously require ``selection on network-adjusted latent factors.'' 
We make two additional assumptions, as follows.

\begin{assumption}[Linear span inclusion]\label{asm:linear_span}
Given a unit $n \in [N]$ and counterfactual treatments $\cfAn \in [D]^{|\cN(n)|}$ of interest, consider the donor set $\cI^{(n)}$. 
We assume the treatment assignment $A$ is such that $\cI^{(n)}$ is non-empty and that $\tilde{\mathbf{u}}_{n, \cN(n)}$ lies in the linear span of $\{\tilde{\mathbf{u}}_{i, \pi_i( \cN(i))}\}_{i \in \cI^{(n)}}$, where  $\{ \pi_i \}_{i \in \cI^{(n)}}$ is defined in Definition \ref{def:node_donors}.
That is, there exists $\boldsymbol{\lambda} \in \bbR^{|\cI^{(n)}|}$ such that 
\begin{align*}
\tilde{\mathbf{u}}_{n, \cN(n)} = \sum_{i \in \cI^{(n)} }  \lambda_{i}  \tilde{\mathbf{u}}_{i, \pi_k( \cN(i))}.
\end{align*} 
\end{assumption}

\begin{assumption}[Subspace inclusion] \label{asm:subspace_inclusion}
Assume that the rowspace of $\bbE \big[  Z_{\post, \cI^{(n)} }  \big| \, \LF, \cO \big]$
lies within the rowspace of $\bbE \big[  Z_{\pre, \cI^{(n)} }  \big| \, \LF, \cO \big]$.
\end{assumption}
We discuss Assumptions \ref{asm:linear_span}-\ref{asm:subspace_inclusion} in Sections \ref{sec:assumptions} and \ref{sec:validity_test}. 
We show that NSI's confidence interval indicates the degree to which Assumption \ref{asm:linear_span} holds, and we provide a way to test for Assumption  \ref{asm:subspace_inclusion} in Section \ref{sec:validity_test}.

Before stating our identification result, 
we first recall identifiability for an arbitrary estimation problem of interest.
In the definition below, $P_\theta \in \cP$ refers to data distribution under model parameters $\theta$, i.e., $P_\theta$ describes how the data behaves under model $\theta$. 
\begin{definition}[Identifiability]\label{def:identifiability}
Let $\theta \in \Theta$ denote the ground-truth model parameters and $\cP = \{ P_{\theta'} : \theta' \in \Theta \}$ denote the set of possible data distributions.
Let $f : \Theta \rightarrow \bbR$ denote the estimand of interest and $P_\theta \in \cP$ denote the data generating distribution parameterized by $\theta$. 
Then, $f(\theta)$ is \emph{identifiable}  if there exists a function $g : \cP \rightarrow \bbR$ such that $f(\theta) = g(P_\theta)$, 
i.e., the estimand can be written as a function of the data distribution. 
\end{definition}
Identifiability implies that if $f(\theta) \neq f(\theta')$, then
$g(P_\theta) \neq g(P_{\theta'})$; otherwise, $f(\theta) = g(P_\theta) = g(P_{\theta'}) = f(\theta')$. 
In other words, the estimand is identifiable if we can compute it \emph{exactly} when given access to the full data distribution, which is necessary for estimation from (noisy) data to be possible.

In our setting, $\theta = \LF$ denotes the latent factors and $f(\theta) = \IPO{n}$ is the target causal estimand, which we note can be written solely as a function of the latent factors.
Recall that the quantities $(A, \cG, \cT_{\pre}, \cT_\post)$ are observed and known.
Let $P_{\theta}$ denote the joint distribution over the matrices of observed outcomes $Z[\cT_\pre,:]$ and $Z[\cT_\post,:]$. 
Note that, by Assumption \ref{asm:model}, the distribution of $Z[\cT_\pre,:]$ and $Z[\cT_\post,:]$ is given by the latent factors $\LF$, the treatment assignment $A$, and the random variables $\epsilon_{t, n}^{(\ba_{\cN(n)})}$.
We now show that  $\IPO{n}$ is identifiable under \eqref{eq:full_model} and Assumptions \ref{asm:model}-\ref{asm:subspace_inclusion}.
\begin{theorem}[Identification]\label{thm:ID}
If Assumptions \ref{assumption:network_sutva}-\ref{asm:subspace_inclusion} hold, then
\begin{align} \label{eq:ITE_identif}
\IPO{n} = \frac{1}{T_{\post}} {\bf 1}^T_{T_\post} ~\bbE [ Z_{\post, \cI^{(n)}} \, |  \, \LF, \cO ] ~\bbE  [ Z_{\pre, \cI^{(n)}} | \LF , \cO]^+ ~\bbE  [ \bz_{\pre, n} | \LF , \cO],
\end{align}
where ${\bf 1}_{T_\post}$ is the all ones vector of length ${T_\post}$, and the set $\cI^{(n)}$ is defined in Definition \ref{def:node_donors}.
This implies that $\IPO{n}$ is identifiable by Definition \ref{def:identifiability}.
\end{theorem}
Under Definition \ref{def:identifiability}, $g$ is given by \eqref{eq:ITE_identif}.
Note only the first moments of $P_\theta = P_{\LF}$ are needed. 
NSI estimates $\IPO{n}$ by replacing the expectations in \eqref{eq:ITE_identif} with the corresponding empirically observed quantities and smoothing out the pseudoinverse using hard singular value thresholding as given by \eqref{eq:NSI_estimator_point}.
For the purposes of the analysis, we denote 
\begin{align}\label{eq:linear_coeff}
\boldsymbol{\alpha}  = \bbE  [ Z_{\pre, \cI^{(n)}} | \LF , \cO]^+ \bbE  [ \bz_{\pre, n} | \LF , \cO].
\end{align}

\subsection{Consistency and asymptotic normality}\label{sec:thm_consistency}
Next, we give conditions under which the NSI estimator achieves finite-sample consistency and asymptotic normality. 
Let $r_{\pre} \in [ r |\cN(n)| ]$ be the rank of $\bbE \big[  Z_{\pre, \cI^{(n)}}  | \, \LF , \cO \big]$, $s_1 \geq \hdots \geq s_{r_{\pre}} \geq 0$ denote its singular values, and $R_{\pre} \in \bbR^{|\cI^{(n)}| \times r_{\pre}}$ denote its right singular vectors.

\begin{assumption}[Sub-Gaussian noise] \label{asm:subG_noise}
Conditioned on $\LF$, we assume that, for all $i \in [N]$, $t \in [T]$, and $\ba \in [D]^N$, $\epsilon^{( \ba_{\cN(i)})}_{t, i}$ are independent, sub-Gaussian random variables with $\text{Var}( \epsilon^{(\ba_{\cN(i)})}_{t, i}  | \LF) = \sigma^2$ and that $\bnorm{ \epsilon^{(\ba_{\cN(i)})}_{t, i} | \, LF}_{\psi_2} \leq \xi \sigma$ for some constant $\xi > 0$.
\end{assumption}

\begin{assumption}[Boundedness] \label{asm:bounded}
We assume that 
$\bbE \big[Y^{(\ba_{\cN(j)})}_{t, i} \big| \,  \LF \big] \in [-1, 1]$ for all $i \in  [N]$, $t \in [T]$.
\end{assumption}

\begin{assumption}[Well-balanced spectrum] \label{asm:rank_balanced_singular_vals}
For parameters $\xi', \xi'' > 0$,
we assume $s_{r_{\pre}} / s_1 \geq \xi'$ and $\bnorm{\bbE \big[  Z_{\pre, \cI^{(n)} }  \big| \, \LF, \cO \big] }_F^2 \geq \xi'' T_{\pre} |\cI^{(n)}| $, where $\cI^{(n)}$ is defined in Definition \ref{def:node_donors}.
\end{assumption}
In Section \ref{sec:experimental_proc}, we prove that in the setting of $d$-regular graphs and where the latent factors are sampled independently from a Gaussian distribution, the parameters $\xi'$ and $\xi''$ are inverse polynomials of $r$ and $d$.
\begin{assumption}[Sufficient number of components]\label{asm:components}
We assume that $\kappa = r_{\pre}$, where $\kappa$ is defined in Section \ref{sec:estimator} and $r_{\pre} \leq r | \cN(n)|$.
\end{assumption}

The following results establish that NSI is consistent and asymptotically normal. 
\begin{theorem}[Finite-sample consistency]
\label{thm:finite_sample_consistency}
Let Assumptions \ref{assumption:network_sutva}-\ref{asm:components} hold. %
Then, 
\begin{align*}
    &\left| 
    \hIPO{n} - \IPO{n}
    \right| 
    \\
    & \hspace{20pt}
    = O_P \left( 
    	\log( T_\pre | \cI^{(n)} | )
    	\left( 
    		\frac{r_\pre^{3/4}}{(\xi''')^{3/2} T_\pre^{1/4} }
    		+
    		\frac{r_\pre^2}{(\xi''')^4 } 
    		{\max} \left(
    			\frac{1}{\sqrt{T_\pre}} , 
    			\frac{1}{ \sqrt{| \cI^{(n)} | } } ,
    			\frac{ \sqrt{| \cI^{(n)} | } }{T_\pre^{3/2}}
    		\right)
    	\right) 
    	\right) ,
\end{align*}
where $\xi''' = \xi' \sqrt{\xi''}$ and $\xi' , \xi''$ are defined in Assumption \ref{asm:rank_balanced_singular_vals}.

\end{theorem}
Theorem \ref{thm:finite_sample_consistency} indicates that, under the stated conditions, 
NSI is consistent. 
Specifically, 
for fixed $r$, $d$, and $r_{\pre} \leq r (d+1)$, 
the estimation error of NSI approaches $0$ as the number of training measurements $T_\pre$ and number of donors $|\cI^{(n)}|$ grow if $T_\pre = \omega(| \cI^{(n)} |^{1/3})$. 
Importantly, 
the number of prediction measurements $T_\post$ need not grow in order for NSI's estimation error to decay to $0$.

Let $\Delta = \hat{\boldsymbol{\alpha}} - \boldsymbol{\alpha}$, the estimation error of learning the linear weights that represent the outcomes of the target units in terms of the donor units (see \eqref{eq:linear_coeff}).
The following result establishes a general result that as long as $\Delta$ is decaying sufficiently quickly for any linear estimator (it does not have to be via principal component regression as we do), the NSI estimator is asymptotically normal. 
While Theorem \ref{thm:finite_sample_consistency} allows NSI to produce the point estimate in Step 1 of Section \ref{sec:estimator}, Theorem \ref{thm:asym_normality} justifies the confidence interval provided in Step 2. 
\begin{theorem}[Asymptotic normality]\label{thm:asym_normality}
Suppose Assumptions \ref{assumption:network_sutva}-\ref{asm:components} hold. 
Suppose  
\ifarxiv
$$\norm{\Delta}_2 = o_P \left(\min \left( 
\frac{\sigma \norm{\boldsymbol{\alpha}}_2}{\sqrt{T_\post | \cI^{(n)} |}} , 
\sqrt{\frac{\norm{\boldsymbol{\alpha}}_2}{\sigma}}
\right) \right).$$
\else
$\norm{\Delta}_2 = o_P \left(\min \left( 
\frac{\sigma \norm{\boldsymbol{\alpha}}_2}{\sqrt{T_\post | \cI^{(n)} |}} , 
\sqrt{\frac{\norm{\boldsymbol{\alpha}}_2}{\sigma}}
\right) \right)$. 
\fi
Then, conditioned on $\LF$ and $\cO$,
\begin{align*}
\frac{\sqrt{T_{\post}}}{\sigma \norm{\balphaperp}_2}
\left(
    \IPO{n}
    - 
    \hIPO{n}
\right)
\stackrel{d}{\rightarrow}
\cN(0, 1) ,
\end{align*}
as $T_{\pre}, T_{\post}, |\cI^{(n)}| \rightarrow \infty$. 
Moreover, the $\hat{\sigma}$ used to construct the NSI confidence interval in Step 3 of Section \ref{sec:estimation_proc} satisfies:
\begin{align*}
| \hat{\sigma}^2 - \sigma^2 |
= 
O_p \left(
    \frac{r_{\pre}}{\sqrt{T_{\pre}}}
    + 
    \frac{r_{\pre}^{2} 
    {\log(T_{\pre} |\cI^{(n)}|)}}{(\xi''')^4 \min(T_{\pre}, |\cI^{(n)}| )}
\right),
\end{align*}
where $\xi''' = \xi' \sqrt{\xi''}$  and $\xi' , \xi''$ are defined in Assumption \ref{asm:rank_balanced_singular_vals}.
\end{theorem}

\subsection{Assumptions \ref{asm:linear_span}, \ref{asm:subspace_inclusion}, and \ref{asm:components}}\label{sec:assumptions}

The key enabling conditions for Theorems \ref{thm:ID}-\ref{thm:asym_normality} are Assumptions \ref{asm:linear_span}, \ref{asm:subspace_inclusion}, and \ref{asm:components}.
In this section, we discuss these assumptions further. 
\ifarxiv 
\\
\else
\medskip
\fi

\noindent
{\bf Assumption \ref{asm:linear_span}.}
Recall from Section \ref{sec:discussion_NSI} that NSI can be interpreted as linearly combining the potential outcomes of donors under appropriately chosen coefficients, denoted by $\hat{\boldsymbol{\alpha}}$. 
The key enabling condition that makes linearly combining donor outcomes valid under our model \eqref{eq:full_model} is Assumption \ref{asm:linear_span}. 
The extent to which Assumption \ref{asm:linear_span} holds can be examined in two ways. 

First, recall that $\hat{\sigma}^2 = \frac{1}{T_{\pre}} \norm{ \bz_{\pre, n}  - Z_{\pre, \cI^{(n)}} \hat{\boldsymbol{\alpha}} }_2^2$ is a measure of how well NSI's linear fit explains the training data.
Recall further that the coefficients $\hat{\boldsymbol{\alpha}}$ are estimates of the coefficients $\boldsymbol{\alpha}$ that are used to combine donor outcomes.
As such, $\hat{\sigma}^2$ can be viewed as a statistic for Assumption \ref{asm:linear_span}, where a large $\hat{\sigma}^2$ suggests Assumption \ref{asm:linear_span} does not hold.
Since NSI's confidence interval scales with $\hat{\sigma}$ (see Step 2 of \ref{sec:estimation_proc}), how well Assumption \ref{asm:linear_span} holds is captured by NSI's confidence interval.

Second, since Assumption \ref{asm:linear_span} requires that unit $n$'s $\tilde{\bu}$-latent factor is contained in the span of the donors' $\tilde{\bu}$-latent factors and $\tilde{\bu}_{n, \cN(n)} \in \bbR^{r |\cN(n)|}$, at least 
$r |\cN(n)|$ donors are needed.
Suppose, for example, that all units' $\tilde{\bu}$-latent factors are drawn i.i.d. from a multivariate Gaussian. 
Then, Assumption \ref{asm:linear_span} holds almost surely if and only if there are at least $r |\cN(n)|$  donors (cf. Lemma \ref{lem:linear_span_gaussian}). 
Given an estimate $\bar{r}$ of $r$, one can therefore perform a simple sanity check that $|\cI^{(n)}| \geq \bar{r}|\cN(n)|$.
\ifarxiv 
\\
\else
\medskip
\fi

\noindent
{\bf Assumption \ref{asm:subspace_inclusion}.}
This condition ensures that the linear coefficients that NSI learns from the training observations generalize to the prediction task. 
In Section \ref{asm:subspace_inclusion}, we provide two validity tests to verify whether Assumption \ref{asm:subspace_inclusion} holds, i.e., whether the observations are sufficiently rich such that a generalizable model can be learned. 
To motivate the need for such tests, below we provide a simple example where Assumption \ref{asm:subspace_inclusion} does not hold.
Suppose that $D = 2$ (the treatments are binary).
Let
\begin{alignat*}{3}
B^{\pre, n}  
&= 
\Big[\, 
\Ind(A[\cN(n), \cT_\pre] = 1)  
\, , 
\quad 
\Ind(A[\cN(n), \cT_\pre] = 2)
\, 
\Big]
\in \{0, 1\}^{ |\cN(n)| \times 2 { T_\pre}}.
\nonumber 
\end{alignat*}
Intuitively, $B_{\pre, n}$ is an indicator matrix that tracks the training treatment assignment over $\cN(n)$.
\begin{proposition}\label{prop:SIA_linear_independence}
Suppose Assumption \ref{asm:model} holds.
Unless $\text{colrank}( B^{\pre, n} ) =| \cN(n) |$, there exist latent factors $\LF$ and target treatment assignments under which Assumption \ref{asm:subspace_inclusion} cannot hold. 
\end{proposition}
Proposition \ref{prop:SIA_linear_independence} shows that the diversity of treatment assignments (as captured by $B^{\pre, n} $) affects the feasibility of Assumption \ref{asm:subspace_inclusion}. 
We unpack this relationship in detail in the next section. 
Before doing so, we present a negative example  in which Assumption \ref{asm:subspace_inclusion} does not hold. 
\begin{example}
Suppose  $\ba^t= \mathbf{1}_N$ for all $t \in \cT_{\pre}$.
As such, $B^{\pre, n}  = [ [1]_{|\cN(n)| \times T_\pre} \, , \,  [0]_{|\cN(n)| \times T_\pre} ]$ and $\text{colrank}(B^{\pre, n} ) = 1  < |\cN(n)|$. 
One can show that Assumption \ref{asm:subspace_inclusion} does \emph{not} hold unless $\tilde{\ba}_{\cN(n)} = \mathbf{1}$ or $\mathbf{2}$. 
The reason Assumption \ref{asm:subspace_inclusion} does not hold is that all of $n$'s neighbors have only been observed under the \emph{same} treatment. 
As such, there is no way for NSI to estimate the potential outcome of $n$ under $\tilde{\ba}_{\cN(n)}^\top= [2, 1, 1, \hdots]$, for example, where only the first neighbor is treated.
It is impossible for NSI (or any estimator, for that matter) to disentangle the spillover of the first neighbor from that of any other neighbor because the training measurements only contain data in which all neighbors receive the same treatment. 
The validity tests in Section \ref{asm:subspace_inclusion} provide a way of testing for whether the treatment assignment and the observations during the training period are rich enough. 
\end{example}
\medskip

{
\noindent
{\bf Assumption \ref{asm:components}.}
This condition requires that the number of components used by NSI, given by $\kappa$ matches the rank $r_\pre$ of $\bbE[ Z_{\pre, \cI^{(n)}} | \LF, A ]$. 
Since $\bbE[ Z_{\pre, \cI^{(n)}} | \LF, A ]$ is unknown in practice, 
one must estimate $r_\pre$, 
which can be done by applying an elbow point (or knee point) method to the spectrum of the observed matrix $Z_{\pre, \cI^{(n)}}$ \citep{zack1977automatic, satopaa2011finding}.
There are other heuristics for setting $\kappa$, such as the universal thresholding method given in \citep{chatterjee2015matrix}. 
Alternatively, suppose we have an estimate $\bar{r}$ of the model ``rank'' $r$, defined in Section \ref{sec:setup}. 
By our  model \eqref{eq:full_model}, 
$r_\pre$ is upper bounded by $r | \cN(n) |$, 
which suggests that one can use the heuristic $\kappa = \bar{r} |\cN(n)|$. 
It also suggests that
one should always set $\kappa$ to be at least $|\cN(n)|$, 
assuming that $r \geq 1$. 
}

%% file: sections/validity_tests.tex
\section{Validity tests}\label{sec:validity_test}
We present two validity tests for Assumption \ref{asm:subspace_inclusion}, one of the key enabling assumptions of Theorems \ref{thm:finite_sample_consistency} and \ref{thm:asym_normality}.
The first test can be performed \emph{before} any data is collected. 
It tests for whether the treatment assignment in the training period is diverse enough relative to the target treatment assignment in the prediction period.
The second test can be performed only \emph{after} the data is collected and, as such, is a relatively stronger test. 
Proofs for this section can be found in Appendix \ref{app:proofs_tests}.

\subsection{Validity test \#1: Pre-Data Collection}
The first test can be run \emph{before} the prediction samples are collected.
\ifarxiv 
\\
\else
\medskip
\fi

\noindent
\textbf{TrainingTreatmentTest.}
This test takes in one hyperparameter $\bar{r} \in \bbN_{> 0}$, which is an estimate of the model ``rank'' $r$ (see Section \ref{sec:lf_model}).
If one does not have a good estimate, $\bar{r}$ can also be an upper bound on $r$.
In order to run this test, we first define several ``masking'' matrices. 
For a given treatment $a \in [D]$, let $B^{\pre}(a) \in \{0, 1\}^{N \times T_{\pre}}$ and $\bb^{\post}(a) \in \{0, 1\}^{N}$ be defined such that their $(i, t)$-th elements are
$
	B_{i t}^{\pre}(a) = \Ind(A^{\pre}_{i t} = a)$
and
$
\tilde{b}_{i}^{\post}(a) = \Ind(\tilde{a}_i  = a) .
$
That is, the $(i, t)$-th entry of $B^{\pre}(a)$ is $1$ if and only if unit $i$ at measurement $t$ receives treatment $a$ under the training treatments $A^{\pre}$.
Similarly, the $i$-th entry of $\tilde{\bb}^{\post}(a)$ is $1$ if and only if unit $i$ is assigned the target counterfactual treatment $a$ under $\tilde{\ba}$.
Let $B^{\pre}$ and $\tilde{B}^{\post}$ be the concatenated matrices across different treatments:
\begin{align*}
B^{\pre} &= [B^{\pre}(1) , B^{\pre}(2) , \hdots, B^{\pre}(D)]
\in \{0, 1\}^{N \times T_{\pre} D}, \\
\tilde{B}^{\post} &= [\tilde{\bb}^{\post}(1) , \tilde{\bb}^{\post}(2) , \hdots, \tilde{\bb}^{\post}(D)]
\in \{0, 1\}^{N \times  D}.
\end{align*}
For the hyperparameter $\bar{r}$, the NSI estimator passes the \textsc{TrainingTreatmentTest} if
\begin{enumerate}
\item $\text{columnspace}(\tilde{B}^{\post} [\cN(n), : ] ) \subseteq \text{columnspace}(B^{\pre}[\cN(n), : ])$, and 
\item for every $t \in \cT_{\pre}$, the treatment $A[\cN(n), t]$ is repeated at least $\bar{r} D$ times in training;
\end{enumerate} 
otherwise, it fails. 
\ifarxiv 
\\
\else
\medskip
\fi

\noindent
\textbf{Connecting the \textsc{TrainingTreatmentTest} to Assumption \ref{asm:subspace_inclusion}.}
The following result formalizes the relationship between the test and  Assumption \ref{asm:subspace_inclusion} under a natural data generating process.
\begin{proposition}\label{prop:SIA_rowspace}
	Suppose Assumption \ref{asm:model} holds and $\bar{r}  = r$. 
	Suppose that $u_{k, n, \ell} \iid p_u$ and  $w_{t, a, \ell} \iid p_w$ for all $k, n \in [N]$, $t \in [T]$, 
	 $a \in [D]$, and $\ell \in [r]$,
	where $p_u$ and $p_w$ are continuous and { non-degenerate  (i.e., the support of $p_u$ or $p_w$ does not have dimension less than $r$)}.
	If  \textsc{TrainingTreatmentTest} is passed, 
	Assumption \ref{asm:subspace_inclusion} holds {almost surely} for any $n$ and target treatment $\tilde{\ba}_{\cN(n)}$ of interest.
\end{proposition}

As such, \textsc{TrainingTreatmentTest} tests whether Assumption \ref{asm:subspace_inclusion} can hold under the training treatments for the given target treatment of interest.  
Intuitively, it requires that the training treatments are sufficiently diverse relative to $\tilde{\ba}_{\cN(n)}$. 
In Section \ref{sec:experimental_proc}, we provide an experiment design that guarantees \textsc{TrainingTreatmentTest} is passed for any $n$ and $\tilde{\ba}_{\cN(n)}$.
Note that the i.i.d. condition in Proposition \ref{prop:SIA_rowspace} can be relaxed, 
as long as the latent factors are always drawn from non-degenerate distributions. 
The condition on latent factors ensures that there is enough variation across units and time such that we can isolate the role that training treatment assignments plays in Assumption \ref{asm:subspace_inclusion} from the role that latent factors play.

\subsection{Validity test \#2: Post-Data Collection}
We now furnish a data-driven check for Assumption \ref{asm:subspace_inclusion} that we call the \textsc{SubspaceInclusionTest}. 
This test can be run only \emph{after} the training and prediction samples are collected as opposed to the \textsc{TrainingTreatmentTest} test, which can be run beforehand.
\ifarxiv 
\\
\else
\medskip
\fi

\noindent
\textbf{SubspaceInclusionTest.}
The test takes in three hyperparameters: $\kappa$, $\kappa'$, and $\gamma$. 
Note that we overload $\kappa$ (which also appears in Section \ref{sec:estimator}) because both instances refer to an estimate of the rank of $\bbE[Z_{\pre, \cI^{(n)}}]$. 
Similarly, let $\kappa'$ denote the estimated rank of $\bbE[Z_{\post, \cI^{(n)}}]$, respectively. (Refer to Appendix \ref{app:preliminaries} for various approaches to selecting parameters  
$\kappa$ and $\kappa'$.)
The third $\gamma \in (0, 1)$ is a tunable parameter, 
where a smaller $\gamma$ results in a stricter test. 

Let $\hat{R}_{\pre} \in \bbR^{|\cI^{(n)}| \times \kappa}$
and $\hat{R}_{\post} \in \bbR^{|\cI^{(n)}| \times \kappa'}$ 
denote the matrices constructed from the top $\kappa$  right singular vectors of $Z_{\pre, \cI^{(n)}}$ and 
 the top $\kappa'$ right singular vectors of $Z_{\post, \cI^{(n)}}$, 
 respectively. 
Let
\begin{align*}
	\hat{\beta} &= \norm{ (\mathbb{I}_{|\cI^{(n)}|} - \hat{R}_{\pre} \hat{R}_{\pre}^\top ) \hat{R}_{\post} }_F^2 .
\end{align*}
Then, 
the NSI estimator passes the \textsc{SubspaceInclusionTest} if $\hat{\beta} \leq (1 - \gamma) \kappa'$;
otherwise, it fails. 
\ifarxiv 
\\
\else
\medskip
\fi

\noindent 
\textbf{SubspaceInclusionTest is a data-driven check for Assumption \ref{asm:subspace_inclusion}.}
Let ${R}_{\post}$ and ${R}_{\pre}$ denote the matrices constructed from the right singular vectors of $\bbE \big[  Z_{\pre, \cI^{(n)} }  \big| \, \LF, \cO \big]$ and $\bbE \big[  Z_{\post, \cI^{(n)} }  \big| \, \LF, \cO \big]$, respectively.
Then, Assumption \ref{asm:subspace_inclusion} can equivalently be stated as requiring that $\text{columnspace} (R_{\post}) \subseteq \text{columnspace}(R_{\pre})$.
Although one cannot directly test for Assumption \ref{asm:subspace_inclusion} since both $\bbE \big[  Z_{\pre, \cI^{(n)} }  \big| \, \LF, \cO \big]$ and $\bbE \big[  Z_{\post, \cI^{(n)} }  \big| \, \LF, \cO \big]$ are not observable due to noise, we now show that \textsc{SubspaceInclusionTest} is a sample-based test for Assumption \ref{asm:subspace_inclusion} using $\hat{R}_\post$ and $\hat{R}_\pre$.
Recall that \textsc{SubspaceInclusionTest} fails if 
$ \hat{\beta} = \norm{ (\mathbb{I} - \hat{R}_{\pre} \hat{R}_{\pre}^\top ) \hat{R}_{\post} }_F^2 \geq (1 - \gamma) \kappa',$
where $\hat{R}_{\pre}$ and $\hat{R}_{\post}$ contain the top $\kappa$ and $\kappa'$ right singular vectors of $Z_{\pre, \cI^{(n)}}$  and $Z_{\post, \cI^{(n)}}$, respectively.
This can be viewed as a test for Assumption \ref{asm:subspace_inclusion} since smaller values of $\hat{\beta}$ indicate the extent to which $\text{columnspace} (\hat{R}_{\post}) \subseteq \text{columnspace}(\hat{R}_{\pre})$.
Indeed, suppose that $R_{\pre} = \hat{R}_{\pre}$, $R_{\post} = \hat{R}_{\post}$, and  Assumption \ref{asm:subspace_inclusion} holds; then, $\hat{\beta} = 0$. 
As the span of $\hat{R}_{\post}$ moves outside of the span of $\hat{R}_{\pre}$, $\hat{\beta}$ increases.
Since $\hat{\beta} = \norm{ (\mathbb{I} - {R}_{\pre} {R}_{\pre}^\top ) {R}_{\post} }_F^2$ is always upper bounded by $r_{\post}$, which is estimated by $\kappa'$, we use the threshold $(1 - \gamma) \kappa'$ such that the test fails if $\hat{\beta} \geq (1 - \gamma) \kappa'$.
A formal analysis of this test remains important future work.
\begin{remark}
The equivalence between Assumption \ref{asm:subspace_inclusion} and $\text{columnspace} (R_{\post}) \subseteq \text{columnspace}(R_{\pre})$ implies that \textsc{LatentFactorTest} would supersede \textsc{TrainingTreatmentTest} \emph{if} $R_{\pre}$ and $R_{\post}$ are known exactly and the prediction samples are already collected. %
However, \textsc{TrainingTreatmentTest} remains useful for two reasons.
First, it can be run even before measurements are collected as it only requires the treatment assignment pattern. 
Second, \textsc{LatentFactorTest} requires estimating $R_{\pre}$ and $R_{\post}$, which \textsc{TrainingTreatmentTest} does not.
\end{remark}

%% file: sections/experimental_procedure.tex
\section{NSI's sample complexity: Experiment design} \label{sec:experimental_proc}
In this section, we propose an experiment design based on graph coloring, under which we can precisely answer the question of how should $T, N$ scale to enable the estimation of $\IPO{n}$ within $\varepsilon \in (0,1)$?
Proofs for this section can be found in Appendix \ref{app:proofs_exp}.

\subsection{Graph  coloring-based experiment design}\label{sec:exp_procedure}

\begin{figure*}[t]
	\centering
	\includegraphics[width=\textwidth]{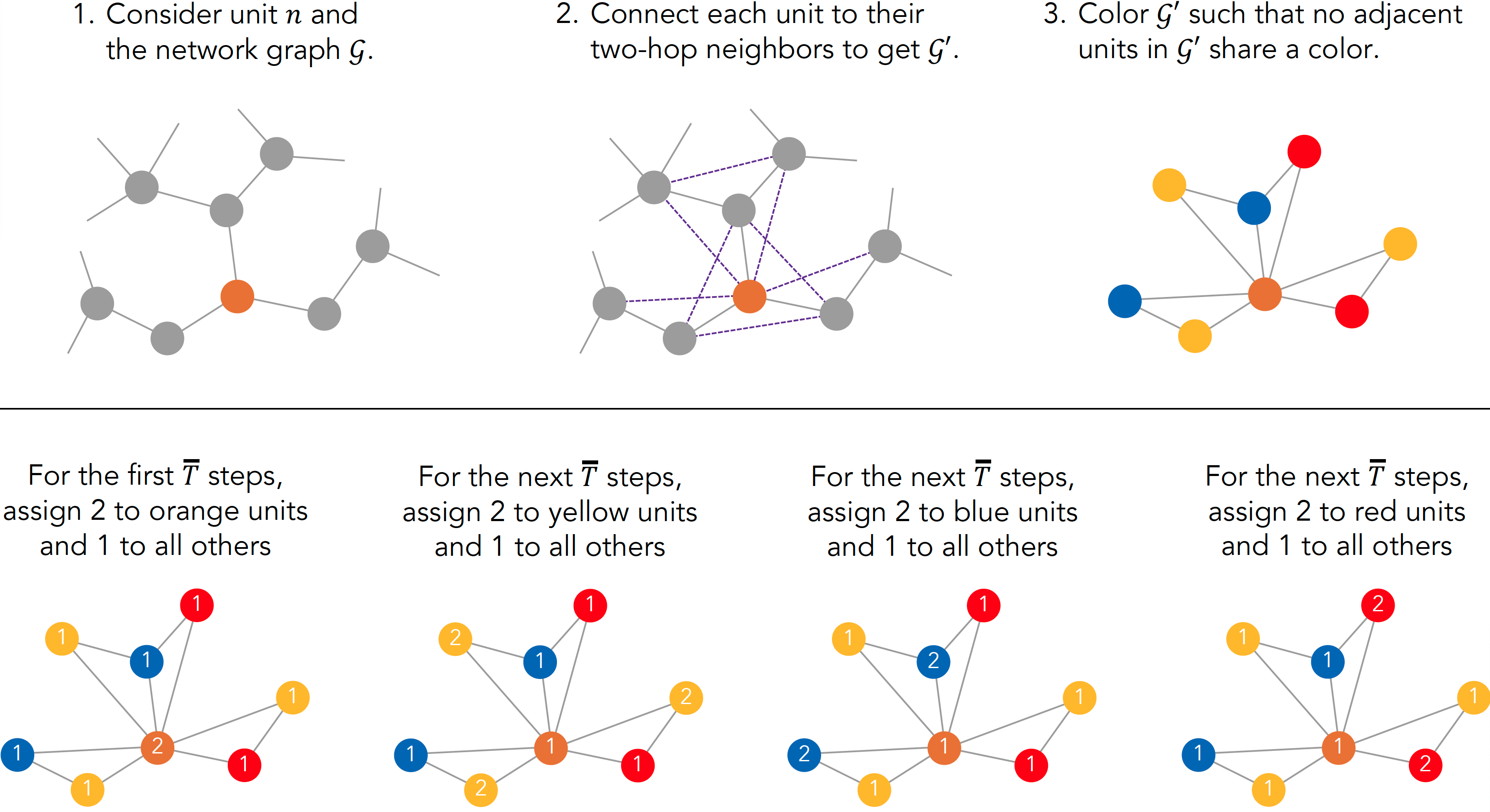}
	\caption{Illustration of experiment design. 
		Consider a unit $n$ and network graph $\cG$ (top left). 
		The experiment design generates the two-hop graph $\cG'$ by connecting every unit to its two-hop neighbors, 
		which translates to adding edges, 
		as given by the purple dotted lines (top center). 
		Next, color $\cG'$ such that no units that are adjacent in $\cG'$ share a color (top right). 
		This coloring is used to generate the training treatment assignments. 
		Specifically, consider $D = 2$. 
		Then, during each $\bar{T}$ training measurements, 
		every unit receives the control treatment $1$, except units of a specific color. 
	}
	\label{fig:exp_design}
\end{figure*}
We begin by describing the experiment design.  
The design assumes access to a subroutine $\textsc{TwoHopColoring}(\cG)$ that outputs $\textsc{NumColors}, \textsc{Coloring}$ and proceeds in two steps: (1) Construct the graph $\cG' = ([N], \cE')$ such that $(i, j) \in \cE \implies (i, j) \in \cE'$ and $(i, j) , (j, k) \in \cE \implies (i, k) \in \cE'$. 
That is, $\cG'$ is constructed by taking $\cG$ and adding edges between every node and its two-hop neighbors.
(2) Perform a coloring on $\cG'$ by 
 labeling the vertices in a graph such that no two adjacent vertices receive the same color, 
 greedily adding colors whenever an existing color cannot be used. 
(Under a color, vertices of the same color form an independent set of $\cG'$.)
Let $\textsc{NumColors}$ denote the number of colors required to color $\cG'$.
Let $\textsc{Coloring} \in [\textsc{NumColors}]^N$ denote the colors assigned to each node (or unit).
As before,
let $\bar{r} \in \bbN_{> 0}$ denote an estimate (or, alternatively, an upper bound) of the model ``rank'' $r$.
Then, the experiment design procedure proceeds as follows. 
\ifarxiv 
\\
\else
\medskip
\fi

\noindent 
\emph{Step 1}.
Let $\textsc{NumColors}, \textsc{Coloring} = \textsc{TwoHopColoring}(\cG)$. 
\ifarxiv 
\\
\else
\medskip
\fi

\noindent 
\emph{Step 2}.
Divide the colors into $T' = \lceil \frac{\textsc{NumColors}}{D - 1} \rceil$ disjoint sets $\{ \textsc{Colors}_\ell : \ell = 0, 1, \hdots, T' - 1 \}$ 
such that $\textsc{Colors}_1$ contains the first $D - 1$ colors, 
$\textsc{Colors}_2$ contains the next $D - 1$ colors, and so on. 

\noindent 
\emph{Step 3}.
Then, for $\ell = 0, 1, \hdots, T' - 1$, 
let $\bc^\ell \in [D]^N$ denote a treatment vector such that 
\begin{align*}
    a_i^\ell &= 
    \begin{cases}
        \textsc{Coloring}_i \, \text{mod} (D - 1) + 2 , & \text{if } \textsc{Coloring}_i \in \textsc{Colors}_\ell ,
        \\
        1 , & \text{otherwise.} 
    \end{cases}
\end{align*}
The intuition behind $\bc^\ell$ is as follows. 
Since $\textsc{Colors}_\ell$ contains at most $D - 1$ colors, 
each color in $\textsc{Colors}_\ell$ can be associated with a different treatment in $\{ 2, 3, \hdots, D\}$. 
Units with one of those colors receive the corresponding treatment. 
That is, any unit $i$ for which $\textsc{Coloring}_i \in  \textsc{Colors}_\ell$
is assigned the corresponding treatment in $\{ 2, 3, \hdots, D\}$. 
All other units receive treatment $1$. 
\ifarxiv 
\\
\else
\medskip
\fi

\noindent 
\emph{Step 4}.
Let $\cT_{\pre}$ be divided into disjoint sets $\{ \cT^\ell_{\pre} : \ell = 0, 1, \hdots, T' - 1  \}$, 
each of length $\bar{T} \geq \bar{r} D$ such that $T_{\pre} = \bar{T} T'$. 
Then,
let
$A^{\pre} \in [D]^{N \times T_{\pre}}$ be defined such that, 
$
    A^{\pre}[:, t] = \bc^{\ell} \quad \forall t \in \cT^\ell_{\pre} 
$
for all $\ell = 0, 1, \hdots, T' - 1$.
For the remainder of this work, we assume $\bar{T} = \bar{r} D$ unless otherwise stated. 
\ifarxiv 
\\
\else
\medskip
\fi

\noindent 
\emph{Step 5}.
Let the prediction treatment of each unit be assigned i.i.d. uniformly at random from the $D$ possible treatments, 
i.e., 
$a^\post_i \iid \text{Unif}(D)$ for all $i \in [N]$.
\ifarxiv 
\\
\else
\medskip
\fi

\noindent{\bf Discussion of experiment design.}
The experiment design described above uses the graph coloring over the two-hop version of $\cG$ to assign treatments. 
The training measurements are divided into $T'$ disjoint sets---that we will call ``periods''---denoted by $\cT^\ell_\pre$ for $\ell = 0, 1, \hdots, T' - 1$. 
The treatment assignment during each period remains constant, i.e., for any $\ell$, 
$\ba^t = \bc^\ell$ for all $t \in \cT^\ell_\pre$. 
 
The experiment design ensures several important properties hold. 
First, all nodes that receive the same color also receive the same treatment at any $t \in \cT_\pre$. 
Second,
nodes that receive the same color have to be at least three hops away from one another, 
which ensures that, for any neighborhood $\cN(n)$ of an arbitrary node $n$, 
no two nodes receive the same non-control treatment at any given $t \in \cT_\pre$.
Third, each node receives the control treatment $1$ for every period, 
except for one period during which it receives a non-control treatment. 
The non-control treatment that a node receives is given by $ \textsc{Coloring}_i \, \text{mod} (D - 1) + 2$, where $\textsc{Coloring}_i$ denotes the color assigned to unit $i$. 
Lastly, each period has length $\bar{T} \geq \bar{r} D$. 
We show that these properties are important to proving Lemma \ref{lem:training_treatments_test_passes} in the next section. 
See \citet{jensen2011graph} for further information on graph colorings.

\subsection{Theoretical guarantees on experiment design} 
We present two results. 
The first establishes that the graph-theoretic experiment design described in Section \ref{sec:exp_procedure} guarantees that \textsc{TrainingTreatmentTest} is passed. 

\begin{lemma}\label{lem:training_treatments_test_passes}
    Suppose Assumption 
    \ref{asm:model} holds. 
    If the training treatments $A^{\pre}$ are assigned as in Section \ref{sec:exp_procedure},
    then \textsc{TrainingTreatmentTest} is passed for any unit $n$ and treatments $\cfAn$. 
\end{lemma}
Importantly, Lemma \ref{lem:training_treatments_test_passes} gives a guarantee for \emph{any} unit $n$ and counterfactual treatments $\cfAn$ of interest, i.e., for all $N D^d$ possible estimands of interest.
That is, the experiment design in Section \ref{sec:exp_procedure} can be used to ensure that \textsc{TrainingTreatmentTest} is passed for any target estimand of interest. 
{Moreover, the experiment design can be applied to any graph of interest (and any valid two-hop coloring, as described in Section \ref{sec:exp_procedure}).}
In Appendix \ref{app:tailor_exp_design}, we discuss how one can adapt our experiment design to be less stringent if one is interested in a \emph{specific} choice of $n$ and $\cfAn$. 

The next result shows that the number of training measurements required under the graph-theoretic experiment design is $O(d^2)$, 
where $d$ denotes the maximum degree of $\cG$.
\begin{lemma}\label{lem:treatment_schedule_complexity}
	The number of training measurements required by the experimental procedure in Section \ref{sec:exp_procedure} is
	$T_{\pre} = \bar{r} D \lceil \frac{\textsc{NumColors}}{D - 1} \rceil$, where
	$\textsc{NumColors} \leq \text{Degree}(\cG') + 1$. 
	As a result, $T_{\pre} \leq \frac{\bar{r} D (d^2 + D)}{D - 1}$. 
\end{lemma}
By Lemma \ref{lem:treatment_schedule_complexity}, the experiment design requires $T_\pre = O(d^2)$ training measurements. 
Since
$r|\cN(n)| = r d$ donors are needed for a given $n$ and $\tilde{\ba}_{\cN(n)}$ of interest (as discussed in Section \ref{sec:assumptions}), 
this result shows that one needs $O(d^3)$ training samples under the experiment design to guarantee that \textsc{TrainingTreatmentTest} passes for a given unit and target treatment of interest. 
That is, with $O(d^3)$ training samples, 
there is enough variation in the training data such that it is possible to generalize to the training data to a given target counterfactual treatment of interest.

\subsection{Data requirement: Generative example}

In this section, 
we examine how much data is needed for the NSI estimator to get within $\varepsilon$ accuracy, 
using regular graphs as an illustrative setting. 

\begin{assumption}\label{asm:regular_graph}
    $\cG$ is a $d$-regular graph.
\end{assumption}

\begin{assumption}\label{asm:unif_lf}
Assume that each $u_{j, i, k} \iid \text{Unif}\hspace{1pt}(-\frac{1}{\sqrt{r d}} , \frac{1}{\sqrt{r d}})$ for all $j, i \in [N]$ and $k \in [r]$. 
Further, each $w_{t, a, k}  \iid \text{Unif}\hspace{1pt}( - \frac{1}{\sqrt{r d}}, \frac{1}{\sqrt{r d}})$ for all $t \in [T]$, $a \in [D]$, and $k \in [r]$. 
\end{assumption}

\begin{proposition}\label{prop:reg_error}
Suppose $\bar{r} = r$.
Suppose Assumptions \ref{asm:model}, \ref{asm:conditional_exo}, 
\ref{asm:subG_noise}, 
\ref{asm:bounded}, 
\ref{asm:components},
\ref{asm:regular_graph},
and
\ref{asm:unif_lf},
hold. 
Suppose that there are at least $\frac{r D (d^2 + D)}{D - 1}$ training measurements assigned according to the experiment design in Section \ref{sec:experimental_proc} and $N = \Omega \left( r^2 d^2 D^{2d + 2}  \right)$ units.
Then, there is a set of units $E$ such that $|E| = N - \Theta(\sqrt{N})$ 
and a method of choosing donors (i.e., units that satisfy Definition \ref{def:node_donors}) such that, for all $n \in E$, 
\begin{align*}
&\left| 
\hIPO{n} - \IPO{n}
\right| 
\\
&\hspace{20pt}= O_P \left( 
r^6 d^{10}
\log \left(  \frac{ T_\pre N }{D^{d+1}} \right)
{\max} \left(
\frac{1}{T_\pre^{1/4}} , 
\frac{ {D^{(d + 1)/2}} }{ N^{1/4} } ,
{\frac{ N^{1/4} }{ D^{(d+1)/2} T_\pre^{3/2}}}
\right)
\right),
\end{align*}
\end{proposition}
For a $d$-regular graph, Proposition \ref{prop:reg_error} suggests that to achieve error of order $\varepsilon$ with high probability, NSI needs $T_\pre = \tilde{\Omega} \left( \frac{ r^{24} d^{40} }{\varepsilon^4}  \right)$.
To understand whether the sample complexity is reasonable, consider a naive alternative. 
For a given unit $n$, there are $D^{|\cN(n)|}$ possible counterfactual treatments that could be applied to the neighborhood $\cN(n)$.
Suppose that we do not impose any structure on the potential outcomes, 
i.e., we do not assume \eqref{eq:full_model}.
Then, to learn how unit $n$ behaves under every possible neighborhood treatment, one would naively need at least one observation per $D^{|\cN(n)|} \leq D^{d+1}$ possible treatments. 
This naive approach would  require $\Omega(D^d)$ samples in order to estimate the potential outcome for a given $n$ and any $\cfAn$ of interest. 
Moreover, to achieve an error of $\varepsilon$, at least $\frac{1}{\varepsilon^2}$ samples are needed per treatment, which implies $T_\pre = \Omega\left(\frac{D^d}{\varepsilon^2}\right)$ under a naive approach.

%% file: sections/simulations.tex
\section{Simulations} \label{sec:experiments}

In this section, 
we present simulation results illustrating the behavior of the NSI estimator and compare it to two related estimators. 
Let $\cG$ be a regular graph with degree $d$, and let the treatments be binary, i.e., $D = 2$.
In each of the experiments below, 
we indicate the graph degree. 
Let the training treatments be assigned according to the experiment design in Section \ref{sec:experimental_proc}.
Further experiments and details are given in Appendix \ref{app:simulations}.
\ifarxiv 
\\
\else
\medskip
\fi

\begin{figure*}[t]
	\centering
	\begin{subfigure}[b]{0.31\textwidth}
		\centering
		\includegraphics[width=\textwidth]{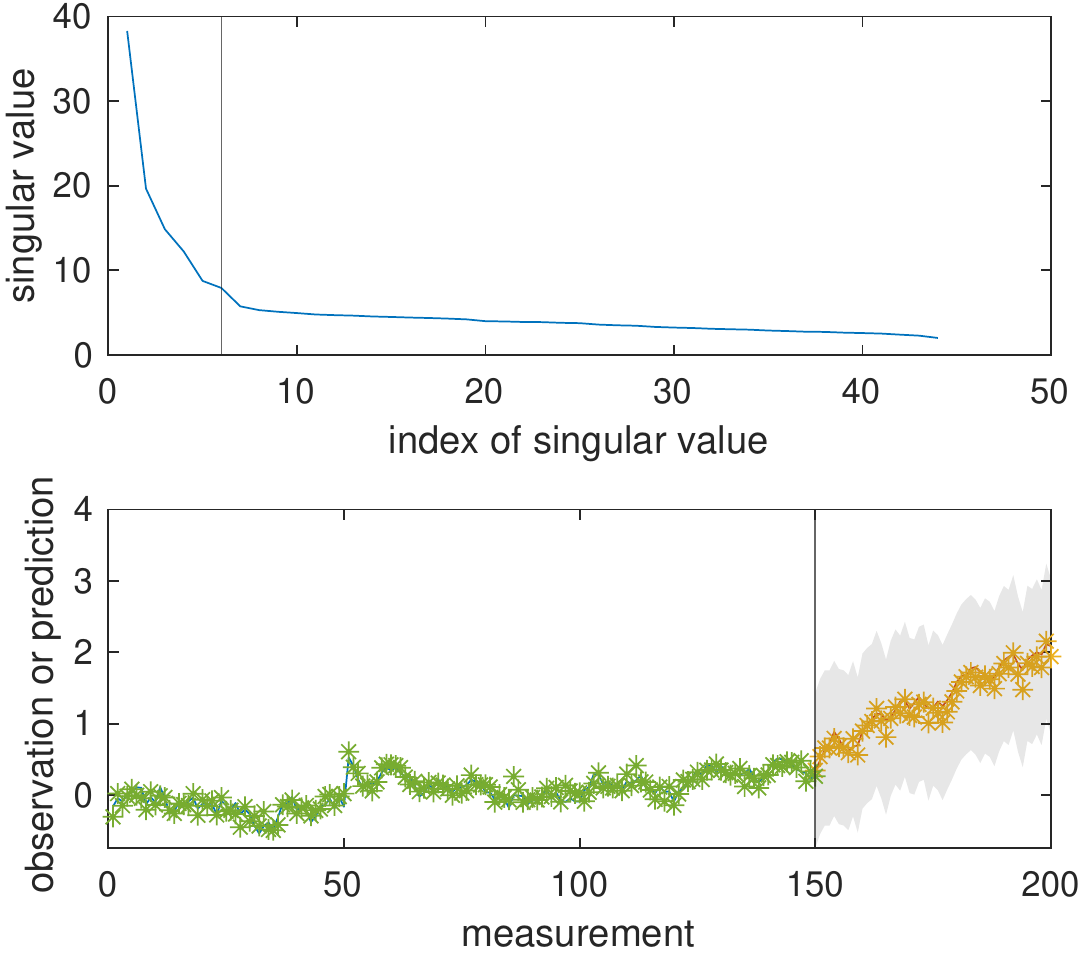}
		\caption{NSI estimates}
		\label{fig:y equals x}
	\end{subfigure}
	\hfill
	\begin{subfigure}[b]{0.328\textwidth}
		\centering
		\includegraphics[width=\textwidth]{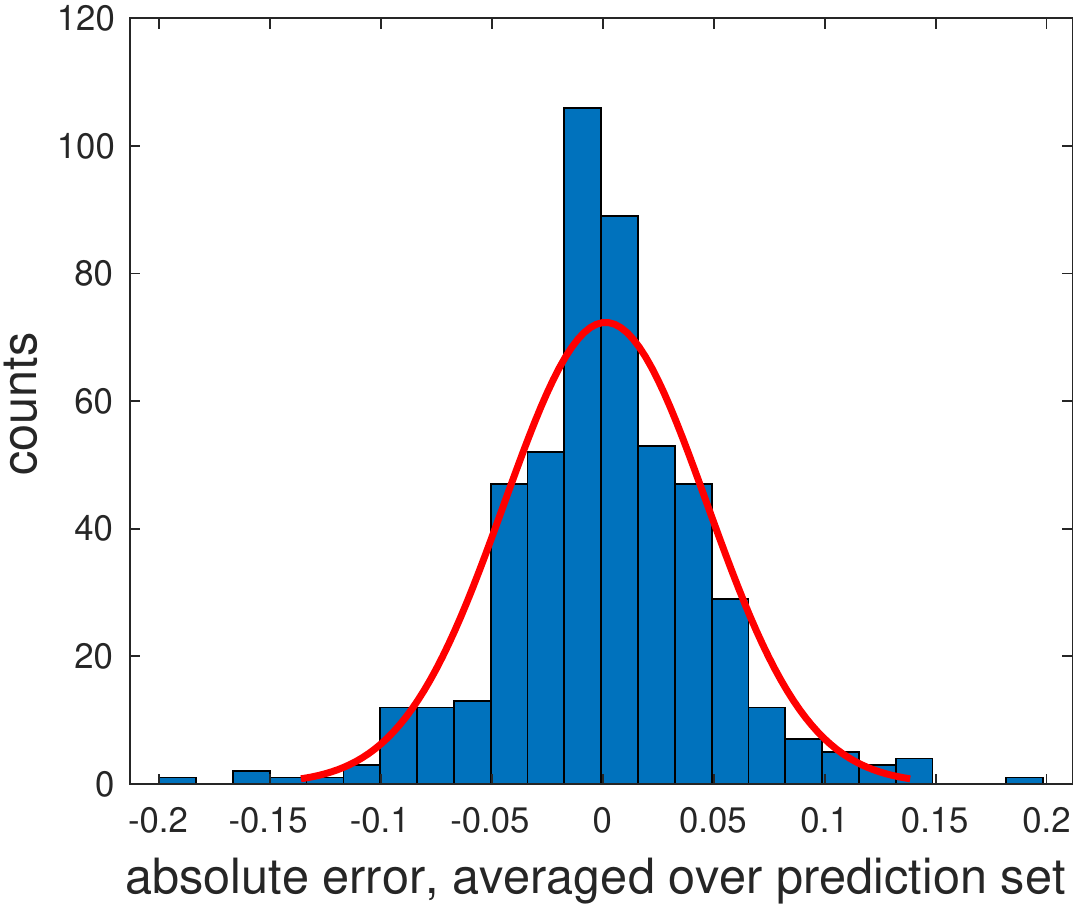}
		\caption{Residuals of NSI estimator}
		\label{fig:three sin x}
	\end{subfigure}
	\hfill
	\begin{subfigure}[b]{0.334\textwidth}
		\centering
		\includegraphics[width=\textwidth]{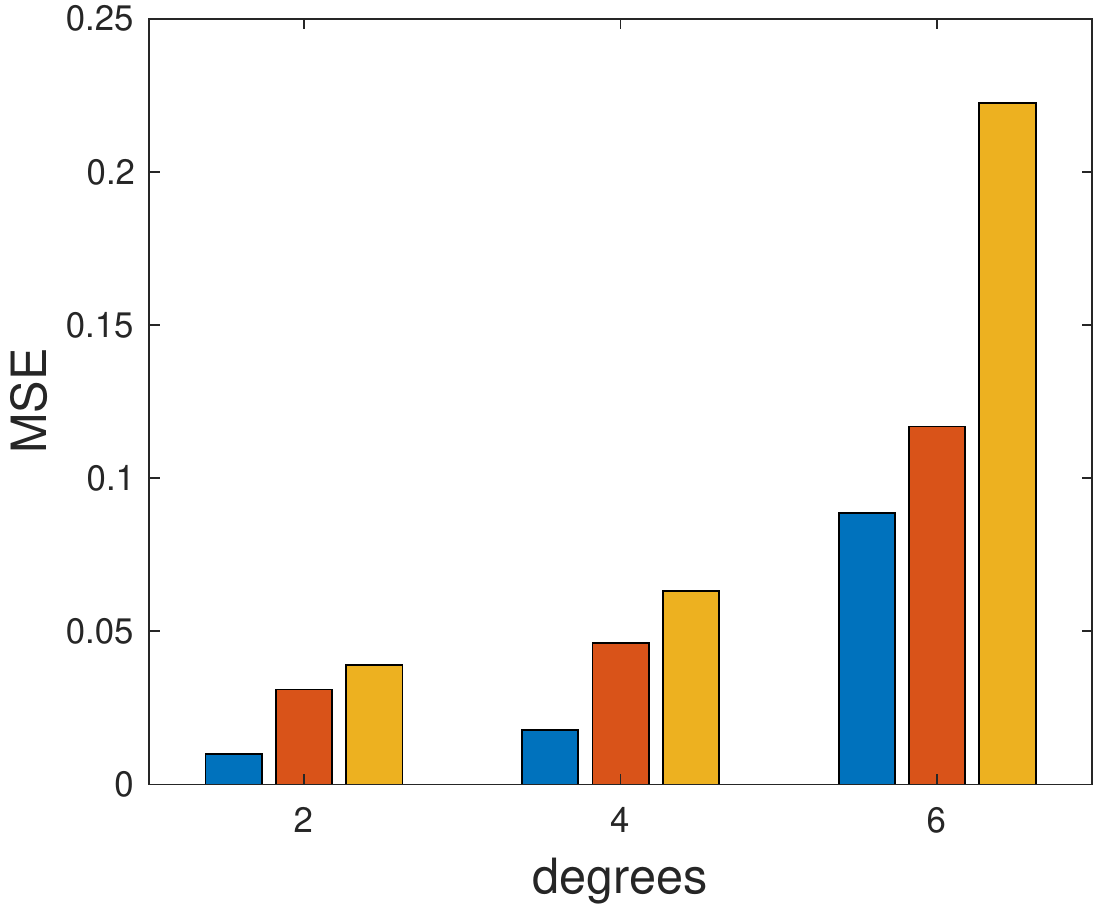}
		\caption{MSE for regular graphs}
		\label{fig:five over x}
	\end{subfigure}
	\caption{Simulation results illustrating the performance of the NSI estimator. 
		(a) considers a specific unit $n$ and counterfactual $\cfAn$ of interest. 
		The top plots the spectrum produced in Step 1 of NSI (see Section \ref{sec:estimation_proc}). 
		The bottom visualizes the pointwise estimates 
		\ifarxiv \else $\widehat{\bbE} \big[ Y_{t, n}^{(\tilde{\ba}_{\cN(n)})} \big]$ \fi produced by NSI for all measurements $t$ in asterisks $*$ with the 95 percent confidence interval in gray. The ground truth potential outcomes \ifarxiv \else ${\bbE} \big[ Y_{t, n}^{(\tilde{\ba}_{\cN(n)})} \big]$\fi are given by the solid lines. 
		(b) plots the residuals $(\hIPO{n} - \IPO{n})$ across 500 simulations. 
		(c) plots the MSE across different hyperparameters. 
		Each group of bars gives the MSE for regular graphs of degree $2$, $4$, and $6$,
		as indicated on the $x$-axis. 
		Within each group of bars, 
		the left (blue) bars are for $N = 1000$, $T_{\pre} = 100$, and $T_{\post} = 50$; 
		the middle (red) bars for $N = 1000$ and $T_{\pre} = T_{\post} = 50$; 
		and the right (yellow) bars for $N = 500$ and $T_{\pre} = T_{\post} = 50$. 
	}
	\label{fig:NSI}
\end{figure*}

\noindent
\textbf{Predictions}.
Note that the NSI estimator \eqref{eq:NSI_estimator_point} can be adapted to produce pointwise estimates
$$\widehat{\bbE} \Big[ Y_{t, n}^{(\tilde{\ba}_{\cN(n)})} \Big] = Z[t, \cI^{(n)}] \hat{\bbE}[Z_{\pre, \cI^{(n)}} | \LF , A ]^+ \bz_{\pre, n} ,$$ 
such that $\hIPO{n} = \frac{1}{T_\post} \sum_{t \in \cT_\post} \widehat{\bbE} \big[ Y_{t, n}^{(\tilde{\ba}_{\cN(n)})} \big]$.
Under this observation,
Figure \ref{fig:NSI}(a) shows an example of the pointwise estimates given by NSI for an example unit $n$. 
Consider the bottom plot.
The solid line gives the ground truth potential outcomes for unit $n$ across measurements $t \in [200]$. 
The pointwise estimates produced by NSI 
are marked by asterisks $*$, with the 95 percent confidence interval in gray. 
The measurements to the left of the vertical line (i.e., in blue and green) correspond to the training set $\cT_{\pre}$ while those to the right (i.e., in red and orange) correspond to the prediction set $\cT_{\post}$. 
The top plot gives the spectrum $\{\hat{s}_\ell\}_{\ell=1}^{q}$ produced in Step 1 of Section \ref{sec:estimation_proc}, where the vertical line marks the hard singular value threshold $\kappa$ that is used in Step 1. 
In Figure \ref{fig:NSI}(a), $\cG$ is a ring graph ($d = 2$) with $N = 1000$ units, $\epsilon^{(\bc_{\cN(i)})}_{\tau, i} \sim \cN(0, 0.1)$, and $r = 2$.

As shown in the bottom plot of Figure \ref{fig:NSI}(a), the predictions closely match the ground-truth values. 
As shown on top, $6$ components are used to construct the estimates. 
Since the network-adjusted rank is $6 $ ($r = 2$ and $|\cN(n)| = 3$), 
the fact that NSI uses $6$ components explains why its estimates are fairly accurate.
We provide similar plots for other units and target treatments in Appendix \ref{app:simulations}.
\ifarxiv 
\\
\else
\medskip
\fi

\noindent
\textbf{Consistency and asymptotic normality}.
Figure \ref{fig:NSI}(b) verifies that the NSI estimates are consistent and asymptotically normal. 
Specifically, we let $\cG$ be a ring graph (i.e., $d = 2$) with $N = 1000$ units, 
$\epsilon^{(\bc_{\cN(i)})}_{\tau, i} \sim \cN(0, 0.1)$, and $r = 2$.
For each simulation, we randomly generate the potential outcomes of all units under \eqref{eq:full_model} (see Appendix \ref{app:simulations} for details).
We run 500 simulations, 
then compute the NSI residuals $(\hIPO{n} - \IPO{n})$ for all units $n \in [N]$ and across all possible counterfactual treatments for each $n$.
That is, we use NSI to estimate $\IPO{n}$ for $\tilde{\ba}_{\cN(n)} = (1, 0, 0)$, 
$\tilde{\ba}_{\cN(n)} = (0, 1, 0)$, $\tilde{\ba}_{\cN(n)} = (1, 1, 0)$, and so on. 
Figure \ref{fig:NSI}(b)  gives a histogram of the NSI residuals. 
A Gaussian distribution is fit to the residuals and given by the red line. 
\ifarxiv 
\\
\else
\medskip
\fi

\noindent
\textbf{MSE trends}. 
Figure \ref{fig:NSI}(c) summarizes the performance of NSI across different parameters. 
The performance is given by the mean-squared error (MSE) across the prediction measurements $\cT_{\post}$, averaged across $50$ units. 
Each group of bars gives the MSE for regular graphs of degree $2$, $4$, and $6$, 
as indicated on the $x$-axis. 
Within each group of bars, 
the left (blue) bars are for $N = 1000$, $T_{\pre} = 100$, and $T_{\post} = 50$; 
the middle (red) bars for $N = 1000$ and $T_{\pre} = T_{\post} = 50$; 
and the right (yellow) bars for $N = 500$ and $T_{\pre} = T_{\post} = 50$. 
Each bar is the average of $200$ simulations with $\epsilon^{(\bc_{\cN(i)})}_{\tau, i} \sim \cN(0, 0.1)$, and $r = 2$.
As expected, the MSE increases with degree ({because, holding $N$ fixed, a higher degree leads to fewer valid donors}), fewer nodes ({which also leads to fewer valid donors}), and fewer training measurements. 
\ifarxiv 
\\
\else
\medskip
\fi

\noindent
\textbf{Comparing to other estimators}.
We also compare the NSI estimator to two others: the SI estimator \citep{agarwal2020synthetic} and a baseline estimator. 
The SI estimator is similar to NSI, but SI assumes that there is no spillover and therefore does not account for network interference.
The baseline estimator finds donor units that satisfy Definition \ref{def:node_donors}, then averages the donor units' observed outcomes. 
We compare the estimators for a ring graph (details given in Appendix \ref{app:simulations}). 
We compare
the estimators for a ring graph under the same parameters as those used in Figure \ref{fig:NSI}(b) averaging
across $200$ simulations, $50$ units, and all possible counterfactual treatments.

The MSEs and R-squared values for the NSI estimator, SI estimator, and baseline estimators are, respectively, \textbf{(0.1174, 0.8735)}, \textbf{(0.2310, 0.8149)}, and \textbf{(3.398, -2.957)}. 
Both the NSI and baseline estimators use donor sets that contain, on average, 41 units. The SI estimator uses donor sets with, on average, 166 units.  
As such, 
even though the SI estimator has more donors, 
the performance of NSI is better than that of SI, which is better than that of the baseline estimator.

%% file: sections/discussion.tex
\section{Conclusion and future work}

There is rising interest in estimating unit-specific potential outcomes. 
In this work, 
we consider the estimation of unit-specific potential outcomes in the presence of spillover, 
i.e., the treatment assigned to one unit affects the outcome of another unit.
We focus on the panel data setting and model spillover as network interference. 

As our main contribution, we provide an estimator that we call Network Synthetic Interventions (NSI). 
In addition to producing point estimates, 
NSI provides confidence intervals. 
We show that, under a low-rank latent-factor model and suitable conditions, 
the NSI estimates are consistent and asymptotically normal. 
We provide two validity tests that determine whether key conditions hold. 
We find that obtaining good estimates under spillover requires that the data is rich enough. 
To this end, 
we provide an experiment design. 

There are many paths for future work. 
Although the method that we provide comes with strong performance guarantees, 
it has strict data requirements, as discussed in Section \ref{sec:assumptions}. 
One path for future work would be to explore whether these requirements can be relaxed. 
Second, we explore unit-specific potential outcome estimation. 
If such fine-grained estimates are not needed,
one could examine whether NSI and its data requirements can be improved for coarser estimands of interest. 
Finally, one compelling path for future work would be to test NSI on real-world datasets.

%% file: sections/app_preliminaries.tex
Let $[X] \defeq \{1,\dots, X\}$ for any positive integer $X$.
For any treatment vector $\ba \in [D]^N$ and some set $S \subseteq [N]$, 
let $\ba_S \in [D]^{|S|}$ denote the vector containing the elements of $\ba$ indexed by $S$. 
Similarly, let $a_i \in [D]$ denote the $i$-th element of $\ba$.
Let $\bbI_x$ denote the $x \times x$ identity matrix 
and $\otimes$ denote the Kronecker product. 
Let $\ind(\cdot)$ denote the indicator function. 
Let $\norm{\cdot}_{\psi_2}$ denote the Orlicz norm. 
Let $O_p$ denote a probabilistic version of big-$O$ notation. 
Formally, 
for any sequence of random vectors $X_n$, 
$X_n = O_p(\chi_n)$ if, for any $\varepsilon > 0$, 
there exists constants $c_\varepsilon$ and $n_\varepsilon$ such that $P(\norm{X_n}_2 > c_\varepsilon \chi_n) < \varepsilon$ for every $n \geq n_\varepsilon$. 
Equivalently, we say that $X_n / \chi_n$ is ``uniformly tight'' or ``bounded in probability.''
Similarly, 
let $o_P$ denote the probabilistic version of little-$o$ notation. 
Formally, for any sequence of random variables $X_n$, 
$X_n = o_P(1)$ if and only if $X_n \stackrel{p}{\rightarrow} 0$.
Let $\tilde{\Omega}$ denote the variation on big-$\Omega$ notation that ignores logarithmic terms such that $\Omega(a(n) \log^k(n)) = \tilde{\Omega}(a(n))$. 
For two sets of indices $S_1 \subseteq [m_1]$ and $S_2 \subseteq [m_2]$ as well as a matrix $\Pi \in \bbR^{m_1 \times m_2}$, 
let $\Pi[S_1, S_2] \in \bbR^{|S_1| \times |S_2|}$ denote the submatrix of $\Pi$ corresponding to the rows indexed by $S_1$ and columns index by $S_2$. 
We use ``$\col$'' as a shorthand for the entire set of indices such that $\Pi[\col, S_2] \in \bbR^{m_1 \times |S_2|}$ and $\Pi[S_1, \col] \in \bbR^{|S_1| \times m_2}$.
Let $\cX^*$ denote the $*$-product space $\cX \times \cX \times \hdots \times \cX$, where the length of the product is not pre-determined. 
Lastly, let $\Pi^+$ denote the pseudo-inverse of $\Pi$.

\begin{remark} \label{rem:SVT}
	The estimation procedure for NSI requires the use of a singular value thresholding (SVT) method.
	Given a list of singular values (also known as a spectrum) $(s_1, s_2, \hdots, s_X)$ where $s_1 \geq s_2 \geq \hdots \geq s_X$, 
	an SVT method determines a threshold $r_{\text{SVT}} \leq X$.
	The singular values $(s_1, s_2, \hdots, s_{r_{\text{SVT}}})$ are preserved, and the remaining are discarded. 
	SVT is often used to ``de-noise'' a matrix using its singular values. 
	That is, 
	one reconstructs a de-noised matrix by keeping the top $r_{\text{SVT}}$ components of that matrix and attributing the remaining components to noise. 
	In this way, one can think of $r_{\text{SVT}}$ as the matrix's estimated rank. 
	
	There are several popular methods for  SVT.
	One could, for instance, use a universal SVT method, such as that given in \citep{chatterjee2015matrix}. 
	There are also popular methods for what is known as elbow (or knee) point detection \citep{zack1977automatic, satopaa2011finding}, 
	which look for the point of maximum curvature along a monotonic curve.
\end{remark}

%% file: sections/app_useful_lemmas.tex
\begin{lemma}\label{lem:unif_full_rank}
	Let $X \in \bbR^{m_1 \times m_2}$ be a random matrix where $X_{i,j} \iid p_x$ is drawn from a continuous, non-degenerate distribution $p_x$. 
	Then, 
	$\text{rank}(X) = \min(m_1, m_2)$ almost surely.
\end{lemma}
\ifarxiv
\begin{proof}
\else
\proof{Proof.}
\fi
Let $V_j = \text{span}(\{ \bx_k  : k = 1 , \hdots, j \})$,
where $\bx_k$ denotes the $k$-th column of $X$. 
Since $V_1$ is a one-dimensional subspace
and $\bx_{k}$ consists of i.i.d. non-degenerate, continuous random variables,
\begin{align*}
	P \left( \bx_2  \in V_1 \right) = 0 .
\end{align*}
In other words, 
with probability $1$, 
$V_2$ is a two-dimensional subspace. 
By induction, $V_j$ is a $j$-dimensional subspace almost surely as long as $j \leq m_1$. For any $j \geq m_1$, 
$V_j$ is an $m_1$-dimensional subspace. 
Therefore, 
$rank(X) = \min(m_1, m_2)$ almost surely. 
\ifarxiv
\end{proof}
\else
\Halmos
\endproof
\fi

\begin{lemma}\label{lem:linear_span_gaussian}
	Suppose Assumptions \ref{assumption:network_sutva}, \ref{asm:model}, 
	and \ref{asm:unif_lf} hold.
	Then, \text{Assumption \ref{asm:linear_span} holds}  almost surely if there are at least $r | \cN(n) |$ donors. 
\end{lemma}
\ifarxiv
\begin{proof}
	\else
	\proof{Proof.}
	\fi
Consider a unit $n$.
Recall that
\begin{align*}
	\tilde{\bu}_{n, \cN(n)} 
	&= [
	\bu_{\cN_1(n), n}^\top \, , \,
	\bu_{\cN_2(n), n}^\top \, , \,
	\hdots \, ,  \,
	\bu_{\cN_{|\cN(n)|}(n), n}^\top 
	]^\top \in \bbR^{r | \cN(n) |}.
\end{align*}
Further, 
recall that linear span inclusion (Assumption \ref{asm:linear_span}) requires that
\begin{align}
	\tilde{\mathbf{u}}_{n, \cN(n)} = \sum_{j \in \cI^{(n)}}  \lambda_{j}  \tilde{\mathbf{u}}_{j, \pi_j( \cN(j) )}, \label{eq:lin_span_proof1}
\end{align} 
for some $\{\lambda_j\}_{j \in \cI^{(n)}}$,
where $\pi_j$ is defined in Definition \ref{def:node_donors}.
To show that linear span inclusion holds, 
suppose that we construct a matrix $U = [\tilde{\mathbf{u}}_{j, \pi_j( \cN(j) )} : j \in \cI^{(n)}] \in \bbR^{r | \cN(n) | \times | \cI^{(n)} |}$.
By Lemma \ref{lem:unif_full_rank}, $\text{rank}(U) = \min(r | \cN(n) |, | \cI^{(n)} |)$ almost surely. 
Since $|\cI^{(n)}| \geq r |\cN(n)|$, 
$\text{rank}(U) = r | \cN(n) |$, 
which implies that \eqref{eq:lin_span_proof1} and therefore that Assumption \ref{asm:linear_span} holds almost surely.
\ifarxiv
\end{proof}
\else
\Halmos
\endproof
\fi

\begin{lemma}\label{lem:prediction_treatment_concentration}
	Consider $\cG$. 
	Suppose that  $\cG$ contains $N'$ disjoint clusters, each of size $M$. 
	Suppose that every unit is assigned a treatment $a \in [D]$ independently and uniformly at random. 
	Let $\bs_i \in [D]^M$ denote the (ordered) sequence of treatments for cluster $i \in [N']$. 
	Let $\bs_0$ denote a reference sequence.  
	Let $B$ denote the number of clusters for which the cluster's treatments $\bs_i$ match the reference treatments $\bs_0$, 
	i.e., $s_i = s_0$. 
	Then, 
	\begin{align*}
		P \left(
		B \geq  \frac{\chi N'}{D^M}
		\right) 
		\leq 
		\exp \left(
		\frac{-2 (\chi - 1)^2 N'}{D^{2M} }
		\right) ,
	\end{align*}
	for any $\chi > 1$, and
	\begin{align*}
		P \left(
		B \leq  \frac{\chi N'}{D^M}
		\right) 
		\leq 
		\exp \left(
		\frac{-2 (\chi - 1)^2 N'}{D^{2M} }
		\right) ,
	\end{align*}
	for any $\chi < 1$. 
\end{lemma}
\ifarxiv
\begin{proof}
	\else
	\proof{Proof.}
	\fi
Let there be $N'$ clusters, each of size $M$. 
Let $\bs_i \in [D]^M$ denote the (ordered) sequence of treatments for cluster $i \in [N']$. 
Let $\bs_0$ denote a reference sequence. 

Let $\bb = (s_1 = s_0, s_2 = s_0, \hdots, s_{N'} = s_0) \in \{0, 1\}^{N'}$. 
Intuitively, 
$\bb$ is a binary vector,
where entry $b_i$ indicates whether $s_i$ matches the reference sequence $s_0$. 
Let $B = \sum_{i = 1}^{N'} b_i$, 
i.e., the number of clusters for which the cluster's treatments $\bs_i$ match the reference treatments $\bs_0$. 

Under the setup (in particular, that units are assigned treatments independently and uniformly at random, and the sequences $s_i$ are over disjoint clusters), $\bb$ is a sequence of i.i.d. Bernoulli random variables and $\bbE[B] = \frac{N'}{D^{M}}$. 
Then, by Hoeffding's inequality, 
\begin{align*}
	P \left(
	B \geq \chi \bbE B 
	\right) 
	&= 
	P \left(
	B - \bbE B \geq (\chi - 1) \bbE B 
	\right) 
	\\
	&\leq 
	\exp \left(
	\frac{-2 (\chi - 1)^2 (\bbE B)^2}{N'}
	\right)
	\\
	&=  \exp \left(
	\frac{-2 (\chi - 1)^2 N'}{D^{2M} }
	\right) ,
\end{align*}
for $\chi > 1$.
\ifarxiv
\end{proof}
\else
\Halmos
\endproof
\fi

\begin{lemma}\label{lem:coloring_num_units}
	Suppose Assumption \ref{asm:regular_graph} holds. 
	For every unit $i \in [N]$, fix an ordering over the neighborhood $\cN(i)$, 
	and let $C_i$ denote the (ordered) colors assigned to $\cN(i)$. 
	We say that a unit $i$ is a ``coloring donor'' for an ego-unit $n \neq i$ if $C_i = C_n$.
	Then, there are at least $N - \Theta(\sqrt{N})$ units with at least $\sqrt{N}$ coloring donors. 
\end{lemma}
\ifarxiv
\begin{proof}
	\else
	\proof{Proof.}
	\fi
First, note that every unit has $d$ neighbors by Assumption \ref{asm:regular_graph}. 
In addition, it is well known that a coloring of $\cG'$  (as defined in Section \ref{sec:exp_procedure}) requires at most $d^2 + 1$ colors. 
Therefore, 
there are  $\rho(d) = (d^2 + 1)^{d+1}$ ways to color each neighborhood. 
Let $\mathcal{O}_1,  \mathcal{O}_2, \hdots, \mathcal{O}_{\rho(d)}$ denote the possible (ordered) colorings and $N_k = | \{ i : C_i = \mathcal{O}_k \} |$ denote the number of units for which the (ordered) neighborhood is colored according to $\mathcal{O}_k$. 
Let $\mathcal{O}^- = \{ k : N_k < \sqrt{N} \}$, 
i.e., $\mathcal{O}^-$ is formed by removing all colorings that match fewer than $\sqrt{N}$ neighborhoods. 

Since at most $\rho(d) \sqrt{N}$ units have colors $\mathcal{O}^-$ and all remaining units have at least $\sqrt{N}$ coloring donors by definition of $\mathcal{O}^-$, 
there are at least $N - \rho(d) \sqrt{N} = N - \Theta(\sqrt{N})$ units with at least $\sqrt{N}$ coloring donors. 
\ifarxiv
\end{proof}
\else
\Halmos
\endproof
\fi

\begin{lemma}\label{lem:rowspace}
	Consider two matrices $X_1 \in \bbR^{m_1 \times m_2}$ and $X_2 \in \bbR^{m_1' \times m_2}$.
	Then, $\text{rowspace}(X_1) \not\subseteq \text{rowspace}(X_2)$
	if and only if there exists a vector $\mathbf{v} \neq \mathbf{0}_{m_1}$ such that $X_2 X_1^\top \mathbf{v} = \mathbf{0}_{m_1'}$ and $X_1^\top \mathbf{v} \neq \mathbf{0}_{m_2}$. 
\end{lemma}
\ifarxiv
\begin{proof}
	\else
	\proof{Proof.}
	\fi
$\text{rowspace}(X_1) \not\subseteq \text{rowspace}(X_2)$ if and only if there exists a vector $\mathbf{v} \neq \mathbf{0}_{m_1}$ such that $X_1^\top \mathbf{v} \neq \mathbf{0}_{m_2}$ is in the null space of $X_2^\top$, 
which is equivalent to $X_2 X_1^\top \mathbf{v} = \mathbf{0}_{m_1'}$. 
\ifarxiv
\end{proof}
\else
\Halmos
\endproof
\fi

\medskip
\noindent 
\textbf{Notation and definitions}. For Lemmas \ref{lem:g1}-\ref{lem:f2}, we  suppress the conditioning on $\LF$ and $\cO$.
Let $\boldsymbol{\epsilon}_{t, \cI^{(n)}}  = [ \epsilon_{t, j}^{( \ba_{\cN(j)}^t )} : j \in \cI^{(n)} ] \in \bbR^{| \cI^{(n)} |}$. 
Let $\boldsymbol{\epsilon}_{\pre, n}  = [ \epsilon_{t, n}^{( \ba_{\cN(n)}^t )} : t \in \cT_{\pre} ] \in \bbR^{T_{\pre}}$. 
Let $\Delta =  \hat{\boldsymbol{\alpha}} - \balphaperp$, 
where $\boldsymbol{\alpha}$ is defined below Theorem \ref{thm:ID}.
Recall that $R_{\pre}$ denotes the matrix containing the right singular vectors of  $\bbE \big[  Z_{\pre, \cI^{(n)} }  \big| \, \LF, \cO \big]$.
Let $Z_{\pre, \cI^{(n)}}^{r_{\pre}} = \sum_{\ell = 1}^{r_\pre} \hat{s}_\ell \hat{\boldsymbol{\mu}}_\ell \hat{\boldsymbol{\nu}}_\ell^\top = \bar{L}_\pre \bar{\Sigma}_\pre \bar{R}_\pre^\top$,
	where $ \hat{s}_\ell$, $ \hat{\boldsymbol{\mu}}_\ell$, and $\hat{\boldsymbol{\nu}}_\ell$ are defined in Section \ref{sec:estimation_proc}.
Let $\cP = R_{\pre} R_{\pre}^\top$ and $\bar{\cP} = \bar{R}_{\pre} \bar{R}_{\pre}^\top$. 
Let $\bar{\cQ} = \bar{L}_\pre \bar{L}_\pre^\top$.

\begin{lemma}[Adapted from \citet{agarwal2023SI}]\label{lem:g1}
    Consider the setup of Theorem \ref{thm:finite_sample_consistency}. 
    Then, 
    \begin{align*}
        \norm{\cP \Delta}_2 = O_P \left( 
            \frac{\sqrt{r_{\pre}}}{ \xi''' T_{\pre}^{ \scriptscriptstyle 1/4} \sqrt{|\cI^{(n)}|}}
            +
            \frac{ r_{\pre}^{3/2}  \sqrt{\log\left(T_{\pre} |\cI^{(n)}| \right) } }{(\xi''')^{5/2} \sqrt{|\cI^{(n)}|} \min\left( \sqrt{T_{\pre}} , \sqrt{|\cI^{(n)}|} 
            	\right)}
            	+ 
            	\frac{r_\pre^2 \sqrt{\log \left(T_{\pre} |\cI^{(n)}| \right) }}{(\xi''')^4 \min \left( T_\pre^{3/2} , |\cI^{(n)}|^{3/2} \right) }
        \right) ,
    \end{align*}
    where $\xi'''$ is defined in Theorem \ref{thm:finite_sample_consistency} and depends only on $r$ and $d$. 
    Furthermore, 
    \begin{align*}
        \norm{
            Z^{r_\pre}_{\pre, \cI^{(n)}} \hat{\boldsymbol{\alpha}} - \bbE \bz_{\pre, n}
        }_2^2
        \leq 
        \norm{
            Z^{r_\pre}_{\pre, \cI^{(n)}}  - \bbE Z_{\pre, \cI^{(n)}}
        }_{2, \infty}^2
        \norm{\balphaperp}_1^2
        + 
        2 \left<
            Z^{r_\pre}_{\pre, \cI^{(n)}}
            \Delta , 
            \boldsymbol{\epsilon}_{\pre, n} 
        \right>
        .
    \end{align*}
\end{lemma}

\begin{lemma}[Adapted from \cite{agarwal2023SI}]\label{lem:g2}
    Let $x_t$ be a sequence of independent, zero-mean sub-Gaussian random variables with variance $\bar{\sigma}^2$. 
    Then, $\frac{1}{H} \sum_{t = 1}^H \gamma_t = O_P(\bar{\sigma}^2 / \sqrt{H})$.
\end{lemma}

\begin{lemma}[Adapted from \cite{agarwal2023SI}]\label{lem:g4}
    Let Assumptions \ref{asm:model}-\ref{asm:conditional_exo} and \ref{asm:subG_noise}-\ref{asm:components} hold. 
    Then, 
    \begin{align*}
        \norm{
            Z_{\pre, \cI^{(n)}}^{r_\pre}
            - 
            \bbE Z_{\pre, \cI^{(n)}}
        }_{2, \infty}
        = 
        O_P \left(  \frac{1}{\xi'''} \left(
            \frac{\sqrt{r_\pre T_\pre \log(T_\pre |\cI^{(n)}|)}}{\min(\sqrt{T_\pre}, |\cI^{(n)}|)}
        \right) \right) ,
    \end{align*}
    where $\xi'''$ is defined in Theorem \ref{thm:finite_sample_consistency} and depends only on $r$ and $d$. 
\end{lemma}

\begin{lemma}[Adapted from \cite{agarwal2023SI}]\label{lem:g5}
    Let Assumptions \ref{asm:model}-\ref{asm:rank_balanced_singular_vals} hold. 
    Then, $Z_{\pre, \cI^{(n)}}^{r_\pre} \hat{\boldsymbol\alpha}
        = 
        \bar{\cQ} ( \bbE \bz_{\pre, n} + \boldsymbol{\epsilon}_{\pre, n} )$ and, for any $\chi > 0$,
        \begin{align*}
            P\left(
                \left< 
                    \bar{\cQ} 
                    \boldsymbol{\epsilon}_{\pre, n} , \boldsymbol{\epsilon}_{\pre, n} 
                \right>
                \geq 
                \sigma^2 r_{\pre}
                + \chi
            \right) 
            \leq \exp \left( 
                - \bar{\xi} \left(
                    \frac{\chi^2}{\sigma^4 r_{\pre}} , 
                    \frac{\chi}{\sigma^2}
                \right)
            \right) ,
        \end{align*}
        for some universal constant $\bar{\xi} > 0$.
    Moreover, given $Z_{\pre, \cI^{(n)}}^{r_\pre}$, 
    \begin{align*}
        \left< 
            Z_{\pre, \cI^{(n)}}^{r_\pre} \Delta , 
            \boldsymbol{\epsilon}_{\pre, n} 
        \right>
        = 
        O_P 
        \left( 
            r_{\pre} + \sqrt{T_\pre}
            + 
            \norm{Z_{\pre, \cI^{(n)}}^{r_\pre} - \bbE Z_{\pre, \cI^{(n)}} }_{2, \infty}
            \norm{\balphaperp}_1
        \right) ,
    \end{align*}
    with respect to the randomness in $\boldsymbol{\epsilon}_{\pre, n}$.
\end{lemma}

\begin{lemma}[Adapted from \cite{agarwal2023SI}]\label{lem:g6}
    Let $\bx \in \bbR^m$ be a random variable with independent, zero-mean sub-Gaussian random coordinates with $\norm{x_i}_{\psi_2} \leq K$ for every $i \in [m]$. 
    Let $\bx' \in \bbR^m$ be another random variable that satisfies $\norm{\bx'}_2 \leq K'$. 
    Then, for any $\chi \geq 0$, 
    $$P\left(\left| \sum_{i = 1}^m x'_i x_i \right| \geq \chi \right) \leq 2 \exp\left(- \frac{\bar{\xi} \chi^2}{ (K K')^2} \right) .$$
\end{lemma}

\begin{lemma}[Adapted from \cite{agarwal2023SI}]\label{lem:f2}    For a given unit $n \in [N]$ and counterfactual treatment $\tilde{\ba}$ of interest, 
    suppose Assumptions \ref{asm:model}-\ref{asm:components} hold. 
    Then, conditioned on $\LF$ and $A$, 
    \begin{align*}
       \norm{ \hat{\boldsymbol{\alpha}}
        - 
        \boldsymbol{\alpha} }_2
        = 
        O_P\left( 
        \frac{\sqrt{\log(T_\pre |\cI^{(n)}|)}}{(\xi''')^{3/2}}
        \left(
        	\frac{r_\pre^{3/4}}{T_\pre^{1/4} | \cI^{(n)} |^{1/2}}
        	+
            \frac{ r^{3/2}_\pre }{(\xi''')^{3/2} \min( \sqrt{T_\pre} , \sqrt{ | \cI^{(n)} | } )}
        \right)
        \right) .
    \end{align*}
\end{lemma}

\begin{remark}
    Lemmas \ref{lem:g1}-\ref{lem:f2} are adapted from \citet{agarwal2020synthetic}. 
    One can check that the assumptions for these lemmas hold in our setting. 
    In particular, the main difference between the assumptions in our work and in \cite{agarwal2020synthetic} is the definition of the ``donor set.''
    
    To see how this affects the assumptions, 
    first note that observation pattern in this work 
    allows for any sequence of treatments during the training  period (referred to as the ``pre-intervention'' period in \cite{agarwal2020synthetic}). 
    In contrast, in \cite{agarwal2020synthetic}, 
    the treatment must be constant across the training measurements, and it is assumed that all units are under treatment $0$ during the pre-intervention (i.e., training) period.
    This difference is reflected in the assumptions via the donor set. 
    Once we adjust the choice of donor set (Definition \ref{def:node_donors}) to suit the network interference setting, the assumptions in \cite{agarwal2020synthetic} can be mapped to ours. 
    
    Second, note that the model in \citep{agarwal2020synthetic} 
    is given by (in their notation) 
    \begin{align}
    	Y_{tn}^{(d)} 
    	&=
    	\left< 
    	u_{t}^{(d)} , 
    	v_n
    	\right> 
    	+ 
    	\varepsilon_{tn}^{(d)} , 
    \end{align}
    where $u_{t}^{(d)}, v_n \in \bbR^r$ are latent factors;
    $\varepsilon_{tn}^{(d)}$ is a zero-mean, independent noise term; 
    and $Y_{tn}^{(d)}$ is the potential outcome of interest. 
    Recall from \eqref{eq:abbrev_model} that our model is given by (in our notation)
    \begin{align}
    	Y_{t , n}^{(\bintv_{\cN(n)})} 
    	= \left< \tilde{\bu}_{n, \cN(n)} ,  \tilde{\bw}_{t, {\bintv_{\cN(n)}}} \right> + \epsilon_{t , n}^{(\bintv_{\cN(n)})} , 
    \end{align}
    where $\tilde{\bu}_{n, \cN(n)} ,  \tilde{\bw}_{t, {\bintv_{\cN(n)}}} \in \bbR^{r |\cN(n)|}$ are latent factors;
    $\epsilon_{t , n}^{(\bintv_{\cN(n)})}$ is a zero-mean, independent noise term; 
    and $Y_{t , n}^{(\bintv_{\cN(n)})}$ is the potential outcome of interest. 
    As such, 
    our setup model is analogous to the model used by \cite{agarwal2020synthetic}, with a change of notation.
    
    We now go through the assumptions one-by-one. 
    As we saw above, Assumption \ref{asm:model} is equivalent to Assumption 2 in \citep{agarwal2020synthetic}, with a change of notation. 
    Furthermore, Assumption 
    \ref{assumption:network_sutva} is automatically satisfied when Assumption \ref{asm:model} holds.
    Assumptions \ref{asm:model} and \ref{asm:conditional_exo} together give Assumption 3 of \citep{agarwal2020synthetic}.
    Similarly, 
    Assumptions \ref{asm:subG_noise} and \ref{asm:bounded}
    map one-to-one to Assumptions 5 and 6 of \citep{agarwal2020synthetic} under the change of notation. 
    Assumptions \ref{asm:linear_span} and \ref{asm:subspace_inclusion} also map one-to-one to Assumptions 4 and 8 under the new definition of a donor set, 
    as given by Definition \ref{def:node_donors}.
    Assumption \ref{asm:rank_balanced_singular_vals} is slightly different than Assumption 7 of \citep{agarwal2020synthetic} in that the constants in this work can depend on model rank $r$ or maximum degree $d$ of $\cG$. 
    In \citep{agarwal2020synthetic}, 
    this change is mainly reflected via Equations (41) and (54).
    In both, the right-hand side should be multiplied by a factor of $1/\xi'''$, where $\xi'''$ is defined in Theorem \ref{thm:finite_sample_consistency}.
    Both these changes are reflected in our Lemmas \ref{lem:g1} and \ref{lem:g4} above.

    Lastly, note that \citep{agarwal2020synthetic} analyze the coefficients $\boldsymbol{\alpha}$ as well as $\boldsymbol{\alpha}_\perp = R_\pre R_\pre^\top \boldsymbol{\alpha}$. 
    In our work, our proofs only utilize $\boldsymbol{\alpha}$ because
    $\bbE  [ Z_{\pre, \cI^{(n)}} | \LF , \cO]^+ = R_{\pre} \Sigma_{\pre}^{-1} L_{\pre}$, 
    which implies that $\boldsymbol{\alpha}_\perp = R_{\pre} R_{\pre}^\top R_{\pre} \Sigma_{\pre}^{-1} L_{\pre} \bbE  [ \bz_{\pre, n} | \LF , \cO] = \bbE  [ Z_{\pre, \cI^{(n)}} | \LF , \cO]^+ \bbE  [ \bz_{\pre, n} | \LF , \cO] = \boldsymbol{\alpha}$.
\end{remark}

\begin{lemma}[Adapted from Lemma 19 by \cite{agarwal2023synthetic}]\label{lem:sc19}
	Suppose that Assumptions \ref{asm:model}, \ref{asm:linear_span}, \ref{asm:bounded}, and \ref{asm:rank_balanced_singular_vals} hold. 
	Then, $\norm{\boldsymbol{\alpha}}_2 \leq \frac{\sqrt{r}}{\xi' \sqrt{\xi'' |\cI^{(n)}|}}$, 
	where $\xi'$ and $\xi''$ are defined in Assumption \ref{asm:rank_balanced_singular_vals} and depend only on the model rank $r$ and maximum degree $d$ of $\cG$. 
\end{lemma} 
\begin{remark}
	Lemma \ref{lem:sc19} is adapted from Lemma 19 of \citep{agarwal2023synthetic}. 
	It is easy to verify that the setup is identical, 
	with Assumptions \ref{asm:model} (which automatically satisfies Assumption \ref{assumption:network_sutva}), 
	\ref{asm:linear_span}, 
	\ref{asm:bounded}, 
	and
	\ref{asm:rank_balanced_singular_vals}
	map to Assumptions 1, 
	3b, 
	6, 
	and
	9, 
	 respectively. 
	 Note that the proof of Lemma 19 only requires Assumption 3a (and not Assumption 3b). 
	 There is one important difference,
	 which is that Assumption \ref{asm:rank_balanced_singular_vals}'s constants $\xi'$ and $\xi''$ can depend on $r$ and $d$. 
	 In \cite{agarwal2023synthetic}, this simply means translates to a factor of $\frac{1}{\xi' \sqrt{\xi''}}$,
	 where $\xi'$ and $\xi''$ are defined in Assumption \ref{asm:rank_balanced_singular_vals}, 
	 as reflected in Lemma \ref{lem:sc19}.
\end{remark}

	\begin{lemma}[Adapted from Theorem 3.1 by \cite{matouvsek2008variants}]\label{lem:johnson_lindenstrauss}
		Let $R \in \bbR^{d_1 \times d_2}$. 
		Let $R_{ij} \iid p_R$, 
		where $\bbE[R_{ij}] = 0$,
		$\text{var}(R_{ij}) = 1$,
		and $p_R$ is a sub-Gaussian distribution. 
		Let $\eta \in (0, 1/2]$, 
		$\delta \in (0, 1)$, 
		$d = C \eta^{-2} \log (2 / \delta)$, 
		where $C$ is a constant that depends on $p_R$. 
		Then, 
		with probability at least $1 - \delta$, 
		\begin{align*}
			(1 - \eta) \norm{\bx}_2 \leq \norm{R\bx}_2 \leq (1 + \eta) \norm{\bx}_2 ,
		\end{align*}
		for all $\bx \in \bbR^{d_2}$.
	\end{lemma}
	
	\begin{lemma}\label{lem:asm_8}
		Suppose Assumption \ref{asm:model}, 
		\ref{asm:conditional_exo}, 
		\ref{asm:regular_graph},
		and \ref{asm:unif_lf} hold. 
		Then, under the experiment design in Section \ref{sec:exp_procedure},
		Assumption \ref{asm:rank_balanced_singular_vals} holds with high probability.
	\end{lemma}
\ifarxiv
\begin{proof}
	\else
	\proof{Proof.}
	\fi
	In this proof, we abbreviate $\cI^{(n)}$ to $\cI$.  
	\medskip
	
	\noindent
	\textbf{Decomposing $\bbE[ Z_{\pre, \cI} | \LF , A ]$.}
	Let $\tilde{\Nc}(j)$ denote $\pi_j(\cN(j))$, 
	where $\pi_j$ is specified in Definition \ref{def:node_donors}, 
	i.e., $\tilde{\Nc}(j)$ corresponds to the permuted neighborhood of donor $j$, 
	where the permutation is fixed under Definition \ref{def:node_donors}.
	In the remainder of this proof, 
	we use the decomposition $\bbE[ Z_{\pre, \cI} | \LF , A ] = \tW U_\cI$, 
	where 
	\begin{align}
		\tW &= 
		\begin{bmatrix}
			\tilde{\bw}_{t, A[\cN(n), t]}^\top : t \in \cT_{\pre}]
		\end{bmatrix}
		\in \bbR^{T_\pre \times r (d + 1)} ,
		\\
		U_{\cI} &= 
		\begin{bmatrix}
			\bu_{\tilde{\cN}_j(\cI_k), \cI_k} : j \leq | \cN(n) |  , k \leq  | \cI | 
		\end{bmatrix}
		\in \bbR^{r (d + 1) \times | \cI |}  \, .
	\end{align}
	\smallskip
	
	\noindent
	\textbf{Reducing the problem to upper bounding the condition number of $\tW$.}
	By Assumption \ref{asm:unif_lf}, 
	the variance of $u_{j, i, k}$ and $w_{t, a, k}$ is $\frac{1}{3 r (d+1)}$.
	Applying Lemma \ref{lem:johnson_lindenstrauss}, 
	\begin{align*}
		(1 - \eta) \norm{\tW^\top \bx}_2 \leq \sqrt{\frac{3 r (d+1) }{|\cI|}} 
		\norm{U_\cI^\top \tW^\top \bx}_2 
		\leq 
		(1 + \eta) \norm{\tW^\top \bx}_2 ,
	\end{align*}	 
	for all $\bx \in \bbR^{T_\pre}$ 
	with probability at least $1 - 2 \exp\left( - \frac{|\cI| \eta^2}{C} \right)$ for $\eta \in (0, 1/2]$ and $|\cI| > \frac{C \log 2}{\eta^2}$.
	
	Let $\phi(X)$ denote the condition number of matrix $X$,
	i.e., the ratio of the largest to $r_X$-th largest singular values of $X$, where $r_X$ denotes the rank of $X$. 
	Let $\cB_b$ denote the unit ball in $\bbR^b$. 
	This  implies that 
	\begin{align*}
		(\phi(\tW U))^2 &= 
		\frac{\max_{\bx \in \cB_{T_\pre}} \norm{U_\cI^\top \tW^\top \bx}_2}{\min_{\bx \in \cB_{T_\pre}} \norm{U_\cI^\top \tW^\top \bx}_2}
		\\
		&\leq 
		\frac{\max_{\bx \in \cB_{T_\pre}} (1 + \eta) \sqrt{\frac{|\cI|}{3 r  (d+1) }} \norm{\tW^\top \bx}_2 }{\min_{\bx \in \cB_{T_\pre}} (1 - \eta) \sqrt{\frac{|\cI|}{3 r  (d+1) }} \norm{\tW^\top \bx}_2}
		\\
		&= 
		\frac{(1 + \eta) \max_{\bx \in \cB_{T_\pre}} \norm{\tW^\top \bx}_2 }{(1 - \eta) \min_{\bx \in \cB_{T_\pre}} \norm{\tW^\top \bx}_2 }
		\\
		&\leq \frac{3}{2} (\phi(\tW))^2 .
	\end{align*}
	Therefore, upper bounding the condition number of $\bbE[Z_{\pre, \cI} | \LF, A]$ comes down to upper bounding the condition number of $\tW$. 
	\medskip
	
	\noindent
	\textbf{Bounding the condition number of $\tW$.}
	The condition number of $\tW$ is given by
	\begin{align}
		\sqrt{\frac{\max_{\bx  \in \cB_{r(d+1)}} \bx^\top \tW^\top \tW \bx }{\min_{\bx  \in \cB_{r(d+1)}} \bx^\top \tW^\top \tW \bx }} ,
	\end{align}
	where we once again take $\max$ and $\min$ over $\bx$ for which $\norm{\bx}_2 = 1$. 
	We therefore study $\tW^\top \tW$. 
	Note that $\tW^\top \tW$ can be split into $(d + 1) \times (d + 1)$ block matrices, where each block matrix is $r \times r$. 
	Let the $(j, k)$-th block matrix be denoted by $X^{j, k} \in \bbR^{r \times r}$. 
	By the definition of $\tW$ and $\tw_{\cdot, \cdot}$, 
	the $(j, k)$-th block matrix can be written as:
	\begin{align*}
		X^{j, k}
		&= 
		\sum_{t \in \cT_\pre}
		\bw_{t, A[\cN_j(n), t]}
		\bw_{t, A[\cN_k(n), t]}^\top .
	\end{align*}
	For the remainder of the proof, 
	we assume $D = 2$ for ease of exposition. 
	However, it is easy to show that our results extend for $D  > 2$. 
	
	Now, note that, 
	under the experiment design in Section \ref{sec:experimental_proc}, three facts hold true if $j \neq k$:
	\begin{enumerate}
		\item $j$ and $k$ receive the treatment $1$ at the same time for exactly $T_\pre - 2 \bar{T}$ measurements; and
		
		\item $j$ receives a non-control treatment and $k$ receives treatment $1$ for exactly $\bar{T}$ time steps; and
		
		\item $k$ receives a non-control treatment and $j$ receives treatment $1$ for exactly $\bar{T}$ time steps.
	\end{enumerate}
	Let $\cT_j$ denote the measurements for which $j$ receives a non-control treatment and $\cT_k$ denote the measurements for which $k$ receives a non-control treatment. Note that $\cT_j$ and $\cT_k$ are disjoint by the experiment design in Section \ref{sec:exp_procedure}.
	Note further that $|\cT_j| = |\cT_k| = \bar{T}$. 
	
	Then, if $j \neq k$, 
	\begin{align*}
		X^{j, k}
		&= 
		\sum_{t \in \cT_\pre \setminus \{\cT_j \cup \cT_k \}}
		\bw_{t, 0}
		\bw_{t, 0}^\top 
		+ 
		\sum_{t \in \cT_j  }
		\bw_{t, 1}
		\bw_{t, 0}^\top 
		+ 
		\sum_{t \in \cT_k}
		\bw_{t, 0}
		\bw_{t, 1}^\top ,
	\end{align*}

	\noindent
	We now make use of two facts. 
	First, since $\bw_{\cdot, \cdot}$ is bounded, it is sub-Gaussian. 
	Second, 
	$\bbE [ \bw_{t, 0}
	\bw_{t, 1}^\top ] = \bbE [ \bw_{t, 1}
	\bw_{t, 0}^\top ] = 0$. 
	Third, 
		$\bbE [ \bw_{t, 0}
	\bw_{t, 0}^\top ] = \bbE [ \bw_{t, 1}
	\bw_{t, 1}^\top ] = \text{Cov}(\bw_{\cdot, \cdot}) = \frac{1}{3r(d+1)} \bbI_{r \times r}$. 
	As such, 
	we can characterize $X^{j, j}$ and $X^{j, k}$ in a high-probability sense, 
	as follows:
	\begin{enumerate}
		\item If $j = k$, $X^{j, k} = \Theta_P \left( \frac{T_\pre}{3 r (d + 1)} \bbI_{r \times r}\right)$.
		
		\item If $j \neq k$, $\sum_{t \in \cT_\pre \setminus \{\cT_j \cup \cT_k \}}
		\bw_{t, 0}
		\bw_{t, 0}^\top = \Theta_P \left( \frac{T_\pre - 2 \bar{T}}{3 r (d + 1)} \bbI_{r \times r} \right)$.
		
		\item If $j \neq k$, the two right-hand sums are $\sum_{t \in \cT_j  }
		\bw_{t, 1}
		\bw_{t, 0}^\top 
		+ 
		\sum_{t \in \cT_k}
		\bw_{t, 0}
		\bw_{t, 1}^\top = \Theta_P \left( [0]_{r \times r}\right)$.
	\end{enumerate}
	Therefore, 
	\begin{align}
		\tW^\top \tW = \Theta_P \left( \frac{2 \bar{T}}{3 r (d + 1)} \bbI_{r(d+1) \times r(d+1)} 
		+ 
		\frac{T_\pre - 2 \bar{T}}{3 r (d+1)} \begin{bmatrix}
			\bbI_{r \times r} & \bbI_{r \times r} & \hdots 
			\\
			\bbI_{r \times r} & \bbI_{r \times r} & \hdots 
			\\
			\vdots & \vdots & \ddots
		\end{bmatrix}
		\right) ,
	\end{align}
	and
	\begin{align}
		\bx^\top \tW^\top \tW \bx &= \Theta_P \left( \frac{2 \bar{T}}{3 r (d + 1)} \norm{\bx}_2^2
		+ 
		\frac{T_\pre - 2 \bar{T}}{3 r (d+1)} \norm{\sum_{k = 1}^{d + 1} \bx_k}_2^2
		\right)
		\\
		&= 
		\Theta_P \left( \frac{2 \bar{T}}{3 r (d + 1)} 
		+ 
		\frac{T_\pre - 2 \bar{T}}{3 r (d+1)} \norm{\sum_{k = 1}^{d + 1} \bx_k}_2^2
		\right) ,
	\end{align}
	where $\bx^\top = [\bx_1^\top, \bx_2^\top, \hdots, \bx_{d+1}^\top]$, 
	and the second equality follows from the fact that we restrict our attention to $\bx$ for which $\norm{\bx}_2 = 1$. 
	Note that $\norm{\sum_{k = 1}^{d + 1} \bx_k}_2^2 = \sum_{\ell = 1}^r ( \sum_{k = 1}^{d + 1} x_{k, \ell})^2 \leq (\sum_{\ell = 1}^r  | \sum_{k = 1}^{d + 1} x_{k, \ell} | )^2  \leq (\sum_{\ell = 1}^r   \sum_{k = 1}^{d + 1} | x_{k, \ell} | )^2 \leq \norm{\bx}_1^2 \leq r(d+1)\norm{\bx}_2^2$.
	Therefore, 
	\begin{align*}
		\phi(\tW) &= 
		\sqrt{\frac{\max_{\bx \in \cB_{r(d+1)}} \bx^\top \tW^\top \tW \bx}{\min_{\bx \in \cB_{r(d+1)}} \bx^\top \tW^\top \tW \bx}}
		\\
		&= O_P \left(   
		\sqrt{1  + \frac{r (d+1) (T_\pre - 2 \bar{T})}{2 \bar{T}}}
		\right)
		\\
		&=
		O_P \left(    \sqrt{1 + \frac{r(d+1)  (d^2 - 1 )}{2 }}
		\right)
		\\
		&=  O_P \left(    \sqrt{1 + \frac{r(d+1)^3 }{2 }}
		\right)
		\\
		&= O_P \left(    \sqrt{1 + 4 r d^3}
		\right) ,
	\end{align*}
	where the second inequality follows from the fact that there are at most $d^2 + 1$ sets of $\bar{T}$ that make up $\cT_\pre$ under the experiment design in Section \ref{sec:exp_procedure}.
	Therefore, the requirement on the condition number in Assumption \ref{asm:rank_balanced_singular_vals} holds with $\xi' = (1 + 4rd^3)^{-1/2}$.
	\medskip
	
	\noindent 
	\textbf{Requirement on Frobenius norm.}
	It remains to show that the second requirement of Assumption \ref{asm:rank_balanced_singular_vals} holds. 
	To do so, note that
	\begin{align}
		\norm{\bbE[ Z_{\pre, \cI} | \LF, A ]}_2^2
		&= 
		\sum_{t,j} (\bbE[ Z_{\pre, \cI} | \LF, A ]_{tj})^2 \nonumber
		\\
		&=
		\sum_{t,j} (
		\tbw_{t, A[\cN(n), t] }^\top 
		\tbu_{j, \cN(j) }
		)^2 \nonumber
		\\
		&=
		\sum_{t,j} \left( \sum_{b = 1}^{d+1}
		\bw_{t, A[\cN_b(n), t] }^\top 
		\bu_{\cN_b(j) , j}
		\right)^2 \nonumber
		\\
		&=
		\sum_{t,j} \left(
		\sum_{a \in [D]}
		\bw_{t, a }^\top 
		\sum_{b = 1}^{d+1}
		\bu_{\cN_b(j) , j}
		\Ind(A[\cN_b(n), t]  = a)
		\right)^2 . \label{eq:asm8_proof_1}
	\end{align}
	Note that 
	\begin{align}
		\bbE& \left[ \left( \sum_{a \in [D]}
		\bw_{t, a }^\top 
		\sum_{b = 1}^{d+1}
		\bu_{\cN_b(j) , j}
		\Ind(A[\cN_b(n), t]  = a)
		\right)^2 \right] \nonumber
		\\
		&\geq 
		\sum_{a \in [D]} \bbE \left[  \left( 
		\bw_{t, a }^\top 
		\sum_{b = 1}^{d+1}
		\bu_{\cN_b(j) , j}
		\Ind(A[\cN_b(n), t]  = a)
		\right)^2 \right]\nonumber
		\\
		&= 
		\sum_{a \in [D]} \bbE \left[  \left( 
		\sum_{k=1}^r
		w_{t, a , k} 
		\sum_{b = 1}^{d+1}
		u_{\cN_b(j) , j, k}
		\Ind(A[\cN_b(n), t]  = a)
		\right)^2 \right] \nonumber
		\\
		&\geq
		\sum_{a \in [D]} 
		\sum_{k=1}^r
		\bbE \left[  
		\left( 
		w_{t, a , k} 
		\sum_{b = 1}^{d+1}
		u_{\cN_b(j) , j, k}
		\Ind(A[\cN_b(n), t]  = a)
		\right)^2 \right] \nonumber
		\\
		&=
		\sum_{a \in [D]} 
		\sum_{k=1}^r
		\bbE \left[  
		w_{t, a , k}^2 
		\left( 
		\sum_{b = 1}^{d+1}
		u_{\cN_b(j) , j, k}
		\Ind(A[\cN_b(n), t]  = a)
		\right)^2 \right] \nonumber
		\\
		&\geq
		\sum_{a \in [D]} 
		\sum_{k=1}^r
		\bbE \left[  
		w_{t, a , k}^2 
		\sum_{b = 1}^{d+1}
		u_{\cN_b(j) , j, k}^2
		\Ind(A[\cN_b(n), t]  = a)^2
		\right] \nonumber
		\\
		&=
		\sum_{a \in [D]} 
		\sum_{k=1}^r
		\bbE \left[  
		w_{t, a , k}^2 \right]
		\sum_{b = 1}^{d+1}
		\bbE \left[  
		u_{\cN_b(j) , j, k}^2
		\right]
		\Ind(A[\cN_b(n), t]  = a) \nonumber
		\\
		&=
		\frac{1}{9 r^2 (d+1)^2}
		\sum_{a \in [D]} 
		\sum_{k=1}^r
		\sum_{b = 1}^{d+1}
		\Ind(A[\cN_b(n), t]  = a) \nonumber
		\\
		&=
		\frac{1}{9 r (d+1)} . \label{eq:asm8_proof_2}
	\end{align}
	Since $w_{t, a, k}$ and $u_{\ell, j, k}$ are bounded, 
	$w^2_{t, a, k}$ and $u^2_{\ell, j, k}$ are sub-Gaussian. 
	As such, 
	combining \eqref{eq:asm8_proof_1} and  \eqref{eq:asm8_proof_2} gives
	\begin{align*}
		\bbE[Z_{\pre, \cI} | \LF, A] = \omega_P \left( \frac{T_\pre |\cI| }{9 r (d+1)} \right) ,
	\end{align*}
	which confirms that Assumption \ref{asm:rank_balanced_singular_vals} holds with high probability for $\xi'' = (9 r (d+1))^{-1}$. 
\ifarxiv
\end{proof}
\else
\Halmos
\endproof
\fi

%% file: sections/app_proofs.tex
\subsection{Proof of Theorem \ref{thm:ID}}

\ifarxiv
\begin{proof}
\else
\proof{Proof.}
\fi
Recall from Definition \ref{def:identifiability} that identifiability requires that the estimand $f(\theta)$ can be written as a function $g(P_\theta)$ of the data distribution $P_\theta$, where $\theta$ are the unknown model parameters. In our setting, the unknown model parameters are the latent factors, thus $\theta = \LF$. The observed dataset consists of the matrices of observed outcomes $Z[\cT_{\pre}, :]$ and $Z[\cT_{\post}, :]$, whose joint distribution, denoted $P_\theta$, is both a function of the unknown parameter $\theta$ and the known and fixed parameters $(A, G, \cT_\pre, \cT_\post)$.

Our estimand  $f(\theta) = \IPO{n}
= \frac{1}{T_{\post}} \sum_{t \in \cT_{\post}} \Ex\Big[Y^{(\tilde{\ba}_{\cN(n)} )}_{t , n} \Big].$
The claim in Theorem \ref{thm:ID} is that $\IPO{n}$ is identifiable as we can write it as a function of expectations of the data, as given by 
\begin{align*}
f(\theta) := \IPO{n} = \frac{1}{T_{\post}} {\bf 1}^T ~\bbE [ Z_{\post, \cI^{(n)}} \, |  \, \LF, \cO ] ~\bbE  [ Z_{\pre, \cI^{(n)}} | \LF , \cO]^+ ~\bbE  [ \bz_{\pre, n} | \LF , \cO] =: g(P_{\theta}).
\end{align*}
To show this claim, we first define some additional notation. Let $U_{\cI^{(n)}}$ be a $r|\cN(n)|\times|\cI^{(n)}|$ matrix where the $j$-th column of $U_{\cI^{(n)}}$ corresponds to the network-adjusted latent factor associated to the $j$-th donor, i.e. $\tilde{\bu}_{i, \cN(v)}$ where unit $i$ is the $j$-th donor in $\cI^{(n)}$. Let $W_\pre$ be a $r|\cN(n)|\times T_{\pre}$ matrix where the $j$-th column of $W_\pre$ corresponds to the network adjusted latent factor and the {\em applied} treatment associated to the $j$-th measurement in the training period $\cT_\pre$, i.e. $\tilde{\bw}_{t, a^t_{\cN(n)}}$ where $t$ is the $j$-th measurement in $\cT_{\pre}$. Similarly let $W_\post$ be a $r|\cN(n)|\times T_{\post}$ matrix where the $j$-th column of $W_\post$ corresponds to the network adjusted latent factor associated to the $j$-th measurement and the {\em counterfactual} treatment $\tilde{\ba}_{\cN(n)}$ in the prediction period $\cT_\post$. 

By conditional exogeneity as stated in Assumption \ref{asm:conditional_exo} and the latent factor model as stated in Assumption \ref{asm:model}, it follows that for any $t,i,\tilde{\ba}$,
\[\bbE [ Y_{t, i}^{(\tilde{\ba})} |  \, \LF, \cO ] = \langle \tilde{\bw}_{t, \tilde{\ba}_{\cN(i)}}, \tilde{\bu}_{i, \cN(i)} \rangle.\]
As a result, we can write the target estimand as
\begin{align*}
\IPO{n} &:= \frac{1}{T_{\post}} \sum_{t \in \cT_{\post}} \Ex\Big[Y^{(\tilde{\ba}_{\cN(n)} )}_{t , n} \Big] 
= \frac{1}{T_{\post}} {\bf 1}^T ~W_\post^T \tilde{\bu}_{n, \cN(n)}.
\end{align*}
By the linear span property as stated in Assumption \ref{asm:linear_span}, there must exist a vector $\boldsymbol{\lambda} \in \bbR^{|\cI^{(n)}|}$ such that $\tilde{\bu}_{n, \cN(n)} = U_{\cI^{(n)}} \boldsymbol{\lambda}$. Along with the latent factor model decomposition and the condition that donors must share the same applied treatment as $n$ in the training period, it also follows that 
$\bbE  [ \bz_{\pre, n} | \LF , \cO] = \bbE  [ Z_{\pre, \cI^{(n)}} | \LF , \cO] \boldsymbol{\lambda}$. 
By substitution,
\begin{align*}
\IPO{n} &= \frac{1}{T_{\post}} {\bf 1}^T ~W_\post^T ~U_{\cI^{(n)}} \boldsymbol{\lambda} 
= \frac{1}{T_{\post}} {\bf 1}^T ~\bbE [ Z_{\post, \cI^{(n)}} \, |  \, \LF, \cO ] \boldsymbol{\lambda},
\end{align*}
where the latter equality follows from the latent factor model, conditional exogeneity, and the construction of the donor set, which enforces that the applied treatments to the donors during the prediction period must match the counterfactual treatment.
By the subspace inclusion property as stated in Assumption \ref{asm:subspace_inclusion}, there must exist a matrix $\Gamma \in \bbR^{T_\post \times T_\pre}$ such that $\bbE [ Z_{\post, \cI^{(n)}} \, |  \, \LF, \cO ] = \Gamma ~\bbE  [ Z_{\pre, \cI^{(n)}} | \LF , \cO]$, such that by substitution
\begin{align*}
\IPO{n} &= \frac{1}{T_{\post}} {\bf 1}^T ~\Gamma ~\bbE  [ Z_{\pre, \cI^{(n)}} | \LF , \cO] \boldsymbol{\lambda}, \\
&\overset{\text{(a)}}{=} \frac{1}{T_{\post}} {\bf 1}^T ~\Gamma ~\bbE  [ Z_{\pre, \cI^{(n)}} | \LF , \cO] ~\bbE  [ Z_{\pre, \cI^{(n)}} | \LF , \cO]^+ ~\bbE  [ Z_{\pre, \cI^{(n)}} | \LF , \cO] \boldsymbol{\lambda}, \\
&\overset{\text{(b)}}{=} \frac{1}{T_{\post}} {\bf 1}^T ~\bbE [ Z_{\post, \cI^{(n)}} \, |  \, \LF, \cO ] ~\bbE  [ Z_{\pre, \cI^{(n)}} | \LF , \cO]^+ ~\bbE  [ \bz_{\pre, n} | \LF , \cO] =: g(P_{\theta}).
\end{align*}
where (a) follows from the property of pseudoinverses, and (b) follows from the construction of $\Gamma$ and $\boldsymbol{\lambda}$ from the linear span and subspace inclusion properties.
\ifarxiv
\end{proof}
\else
\Halmos
\endproof
\fi

\subsection{Proof of Theorem \ref{thm:finite_sample_consistency}} \label{sec:app_thm_consistency}

\ifarxiv
\begin{proof}
\else
\proof{Proof.}
\fi 
In this proof, we suppress the conditioning on $\LF$ and $\cO$. 
Let $\Delta =  \hat{\boldsymbol{\alpha}} - \balphaperp$, 
where $\boldsymbol{\alpha}$ is defined in Section \ref{sec:id}.
Let $\boldsymbol{\epsilon}_{t, \cI^{(n)}}  = \left[ \epsilon_{t, j}^{( \ba_{\cN(j)}^t )} : j \in \cI^{(n)} \right] \in \bbR^{| \cI^{(n)} |}$. 
Lastly, recall that $R_{\post}$ and $R_{\pre}$ denote the matrices containing the right singular vectors of $\bbE \big[  Z_{\post, \cI^{(n)} }  \big| \, \LF, \cO \big]$ and $\bbE \big[  Z_{\pre, \cI^{(n)} }  \big| \, \LF, \cO \big]$, respectively.

	By \eqref{eq:estimand}, \eqref{eq:NSI_estimator_point}, the definition of $\boldsymbol{\alpha}$ in Section \ref{sec:id} and the definition of $\hat{\boldsymbol{\alpha}}$ in Section \ref{sec:id},
\begin{align*}
	\hIPO{n} - &\IPO{n}
	\\
	&= 
	\frac{1}{T_{\post}}
	\sum_{t \in \cT_{\post}}
	\left(
		\left< Z[t, \cI^{(n)}] , \hat{\boldsymbol{\alpha}} \right>
		- 
		\left< \bbE Z[t, \cI^{(n)}]  , \balphaperp \right>
	\right)
	\\
	&= 
	\frac{1}{T_{\post}}
	\sum_{t \in \cT_{\post}}
	\left(
	\left< \bbE Z[t, \cI^{(n)}]  ,
	\Delta \right>
	-
	\left< \bbE Z[t, \cI^{(n)}]  ,
	\hat{\boldsymbol{\alpha}}  \right>
	+ 
	\left< \bbE Z[t, \cI^{(n)}] +  \boldsymbol{\epsilon}_{t, \cI^{(n)}} , \hat{\boldsymbol{\alpha}} \right>
	\right)
	\\
	&= 
	\frac{1}{T_{\post}}
	\sum_{t \in \cT_{\post}}
	\left(
	\left< \bbE Z[t, \cI^{(n)}]  ,
	\Delta \right>
	+ 
	\left<  \boldsymbol{\epsilon}_{t, \cI^{(n)}} , \hat{\boldsymbol{\alpha}} \right>
	\right)
	\\
	&= 
	\frac{1}{T_{\post}}
	\sum_{t \in \cT_{\post}}
	\left(
	\left< \bbE Z[t, \cI^{(n)}]  ,
	\Delta \right>
	+ 
	\left<  \boldsymbol{\epsilon}_{t, \cI^{(n)}}, \balphaperp
	\right>
	+
	\left<  \boldsymbol{\epsilon}_{t, \cI^{(n)}},
	\Delta
	 \right>
	\right) ,
\end{align*}
where the second equality follows from $Z[t, \cI^{(n)}]  = \bbE Z[t, \cI^{(n)}]  + \boldsymbol{\epsilon}_{t, \cI^{(n)}}$.

Let $\cP = R_{\pre} R_{\pre}^\top$. 
By Assumption \ref{asm:subspace_inclusion}, 
$R_\post = \cP R_{\post}$, 
which implies 
$\bbE[Z_{\post, \cI^{(n)}} ] = \bbE[Z_{\post, \cI^{(n)}} ] \cP$. 
Therefore, 
\begin{align}
\hIPO{n} - \IPO{n} =
	\frac{1}{T_{\post}}
	\sum_{t \in \cT_{\post}}
	\left(
	\left< \bbE Z[t, \cI^{(n)}]  ,
	\cP \Delta \right>
	+ 
	\left<  \boldsymbol{\epsilon}_{t, \cI^{(n)}} , \balphaperp
	\right>
	+
	\left<  \boldsymbol{\epsilon}_{t, \cI^{(n)}} ,
	\Delta
	\right>
	\right) . \label{eq:thm_three_terms}
\end{align}
The three terms on the right-hand side can be bounded using Lemmas \ref{lem:g1}, \ref{lem:g2}, and \ref{lem:f2}, as follows.
\\

\noindent 
\textbf{Bounding the three terms in \eqref{eq:thm_three_terms}}.
To bound the first term in \eqref{eq:thm_three_terms}, observe that
\begin{align*}
	\left< \bbE Z[t, \cI^{(n)}]  ,
	\cP \Delta \right>
&	\leq 
	\norm{\bbE Z[t, \cI^{(n)}]  }_2 
	\norm{\cP \Delta}_2 
	\leq 
	\norm{\cP \Delta}_2 \sqrt{| \cI^{(n)} |} 
	\\
	&\implies 
	\frac{1}{T_{\post}}
	\sum_{t \in \cT_{\post}}
	\left< \bbE Z[t, \cI^{(n)}]  ,
	\cP \Delta \right>
	\leq 
	\norm{\cP \Delta}_2  \sqrt{| \cI^{(n)} |} ,
\end{align*} 
where the first inequality follows from the Cauchy-Schwartz inequality and the second inequality from Assumption \ref{asm:bounded}.
One can then upper bound $\norm{\cP \Delta}_2$ using Lemma \ref{lem:g1} to get
\begin{align*}
		\frac{1}{T_{\post}}
	\sum_{t \in \cT_{\post}} &
	\left< \bbE Z[t, \cI^{(n)}]  ,
	\cP \Delta \right>
	\\
	&=
	O_P \left(
	\frac{\sqrt{r_{\pre}}}{ \xi''' T_{\pre}^{ \scriptscriptstyle 1/4} }
	+
	\frac{r_{\pre}^{3/2}  \sqrt{\log\left(T_{\pre} |\cI^{(n)}| \right) } }{(\xi''')^{5/2} \min\left( \sqrt{T_{\pre}} , \sqrt{|\cI^{(n)}|} 
		\right)}
	+ 
	\frac{r_\pre^2 \sqrt{ |\cI^{(n)}| \log \left(T_{\pre} |\cI^{(n)}| \right) }}{(\xi''')^4 \min \left( T_0^{3/2} , N_d^{3/2} \right) }
	\right) ,
\end{align*}
where $\xi'''$ is defined in Theorem \ref{thm:finite_sample_consistency}.

To bound the second term in \eqref{eq:thm_three_terms}, 
observe that 
\begin{align*}
	\bbE \left[ \left< \boldsymbol{\epsilon}_{t, \cI^{(n)}} , \balphaperp
	\right> \right] = 0 , 
	\qquad 
	\text{Var}(
		\boldsymbol{\epsilon}_{t , \cI^{(n)}} , 
		\balphaperp
	)
	= 
	\sigma^2 \norm{\boldsymbol{\alpha}}_2^2 ,
\end{align*}
for all $t \in \cT_{\post}$ by
by Assumptions \ref{asm:model} and \ref{asm:subG_noise}. 
Furthermore, 
Assumption \ref{asm:subG_noise} gives that $	\left<  \boldsymbol{\epsilon}_{t, \cI^{(n)}}, \balphaperp
\right>$ are independent across $t$. 
By Lemmas \ref{lem:g2} and \ref{lem:sc19}, 
\begin{align*}
	\frac{1}{T_{\post}} 
	\sum_{t \in \cT_{\post}}
		\left<  \boldsymbol{\epsilon}_{t, \cI^{(n)}} , \balphaperp
	\right>
	= O_P 
	\left( 
		\frac{\norm{\balphaperp}_2}{\sqrt{T_{\post}}}
	\right) = 
	O_P \left(
		\frac{\sqrt{r_\pre}}{ \xi' \sqrt{\xi'' T_\post |\cI^{(n)}|}}
	\right) .
\end{align*}
Lastly, to bound the third term in \eqref{eq:thm_three_terms}, 
we define the following events
\begin{align*}
	E_1 &= \left\{
		\norm{\Delta}_2 = O \left( 
		\frac{\sqrt{\log(T_\pre |\cI^{(n)}|)}}{(\xi''')^{3/2}}
		\left(
		\frac{r_\pre^{3/4}}{T_\pre^{1/4} | \cI^{(n)} |^{1/2}}
		+
		\frac{ r^{3/2}_\pre }{(\xi''')^{3/2} \min( \sqrt{T_\pre} , \sqrt{ | \cI^{(n)} | } )}
		\right)
		\right) 
	\right\} ,
	\\
	E_2 &= \Bigg\{
			\frac{1}{T_{\post}}
			\sum_{t \in \cT_{\post}}
			\left<
				\boldsymbol{\epsilon}_{t, \cI^{(n)}} , \Delta
			\right>
			\\
			& \hspace{50pt} = O 
			\left( 
			\frac{\sqrt{\log(T_\pre |\cI^{(n)}|)}}{\sqrt{T_{\post}} (\xi''')^{3/2}}
			\left(
			\frac{r_\pre^{3/4}}{T_\pre^{1/4} | \cI^{(n)} |^{1/2}}
			+
			\frac{ r^{3/2}_\pre }{(\xi''')^{3/2} \min( \sqrt{T_\pre} , \sqrt{ | \cI^{(n)} | } )}
			\right)
			\right) 
		\Bigg\} .
\end{align*}
Noting that $\left< 
\boldsymbol{\epsilon}_{t , \cI^{(n)}} , \Delta
\right>$ are independent across $t$, 
Lemmas \ref{lem:f2} and \ref{lem:g2} imply that $E_1$ and $E_2 | E_1$ occur with high probability, which also implies that $E_2$ occurs with high probability and therefore that
\begin{align*}
	\frac{1}{T_{\post}} 
	\sum_{t \in \cT_{\post}}
	\left<  \boldsymbol{\epsilon}_{t, \cI^{(n)}} ,
	\Delta
	\right>
	= O_P
	\left( 
	\frac{\sqrt{\log(T_\pre |\cI^{(n)}|)}}{\sqrt{T_{\post}} (\xi''')^{3/2}}
	\left(
	\frac{r_\pre^{3/4}}{T_\pre^{1/4} | \cI^{(n)} |^{1/2}}
	+
	\frac{ r^{3/2}_\pre }{(\xi''')^{3/2} \min( \sqrt{T_\pre} , \sqrt{ | \cI^{(n)} | } )}
	\right)
	\right) .
\end{align*}
Note that we can safely assume that $\xi''' < 1$
because if there exists a $\xi''' \geq 1$, 
letting $\xi'''$ take some value less than $1/\xi'''$ will always satisfy Assumption \ref{asm:rank_balanced_singular_vals}.
Together, the three bounds and the observation above give the theorem result.
\ifarxiv
\end{proof}
\else
\Halmos
\endproof
\fi

\subsection{Proof of Theorem \ref{thm:asym_normality}}

\ifarxiv
\begin{proof}
\else
\proof{Proof.}
\fi
	In this proof, we suppress the conditioning on $\LF$ and $\cO$. 
	Let $\Delta = \hat{\boldsymbol{\alpha}} - \balphaperp$,
	where $\boldsymbol{\alpha}$ is defined below Theorem \ref{thm:ID}.
	Let $\boldsymbol{\epsilon}_{\pre, n} = [\epsilon_{\tau, n}^{(\ba^\tau_{\cN(n)})} : \tau \in \cT_\pre]$. 
	Let $\boldsymbol{\epsilon}_{t, \cI^{(n)}}  = [ \epsilon_{t, j}^{( \ba_{\cN(j)}^t )} : j \in \cI^{(n)} ] \in \bbR^{| \cI^{(n)} |}$. 
	Recall that $R_{\pre}$ denotes the matrix containing the right singular vectors of $\bbE \big[  Z_{\pre, \cI^{(n)} }  \big| \, \LF, \cO \big]$.
    Let $Z_{\pre, \cI^{(n)}}^{r_{\pre}} = \sum_{\ell = 1}^{r_\pre} \hat{s}_\ell \hat{\boldsymbol{\mu}}_\ell \hat{\boldsymbol{\nu}}_\ell^\top = \bar{L}_\pre \bar{\Sigma}_\pre \bar{R}_\pre^\top$,
    	where $ \hat{s}_\ell$, $ \hat{\boldsymbol{\mu}}_\ell$, and $\hat{\boldsymbol{\nu}}_\ell$ are defined in Section \ref{sec:estimation_proc}.
    Let $\cP = R_{\pre} R_{\pre}^\top$ and $\bar{\cQ} = \bar{L}_\pre \bar{L}_\pre^\top$.
	\\
	
	\noindent 
	\textbf{Asymptotic normality}. 
	We begin by establishing that, conditioned on $\LF$ and $\cO$, 
	\begin{align*}
		\frac{\sqrt{T_{\pre}}}{\sigma \norm{\balphaperp}_2}
		\left(
		\IPO{n}
		- 
		\hIPO{n}
		\right)
		\stackrel{d}{\rightarrow}
		\cN(0, 1) .
	\end{align*}
	We use an equation from the proof of Theorem \ref{thm:finite_sample_consistency}. 
	By \eqref{eq:thm_three_terms}, 
	we have
	\begin{align}
		\hIPO{n} - \IPO{n}
		&= 
		\frac{1}{{T_\post}}
		\sum_{t \in \cT_{\post}}
		\left(
		\left< \bbE Z[t, \cI^{(n)}]  ,
		\cP \Delta \right>
		+ 
		\left<  \boldsymbol{\epsilon}_{t, \cI^{(n)}} , \balphaperp
		\right>
		+
		\left<  \boldsymbol{\epsilon}_{t, \cI^{(n)}} ,
		\Delta
		\right>
		\right)  \label{eq:thm_normal_1}
	\end{align}
	We now analyze each of the three terms on the right-hand side of \eqref{eq:thm_normal_1}.

	To characterize the first term in \eqref{eq:thm_normal_1}, observe that
	\begin{align*}
		\left< \bbE Z[t, \cI^{(n)}]  ,
		\cP \Delta \right>
		&	\leq 
		\norm{\bbE Z[t, \cI^{(n)}]  }_2 
		\norm{\cP \Delta}_2 
		\leq 
		\norm{\cP \Delta}_2 \sqrt{| \cI^{(n)} |} 
		\\
		&\implies
		\frac{1}{\sigma \norm{\boldsymbol{\alpha}}_2 \sqrt{T_\post}} 
		\sum_{t \in \cT_{\post}}
		\left< \bbE Z[t, \cI^{(n)}]  ,
		\cP \Delta \right>
		\leq 
		\norm{\cP \Delta}_2  	\frac{\sqrt{T_\post | \cI^{(n)} | }}{\sigma \norm{\boldsymbol{\alpha}}_2 } ,
	\end{align*} 
	where the first inequality follows from the Cauchy-Schwartz inequality and the second inequality from Assumption \ref{asm:bounded}.
	Then,
	{
	\begin{align*}
			\sum_{t \in \cT_{\post}}
		\left< \bbE Z[t, \cI^{(n)}]  ,
		\cP \Delta \right> 
		&\leq 
			\sum_{t \in \cT_{\post}}
		\norm{	\cP \Delta }_1
		\leq 		\sum_{t \in \cT_{\post}}
		\norm{	\cP \Delta }_2 \sqrt{|\cI^{(n)}|}
		\leq 		\norm{	\cP \Delta }_2 T_{\post} \sqrt{|\cI^{(n)}|} .
	\end{align*}
	Therefore, 
	\begin{align*}
		\frac{1}{\sigma \norm{\boldsymbol{\alpha}}_2 \sqrt{T_\post}} 
		\sum_{t \in \cT_{\post}}
		\left< \bbE Z[t, \cI^{(n)}]  ,
		\cP \Delta \right>
		&\leq 
		\frac{	\norm{	\cP \Delta }_2 \sqrt{T_\post |\cI^{(n)}|}  }{\sigma \norm{\boldsymbol{\alpha}}_2 } 
		= o_P(1) ,
	\end{align*}
	under the condition $\norm{\Delta}_2 = o_P\left( 
	\frac{\sigma \norm{\boldsymbol{\alpha}}_2}{\sqrt{T_\post | \cI^{(n)} |}}
	\right)$, 
	as given in the theorem statement. }
	
	To characterize the second term in \eqref{eq:thm_normal_1}, 
	observe that 
	\begin{align*}
		\bbE \left[ \left< \boldsymbol{\epsilon}_{t, \cI^{(n)}} , \balphaperp
		\right> \right] = 0 , 
		\qquad 
		\text{Var}(
		\boldsymbol{\epsilon}_{t , \cI^{(n)}} , 
		\balphaperp
		)
		= 
		\sigma^2 \norm{\boldsymbol{\alpha}}_2^2 ,
	\end{align*}
	for all $t \in \cT_{\post}$ by
	Assumptions \ref{asm:model} and \ref{asm:subG_noise}. 
	By the Lindelberg-L\'{e}vy Central Limit Theorem,
	\begin{align}
		\sqrt{T_{\post}} \left(
			\frac{\frac{1}{T_\post} \sum_{t \in \cT_\post} \left<  \boldsymbol{\epsilon}_{t, \cI^{(n)}} , \balphaperp
				\right> }{ \sqrt{ \text{Var}( \left<  \boldsymbol{\epsilon}_{t, \cI^{(n)}} , \balphaperp
				\right> ) } }
		\right) 
		& \stackrel{d}{\rightarrow} \cN(0, 1) 
		\\
		\implies 
		\frac{1}{ \sigma \norm{\balphaperp}_2 \sqrt{T_\post} }
		\sum_{t \in \cT_{\post}} \left<  \boldsymbol{\epsilon}_{t, \cI^{(n)}} , \balphaperp
		\right> 
		& \stackrel{d}{\rightarrow} \cN(0, 1) 
	\end{align}
	To characterize the third term in \eqref{eq:thm_normal_1},
	note that $\left< 
	\boldsymbol{\epsilon}_{t , \cI^{(n)}} , \Delta
	\right>$ are independent across $t$ and mean-zero for $t \in \cT_\post$. 
	We define two events:
	\begin{align*}
		E_1 &= \left\{ 
			\norm{\Delta}_2 = o\left(
				\sqrt{\frac{\norm{\boldsymbol{\alpha}}_2}{\sigma}}
			\right)
		\right\} ,
		\\
		E_2 &= \left\{
			\frac{1}{T_{\post}}
			\sum_{t \in \cT_{\post}}
			\left<
			\boldsymbol{\epsilon}_{t, \cI^{(n)}} , \Delta
			\right>
			= 
			o\left( 
				\frac{\norm{\boldsymbol{\alpha}}_2 \sigma}{\sqrt{T_\post}}
			\right)
		\right\}
	\end{align*}
	By the theorem statement, 
	$E_1$ holds with high probability. 
	Moreover, Lemma \ref{lem:g2} implies that $E_2 | E_1$ holds with high probability since, 
	conditioned on $E_1$, 
	$\left<
	\boldsymbol{\epsilon}_{t, \cI^{(n)}} , \Delta
	\right>$ 
	is sub-Gaussian with variance upper bounded by $\norm{\boldsymbol{\alpha}}_2 \sigma$.
	Therefore, $E_2$ holds with high probability, 
	which implies that 
	\begin{align*}
		\frac{1}{\sigma \norm{\balphaperp}_2 \sqrt{T_\post}}
		\sum_{t \in \cT_{\post}}
		\left<  \boldsymbol{\epsilon}_{t, \cI^{(n)}} ,
		\Delta
		\right>
		= 
		o_P
		\left( 
			1
		\right) ,
	\end{align*}
	Combining all three terms, 
	\begin{align}
		\frac{\sqrt{T_\post}}{\sigma \norm{\boldsymbol{\alpha}}_2}
		\left(
			\frac{1}{T_\post}
			\sum_{t \in \cT_{\post}}
			\left(
			\left< \bbE Z[t, \cI^{(n)}]  ,
			\cP \Delta \right>
			+ 
			\left<  \boldsymbol{\epsilon}_{t, \cI^{(n)}} , \balphaperp
			\right>
			+
			\left<  \boldsymbol{\epsilon}_{t, \cI^{(n)}} ,
			\Delta
			\right>
			\right) 
		\right)
		&\stackrel{d}{\rightarrow} \cN(0, 1)
		\\
		\implies 
		\frac{\sqrt{T_\post}}{\sigma \norm{\boldsymbol{\alpha}}_2}
		\left(
		\hIPO{n} - \IPO{n}
		\right)
		&\stackrel{d}{\rightarrow}  \cN(0, 1) ,
	\end{align}
	as stated in the theorem. 
	\\
	
	\noindent 
	\textbf{Convergence of variance}.
	By the definition of $\hat{\boldsymbol{\alpha}}$ in Section \ref{sec:estimation_proc} and Assumption \ref{asm:components},
    we know that
	$Z_{\pre, \cI^{(n)}} \hat{\boldsymbol{\alpha}} = Z^{r_\pre}_{\pre, \cI^{(n)}} \hat{\boldsymbol{\alpha}}$. 
	Then by the definition of $\hat{\sigma}^2$ in Section \ref{sec:estimation_proc},
	\begin{align}
		| \hat{\sigma}^2 - \sigma^2 |
		&= 
		\left| \frac{1}{T_{\pre}}
		\norm{
			\bz_{\pre, n}  - Z_{\pre, \cI^{(n)}}^{r_\pre} \hat{\boldsymbol{\alpha}} 
		}_2^2 - \sigma^2 \right| \nonumber
		\\
		&= \left| \frac{1}{T_{\pre}}
		\norm{
			\bbE[ \bz_{\pre, n} ] - Z_{\pre, \cI^{(n)}}^{r_\pre} \hat{\boldsymbol{\alpha}}
		}_2^2
		+ 
		\left( \frac{1}{T_{\pre}} \norm{\boldsymbol{\epsilon}_{\pre, n}}_2^2 - \sigma^2 \right)
		+  \frac{2}{T_{\pre}}
		\left<
			\boldsymbol{\epsilon}_{\pre, n} , 
				\bbE[\bz_{\pre, n} ] - Z_{\pre, \cI^{(n)}}^{r_\pre} \hat{\boldsymbol{\alpha}} ]
		\right> \right|  \nonumber
		\\
		&\leq \frac{1}{T_{\pre}}
		\norm{
			\bbE[ \bz_{\pre, n} ] - Z_{\pre, \cI^{(n)}}^{r_\pre} \hat{\boldsymbol{\alpha}}
		}_2^2
		+ 
		\left| \frac{1}{T_{\pre}} \norm{\boldsymbol{\epsilon}_{\pre, n}}_2^2 - \sigma^2 \right|
		+  \frac{2}{T_{\pre}}
		\left|
		\left<
		\boldsymbol{\epsilon}_{\pre, n} , 
		\bbE[\bz_{\pre, n}] - Z_{\pre, \cI^{(n)}}^{r_\pre} \hat{\boldsymbol{\alpha}} ]
		\right> \right|
		. \label{eq:thm_normal_5}
	\end{align}
	We upper bound these three terms next. 
	\\
	
	\noindent 
	The first term of \eqref{eq:thm_normal_5} can be upper bounded using Lemmas \ref{lem:g1}, \ref{lem:g4}, and \ref{lem:g5}.
	First, by Lemma \ref{lem:g1},
	\begin{align*}
		\frac{1}{T_{\pre}} \norm{
			\bbE [\bz_{\pre, n} ] - Z_{\pre, \cI^{(n)}}^{r_\pre} \hat{\boldsymbol{\alpha}}
		}_2^2
		&\leq 
		\frac{1}{T_{\pre}} \norm{
			Z^{r_\pre}_{\pre, \cI^{(n)}}  - \bbE Z_{\pre, \cI^{(n)}}
		}_{2, \infty}^2
		\norm{\balphaperp}_1^2
		+ 
		\frac{2}{T_{\pre}} \left<
		Z^{r_\pre}_{\pre, \cI^{(n)}}
		\Delta , 
		\boldsymbol{\epsilon}_{\pre, n} 
		\right> ,
	\end{align*}
	which by Lemma \ref{lem:g5},
	\begin{align}
		= O_P 
		\left(
				\frac{1}{T_{\pre}} \norm{
				Z^{r_\pre}_{\pre, \cI^{(n)}}  - \bbE Z_{\pre, \cI^{(n)}}
			}_{2, \infty}^2
			\norm{\balphaperp}_1^2
			+
			\frac{2 r_\pre }{T_{\pre}} + \frac{2}{ \sqrt{T_\pre} }+ \frac{2}{T_{\pre}} \norm{Z_{\pre, \cI^{(n)}}^{r_\pre}  - \bbE Z_{\pre, \cI^{(n)}} }_{2, \infty} \norm{\boldsymbol{\alpha}}_1
		\right) . \nonumber
	\end{align}
	By Lemma \ref{lem:g4},
	\begin{align*}
		\frac{1}{T_{\pre}} & \norm{
			\bbE [\bz_{\pre, n} ] - Z_{\pre, \cI^{(n)}}^{r_\pre} \hat{\boldsymbol{\alpha}}
		}_2^2
		\\
		&=
		O_P \Bigg(
			\frac{\norm{\balphaperp}_1^2 r_\pre \log ( T_\pre | \cI^{(n)} | ) }{(\xi''')^2 \min ( T_\pre , | \cI^{(n)} | )  } 
			+
			\frac{r_\pre }{T_{\pre}} + \frac{1}{ \sqrt{T_\pre} } 
			+
			 \frac{\norm{\boldsymbol{\alpha}}_1 \sqrt{r_\pre \log( T_\pre | \cI^{(n)} | ) } }{\xi''' \sqrt{T_{\pre}}  \min ( \sqrt{T_\pre} , \sqrt{|\cI^{(n)}|} ) } 
		\Bigg) ,
	\end{align*}
		where $\xi'''$ is defined in Theorem \ref{thm:asym_normality}.
	Note that $\norm{\boldsymbol{\alpha}}_1 \leq \sqrt{| \cI^{(n)} |} \norm{\boldsymbol{\alpha}}_2$. 
	Therefore, by Lemma \ref{lem:sc19},
	\begin{align*}
			\frac{1}{T_{\pre}} \norm{
			\bbE [\bz_{\pre, n} ] - Z_{\pre, \cI^{(n)}}^{r_\pre} \hat{\boldsymbol{\alpha}}
		}_2^2
		&=
		O_P \Bigg(
		\frac{\norm{\balphaperp}_2^2 | \cI^{(n)} | r_\pre \log ( T_\pre | \cI^{(n)} | ) }{(\xi''')^2 \min ( T_\pre , | \cI^{(n)} | )  } 
		\\
		&\hspace{60pt} +
		\frac{ r_\pre }{T_{\pre}} + \frac{1}{ \sqrt{T_\pre} } 
		+
		\frac{ \norm{\boldsymbol{\alpha}}_2 \sqrt{r_\pre | \cI^{(n)} | \log( T_\pre | \cI^{(n)} | ) } }{\xi''' \sqrt{T_{\pre}}  \min ( \sqrt{T_\pre} , \sqrt{|\cI^{(n)}|} ) } 
		\Bigg) 
		\\
		&=
		O_P \Bigg(
		\frac{r r_\pre \log ( T_\pre | \cI^{(n)} | ) }{(\xi''')^4 \min ( T_\pre , | \cI^{(n)} | )  } 
		\\
		&\hspace{60pt} +
		\frac{r_\pre }{T_{\pre}} + \frac{1}{ \sqrt{T_\pre} } 
		+
		\frac{ \sqrt{r r_\pre  \log( T_\pre | \cI^{(n)} | ) } }{  (\xi''')^2 \sqrt{ T_{\pre}}  \min ( \sqrt{T_\pre} , \sqrt{|\cI^{(n)}|} ) } 
		\Bigg) 
		\\
		&=
		O_P \Bigg(
		\frac{r^2_\pre \log ( T_\pre | \cI^{(n)} | ) }{(\xi''')^4 \min ( T_\pre , | \cI^{(n)} | )  } 
		\\
		&\hspace{60pt} +
		\frac{ r_\pre }{T_{\pre}} + \frac{1}{ \sqrt{T_\pre} } 
		+
		\frac{\sqrt{r^2_\pre  \log( T_\pre | \cI^{(n)} | ) } }{  (\xi''')^2 \sqrt{ T_{\pre}}  \min ( \sqrt{T_\pre} , \sqrt{|\cI^{(n)}|} ) } 
		\Bigg) 
		,
	\end{align*}
	which, after grouping terms, implies that
	\begin{align}
		\frac{1}{T_{\pre}} \norm{
			\bbE [\bz_{\pre, n} ] - Z_{\pre, \cI^{(n)}}^{r_\pre} \hat{\boldsymbol{\alpha}}
		}_2^2
		=
		O_P \Bigg(
		\frac{r^2_\pre \log ( T_\pre | \cI^{(n)} | ) }{(\xi''')^4 \min ( T_\pre , | \cI^{(n)} | )  } 
		+ \frac{ r_\pre}{ \sqrt{T_\pre} }
		\Bigg) 
		\label{eq:thm_normal_6} .
	\end{align}
	The second term in \eqref{eq:thm_normal_5} can be bounded using Assumption \ref{asm:subG_noise} and Lemma \ref{lem:g2} to obtain
	\begin{align}
		\left| \frac{1}{T_{\pre}} \norm{\boldsymbol{\epsilon}_{\pre, n}}_2^2 - \sigma^2 \right| = O_P \left( T_{\pre}^{-1/2} \right)
		\label{eq:thm_normal_7} .
	\end{align}
	The third term in \eqref{eq:thm_normal_5} can be bounded
	using Lemma \ref{lem:g5} to obtain
	\begin{align}
		\left<
			\boldsymbol{\epsilon}_{\pre, n} ,   
			\bbE[ \bz_{\pre, n} ] - Z_{\pre, \cI^{(n)}}^{r_\pre} \hat{\boldsymbol{\alpha}} 
		\right>
			= 
			\left<
				\boldsymbol{\epsilon}_{\pre, n}  , 
				\bbE[ \bz_{\pre, n} ] 
			\right>
			- 
			\left<
				\boldsymbol{\epsilon}_{\pre, n} , 
				\bar{\cQ} \bbE[ \bz_{\pre, n} ] 
			\right>
			- 	
			\left<
				\boldsymbol{\epsilon}_{\pre, n}  , 
				\bar{\cQ} \boldsymbol{\epsilon}_{\pre, n} 
			\right> . 
			\label{eq:thm_normal_8}
	\end{align}
	Note that $\norm{\bar{\cQ}}_{op} \leq 1$ and, by Assumption \ref{asm:bounded}, 
	$\norm{\bar{\cQ} \bbE[ \bz_{\pre, n}] }_2 \leq \norm{\bbE [\bz_{\pre, n}] }_2 \leq \sqrt{T_\pre}$.
	By Lemma \ref{lem:g6} and Assumption \ref{asm:subG_noise}, 
	for any $\eta > 0$, 
	\begin{align*}
		P( \left<
			\boldsymbol{\epsilon}_{\pre, n} , 
			\bbE[\bz_{\pre, n}] \geq \xi
		\right> )
		&\leq 
		\exp \left( -
			\frac{\xi \eta^2}{T_\pre \sigma^2}
		\right) , 
		\\
		P( \left<
		\boldsymbol{\epsilon}_{\pre, n} , 
		\bar{\cQ}
		\bbE[\bz_{\pre, n}] \geq \xi
		\right> )
		&\leq 
		\exp \left( -
		\frac{\xi \eta^2}{T_\pre \sigma^2}
		\right) ,
	\end{align*}
	which imply that
	\begin{align*}
		\left<
		\boldsymbol{\epsilon}_{\pre, n} , 
		\bbE[\bz_{\pre, n}]
		\right> &= O_P(\sqrt{T_\pre}) ,
		\\
		 \left<
		\boldsymbol{\epsilon}_{\pre, n} , 
		\bar{\cQ}
		\bbE[\bz_{\pre, n}]
		\right>  &= O_P(\sqrt{T_\pre}) .
	\end{align*}
	From Lemma \ref{lem:g5}, for any $\eta > 0$,
	\begin{align*}
		P(\left<
			\boldsymbol{\epsilon}_{\pre, n} , 
			\bar{\cQ} \boldsymbol{\epsilon}_{\pre, n}
		\right> \geq \sigma^2 r_{\pre} + \eta )
		\leq 
		\exp \left(
			- \xi \min\left(
				\frac{\eta^2}{\sigma^4 r_\pre} , 
				\frac{\eta}{\sigma^2}
			\right)
		\right) \implies \left<
		\boldsymbol{\epsilon}_{\pre, n} , 
		\bar{\cQ} \boldsymbol{\epsilon}_{\pre, n}
		\right> = O_P(r_\pre) .
	\end{align*}
	Therefore, by \eqref{eq:thm_normal_8}, 
	\begin{align}
		\frac{2}{T_\pre} \left|
			\left<
			\boldsymbol{\epsilon}_{\pre, n} ,   
			\bbE[ \bz_{\pre, n} ] - Z_{\pre, \cI^{(n)}}^{r_\pre} \hat{\boldsymbol{\alpha}} 
			\right>
		\right|	= O_P \left( \frac{1}{\sqrt{T_\pre}} + \frac{r_\pre}{T_\pre} \right)
		\label{eq:thm_normal_9}
	\end{align}
	Together \eqref{eq:thm_normal_6}, \eqref{eq:thm_normal_7}, and \eqref{eq:thm_normal_9} give the desired result.
\ifarxiv
\end{proof}
\else
\Halmos
\endproof
\fi

\subsection{Proof of Proposition \ref{prop:SIA_linear_independence}}

In order to prove Proposition \ref{prop:SIA_linear_independence}, 
we prove another result that subsumes Proposition \ref{prop:SIA_linear_independence}.

We first introduce some notation. 
First, 
we assume that $D = 2$ for ease of exposition.
The proof can be straightforwardly extended for $D > 2$. 
Consider a unit $n \in [N]$ and counterfactual, prediction treatments of interest $\cfAn \in \{1 , 2\}^{|\cN(n)|}$.
We use $\cI$ as a shorthand for $\cI^{(n)}$
and let $\cI_j$ refer to the $j$-th donor in the donor set $\cI$.

Recall that 
$B^{\pre}(a) \in \{0, 1\}^{N \times T_{\pre}}$ 
and 
$\bb^{\post}(a) \in \{0, 1\}^{N}$ are defined such that
\begin{align*}
	B_{i t}^{\pre}(a) = \Ind(A^{\pre}_{i t} = a) 
	\qquad 
	\text{and}
	\qquad 
	\tilde{b}_{i}^{\post}(a) = \Ind(\tilde{a}_i  = a) ,
\end{align*}
and $B^{\pre}$ and $\tilde{B}^{\post}$ be the concatenated matrices across different treatments, 
i.e., 
\begin{align*}
	B^{\pre} &= [B^{\pre}(1) , B^{\pre}(2) , \hdots, B^{\pre}(D)]
	\in \{0, 1\}^{N \times T_{\pre} D} ,
	\\
	\tilde{B}^{\post} &= [\tilde{\bb}^{\post}(1) , \tilde{\bb}^{\post}(2) , \hdots, \tilde{\bb}^{\post}(D)]
	\in \{0, 1\}^{N \times  D} .
\end{align*}

Finally, 
without loss of generality, 
let us re-order the training measurements such that the treatment assignments over $\cN(n)$ are grouped together, i.e., 
\begin{align*}
	A[\cN(n), \cT_{\pre}]
	&= 
	[
	\bc^1_{\cN(n)} ,  \hdots ,  \bc^1_{\cN(n)} \,\, \vline \,\, \bc^2_{\cN(n)} , \hdots , \bc^2_{\cN(n)} \,\, \vline \,\, \hdots \,\, \vline \,\, \bc^K_{\cN(n)} ,  \hdots, \bc^K_{\cN(n)}
	] ,
\end{align*}
where $K$ is the number of distinct training treatment vectors. 
Let $\cT_1$ denote the first $T_1$ measurements such that $A[\cN(n) , \tau] = \bc^1_{\cN(n)}$ for all $\tau \in \cT_1$, 
$\cT_2$ denote the next $T_2$ measurements such that $A[\cN(n) , \tau] = \bc^2_{\cN(n)}$ for all $\tau \in \cT_2$,
and so on through $\cT_K$.
\medskip

\noindent
\textbf{Matrix representation of $\bbE[Z_{\pre, \cI} | \LF, \cO]$}.
Recall that for unit $n \in [N]$, 
measurement $t \in [T]$, 
and treatments $\ba \in [D]_0^{N}$,
\begin{align}
	Y_{t , n}^{(\ba_{\cN(n)})} &= 
	\sum_{k \in \cN(n)}
	\left< 
	\bu_{k, n} , 
	\bw_{t, a_k} 
	\right> 
	+ \epsilon_{t , n}^{(\bintv_{\cN(n)})}, 
\end{align}
where $n \in \cN(n)$. 
Under $D = 2$, 
\begin{align}
	Y_{t , n}^{(\ba_{\cN(n)})} - \epsilon_{tn}^{(\bintv_{\cN(n)})}
	&= 
	\sum_{k \in \cN(n)}
	\ind ( a_k = 1 )
	\bu_{k, n}^\top 
	\bw_{t, 1}
	+ 
	\sum_{k \in \cN(n)}
	\ind ( a_k = 2 )
	\bu_{k, n}^\top 
	\bw_{t, 2} \nonumber
	\\
	&= 
	\begin{bmatrix}
		\sum_{k \in \cN(n)}
		\ind ( a_k = 1 )
		\bu_{k, n}^\top
		& 
		,
		&
		\sum_{k \in \cN(n)}
		\ind ( a_k = 2 )
		\bu_{k, n}^\top 
	\end{bmatrix}
	\begin{bmatrix}
		\bw_{t, 1}
		\\
		\bw_{t, 2}
	\end{bmatrix} ,
	\label{eq:model_decomp}
\end{align}
We will use this decomposition to rewrite $\bbE[Z_{\pre, \cI} | \LF, \cO]$
as a product of matrices. 

First, 
recall that
\begin{align}
	Z_{\pre, \cI} = 
	\begin{bmatrix}
		Z_{t, \cI_j} : t \in \cT_{\pre} , j \leq  | \cI |  
	\end{bmatrix}
	\in \bbR^{T_{\pre}  \times |\cI|} .
\end{align}
Second, let $\tilde{\Nc}(j)$ denote $\pi_j({\cN}(j))$, 
where $\pi_j$ is specified in Definition \ref{def:node_donors}, 
i.e., $\tilde{\Nc}(j)$ corresponds to the permuted neighborhood of donor $j$, 
where the permutation is fixed under Definition \ref{def:node_donors}.
Let 
\begin{align*}
	U_{\cI} &= 
	\begin{bmatrix}
		\bu_{\tilde{\cN}_j(\cI_k), \cI_k} : j \leq | \cN(n) |  , k \leq  | \cI | 
	\end{bmatrix}
	\in \bbR^{r |{\cN}(n)| \times | \cI |}  \, ,
	\\[8pt]
	H_{\pre}^\ell 
	&= 
	\begin{bmatrix}
		\ind( c^\ell_{\cN_1(n)} = 1) , & 
		\ind(c^\ell_{\cN_2(n)} = 1) , & 
		\hdots  , & 
		\ind(c^\ell_{\cN_{|\cN(n)|}(n)} = 1)
		\\[4pt]
		\ind( c^\ell_{\cN_1(n)} = 2) , & 
		\ind( c^\ell_{\cN_2(n)} = 2) , & 
		\hdots , & 
		\ind( c^\ell_{\cN_{|\cN(n)|}(n)} = 2)
	\end{bmatrix}
	\in \{0, 1\}^{2 \times |\cN(n)|} \, 
\end{align*}
Let $H_\pre \in \{0 , 1 \}^{  2 K \times | \cN(n) |}$ be constructed by stacking $H^1_\pre$, $H^2_\pre, \hdots, H^K_\pre$ on top of one another. 
Let 
\begin{align*}
	W^j = [( \bw_{\tau, 1}^\top , \bw_{\tau, 2}^\top ) : \tau \in \cT_j ]
	\in 
	\bbR^{T_j \times 2 r} ,
\end{align*}
and $W_{\pre} \in \bbR^{T_{\pre} \times 2 r K}$ be the block-diagonal matrix with matrices $W^1, W^2, \hdots , W^K$ along the diagonal. 

Then, by the decomposition in \eqref{eq:model_decomp} and Definition \ref{def:node_donors},
\begin{align}
	\bbE[Z_{\pre, \cI} | \LF, \cO]
	&= 
	W_{\pre}
	\left(
	H_{\pre}
	\otimes 
	\bbI_r
	\right)
	U_{\cI} .
	\label{eq:training_decomposition}
\end{align}
\noindent 
\textbf{Matrix representation of $\bbE[Z_{\post, \cI} | \LF, \cO]$}.
Using the same reasoning as above, 
one can write 
\begin{align}
	\bbE[Z_{\post, \cI} | \LF, \cO]
	&= 
	W_{\post}
	\left(
	H_{\post}
	\otimes 
	\bbI_r
	\right)
	U_{\cI} , \label{eq:prediction_decomposition}
\end{align}
where
\begin{align*}
	H_{\post}
	&= 
	\begin{bmatrix}
		\ind( a^{\post}_{\cN_1(n)} = 1) & 
		\ind(a^{\post}_{\cN_2(n)} = 1)& 
		\hdots & 
		\ind(a^{\post}_{\cN_{|\cN(n)|}(n)} = 1)
		\\
		\ind( a^{\post}_{\cN_1(n)} = 2) & 
		\ind( a^{\post}_{\cN_2(n)} = 2)& 
		\hdots & 
		\ind( a^{\post}_{\cN_{|\cN(n)|}(n)} = 2)
	\end{bmatrix}  \otimes \mathbf{1}_{T_\post}
	\in
	\{0 , 1 \}^{  2 T_\post \times | \cN(n) |} ,
	\\[10pt]
	W_{\post}^\tau &= [ \bw_{\tau, 1}^\top ,  \bw_{\tau, 2}^\top ]
	\in 
	\bbR^{1 \times 2 r} ,
\end{align*}
and $W_{\post} \in \bbR^{T_\post \times 2 r T_{\post}}$ denote the block-diagonal matrix with $W_{\post}^{T_{\pre} + 1} , W_{\post}^{T_{\pre} + 2} , \hdots, W_{\post}^{T}$ along the diagonal. 

Recall that Assumption \ref{asm:subspace_inclusion} is that the rowspace of $\bbE[Z_{\post, \cI^{(n)}} | \LF , A]$ is contained within the rowspace of $\bbE[Z_{\pre, \cI^{(n)}} | \LF , A]$.
\medskip

\begin{lemma}\label{lem:SIA_training_fails}
	Suppose Assumption \ref{asm:model} holds.
	If $$\text{columnspace}(\tilde{B}^{\post} [\cN(n), : ] ) \not\subseteq \text{columnspace}(B^{\pre}[\cN(n), : ]),$$
	then there exist latent factors $\LF$ under which Assumption \ref{asm:subspace_inclusion} cannot hold. 
\end{lemma}

\ifarxiv
\begin{proof}
\else
\proof{Proof.}
\fi
Our goal is to show that there exist latent factors $\LF$ such that, 
if 
$\text{columnspace}(\tilde{B}^{\post} [\cN(n), : ] ) \not\subseteq \text{columnspace}(B^{\pre}[\cN(n), : ])$, 
then Assumption \ref{asm:subspace_inclusion} does not hold. 
Since 
$\rs(H_{\post}) = \rs(\tilde{B}^{\post} [\cN(n), : ]^\top)  $ 
and 
$\rs(H_{\pre}^\top) = \rs( \tilde{B}^{\pre} [\cN(n), : ]^\top ) $, 
$\text{columnspace}(\tilde{B}^{\post} [\cN(n), : ] ) \not\subseteq \text{columnspace}(B^{\pre}[\cN(n), : ])$ is equivalent to  
$\rs( H_\post ) \not\subseteq \rs( H_\pre )$, 
which is equivalent to 
$\rs( H_\post \otimes \bbI_r ) \not\subseteq \rs( H_\pre \otimes \bbI_r )$. 

By Lemma \ref{lem:rowspace}, 
there exists a vector $\bv \neq \mathbf{0}_{2 r T_\post}$ such that 
$(H_\post \otimes \bbI_r)^\top \bv \neq \mathbf{0}_{r | \cN(n) | }$ 
and
\begin{align}
	(H_\pre \otimes \bbI_r ) (H_\post \otimes \bbI_r)^\top \bv = \mathbf{0}_{2 K} 
	. 
	\label{eq:training_test_proof_1}
\end{align}
Since  $(H_\post \otimes \bbI_r)^\top \bv \neq \mathbf{0}_{r | \cN(n) | }$, 
there must exist $\bu$-latent factors 
such that, for the same $\bv$, 
$U_\cI^\top (H_\post \otimes \bbI_r)^\top \bv \neq \mathbf{0}_{|\cI|}$. 
Suppose that $U_\cI$ reflects these latent factors. 
Then, \eqref{eq:training_test_proof_1} implies  
\begin{align}
	(H_\pre \otimes \bbI_r ) U_\cI U_\cI^\top (H_\post \otimes 	\bbI_r)^\top \bv = \mathbf{0}_{2 K} 
	. 
	\label{eq:training_test_proof_2}
\end{align}
Let $\bv' = (W_\post W_\post^\top)^{-1} W_\post \bv$.
Let the $\bw$-latent factors be defined such that $\bv' \neq \mathbf{0}_{T_\post}$. 
Then,
\eqref{eq:training_test_proof_2} implies  \begin{align}
	&W_\pre (H_\pre \otimes \bbI_r ) U_\cI U_\cI^\top (H_\post \otimes 	\bbI_r)^\top \bv = \mathbf{0}_{2 K} ,
	\nonumber 
	\\
	\implies 
	&W_\pre (H_\pre \otimes \bbI_r ) U_\cI U_\cI^\top (H_\post \otimes 	\bbI_r)^\top W_\post^\top \bv' = \mathbf{0}_{2 K} 
	. 
	\label{eq:training_test_proof_3}
\end{align}
By Lemma \ref{lem:rowspace}, 
\eqref{eq:training_test_proof_3} implies that $\rs(W_\post (H_\post \otimes \bbI_r ) U_\cI ) \not\subseteq \rs(W_\pre (H_\pre \otimes \bbI_r ) U_\cI)$, 
which implies that $\rs(\bbE  [ Z_{\post, \cI} | \LF , \cO] ) \not\subseteq \rs(\bbE  [ Z_{\pre, \cI} | \LF , \cO])$. 
Therefore, 
then Assumption \ref{asm:subspace_inclusion} does not hold, 
as claimed. 
\ifarxiv
\end{proof}
\else
\Halmos
\endproof
\fi
\medskip

Proposition \ref{prop:SIA_linear_independence} follows immediately from Lemma \ref{lem:SIA_training_fails}. 
First note that $B^{\pre, n} = B^{\pre}[\cN(n), :]$.
Second, if $\text{colrank}(B^{\pre, n}) < |\cN(n)|$, 
then there exists a $\tilde{\ba}$ such that 
 $\text{columnspace}(\tilde{B}^{\post} [\cN(n), : ] ) \not\subseteq \text{columnspace}(B^{\pre}[\cN(n), : ])$.
By Lemma \ref{lem:SIA_training_fails}, 
if $\text{colrank}(B^{\pre, n}) < |\cN(n)|$, 
then there exists a target treatment $\tilde{\ba}$  and latent factors $\LF$ such that Assumption \ref{asm:subspace_inclusion} does not hold, 
as claimed.

%% file: sections/app_valid_test_proofs.tex
\subsection{Proof of Proposition \ref{prop:SIA_rowspace}}

\ifarxiv
\begin{proof}
\else
\proof{Proof.}
\fi
The {subspace inclusion assumption} (SIA), or Assumption \ref{asm:subspace_inclusion}, requires that $$\rs(\bbE  [ Z_{\post, \cI} | \LF , \cO] ) \subseteq \rs(\bbE  [ Z_{\pre, \cI} | \LF , \cO]).$$ 
By \eqref{eq:training_decomposition} and \eqref{eq:prediction_decomposition}, 
this is equivalent to requiring that,
for the given $n$ and $\tilde{\ba}_{\cN(n)}$ and for every $i \in [T_{\post}]$, there exists some $\boldsymbol{\phi} \in \bbR^{T_{\pre}}$ such that
\begin{align}
	\mathbf{e}_i^\top
	W_{\post} ( H_{\post} \otimes \bbI_r) U_{\cI}
	&= 
	\boldsymbol{\phi}^\top W_{\pre} ( H_{\pre} \otimes \bbI_r)  U_{\cI} . \label{eq:prop_2_1}
\end{align}
{Under the assumption on latent factors, 
as long as $|\cI^{(n)}| \geq r |\cN(n)|$, 
then \eqref{eq:prop_2_1} holds if and only if }
there exists some $\boldsymbol{\phi} \in \bbR^{T_{\pre}}$ such that
\begin{align}
	\mathbf{e}_i^\top
	W_{\post} ( H_{\post} \otimes \bbI_r)
	= \boldsymbol{\phi}^\top W_{\pre} (H_{\pre} \otimes \bbI_r) . 
\end{align}  
Therefore, 
subspace inclusion requires that  $\rs(W_{\post} ( H_{\post} \otimes \bbI_r)) \subseteq \rs( W_{\pre} ( H_{\pre} \otimes \bbI_r) )$.

To conclude the proof, we  use  several facts. 
First, 
$\rs( W_{\post} ( H_{\post} \otimes \bbI_r) ) \subseteq \rs(H_{\post}\otimes \bbI_r)$.
Second, by Lemma \ref{lem:unif_full_rank},
each $W^j$ is has linearly independent columns almost surely (since $T_j \geq 2r$ by the second condition of \textsc{TrainingTreatmentTest}) and $W_{\pre}$ therefore also has linearly independent columns almost surely. 
As such, 
$\rs(W_{\pre} ( H_{\pre} \otimes \bbI_r)) = \rs(H_{\pre} \otimes \bbI_r)$ almost surely.

Therefore, 
\begin{align*}
	\rs(H_{\post}) &\subseteq \rs(H_{\pre})
	\\
	{\iff} \rs(H_{\post} \otimes \bbI_r) &\subseteq \rs(H_{\pre} \otimes \bbI_r) 
	\\
	\implies
	\rs( W_{\post} ( H_{\post} \otimes \bbI_r) ) &\subseteq \rs(H_{\pre} \otimes \bbI_r) 
	\\
		{\iff}\rs(W_{\post} ( H_{\post} \otimes \bbI_r)) &\subseteq \rs( W_{\pre} ( H_{\pre} \otimes \bbI_r) ) ,
\end{align*}
i.e., subspace inclusion holds if $\rs(H_{\post}) \subseteq \rs(H_{\pre})$. 
This condition is equivalent to the first condition in \textsc{TrainingTreatmentTest} because
because 
$H_{\post}^\top = \tilde{B}^{\post} [\cN(n), : ]  $ and
$\text{columnspace}(H_{\pre}^\top) = \text{columnspace}( B^{\pre}[\cN(n), : ] )$. 
As such, 
given the two conditions in \textsc{TrainingTreatmentTest}, 
Assumption \ref{asm:subspace_inclusion} holds almost surely. 
\ifarxiv
\end{proof}
\else
\Halmos
\endproof
\fi

%% file: sections/app_proofs_exp_design.tex
\subsection{Proof of Lemma \ref{lem:training_treatments_test_passes}}

Recall that, 
for a given treatment $a \in [D]$, 
$B^{\pre}(a) \in \{0, 1\}^{N \times T_{\pre}}$ 
and 
$\bb^{\post}(a) \in \{0, 1\}^{N}$ are defined such that their $(i, t)$-th elements are given by
\begin{align*}
	B_{i t}^{\pre}(a) = \Ind(A^{\pre}_{i t} = a) 
	\qquad 
	\text{and}
	\qquad 
	\tilde{b}_{i}^{\post}(a) = \Ind(\tilde{a}_i  = a) .
\end{align*}
That is, 
the $(i, t)$-th entry of $B^{\pre}(a)$ is $1$ if and only if unit $i$ at measurement $t$ receives treatment $a$ under the training treatments $A^{\pre}$.
Similarly, 
the $i$-th entry of $\tilde{\bb}^{\post}(a)$ is $1$ if and only if unit $i$ is assigned counterfactual treatment $a$ under $\tilde{\ba}$.
Further recall that 
\begin{align*}
	B^{\pre} &= [B^{\pre}(1) , B^{\pre}(2) , \hdots, B^{\pre}(D)]
	\in \{0, 1\}^{N \times T_{\pre} D} ,
	\\
	\tilde{B}^{\post} &= [\tilde{\bb}^{\post}(1) , \tilde{\bb}^{\post}(2) , \hdots, \tilde{\bb}^{\post}(D)]
	\in \{0, 1\}^{N \times  D} .
\end{align*} 
Before proving Lemma \ref{lem:training_treatments_test_passes},
we first introduce a lemma.

\begin{lemma}\label{lem:same_color}
	Under the experiment design in Section \ref{sec:exp_procedure},
	$B^{\pre}[i, :] = B^{\pre}[j, :]$ if and only if units $i$ and $j$ have been assigned the same color. 
\end{lemma}
\ifarxiv
\begin{proof}
\else
\proof{Proof.}
\fi
Observe that Step 3 in the experiment design identifies which units received certain colors and assigns those units a treatment other than $1$. 
That is, for a given iteration $\ell$ in the for loop, 
every unit is  assigned treatment $1$ except for units of certain colors. 
Moreover, Step 3 never examines the same color twice. 
Therefore, since each unit $i$ has only one color, 
$c^\ell_i \neq 1$ for exactly one value of $\ell$,
whose value is determined by the color given to unit $i$. 
One can conclude that 
$A^{\pre}[i, :] = A^{\pre}[j, :]$ if and only if unit $i$ and unit $j$ receive the same color. 
From the definitions of $B^{\pre}$, 
it therefore follows that 	$B^{\pre}[i, :] = B^{\pre}[j, :]$ if and only if units $i$ and $j$ have been assigned the same color. 
\ifarxiv
\end{proof}
\else
\Halmos
\endproof
\fi

\medskip

\noindent 
We now provide a proof of Lemma \ref{lem:training_treatments_test_passes}. 
\medskip

\ifarxiv
\begin{proof}
\else
\proof{Proof.}
\fi 
We show that the two conditions in \textsc{TrainingTreatmentTest} hold when the training treatments $A^\pre$ are assigned as
described in Section \ref{sec:exp_procedure}.
\medskip

\paragraph{First requirement of \textsc{TrainingTreatmentTest}.}
We begin by proving that $$\text{columnspace}( \tilde{B}^{\post} [\cN(n), : ] ) \subseteq \text{columnspace}( B^{\pre}[\cN(n), : ] ),$$ for all possible $\tilde{B}^{\post} [\cN(n), : ] $ when $A^{\pre}$ is generated using the experiment design in Section \ref{sec:exp_procedure}. 	
It suffices to prove that $B^{\pre}[\cN(n), : ]$ has full row-rank.
We make use of the following three facts. 

First, the proposed procedure ensures that no two units in the same neighborhood ever receive the same color. 
This is guaranteed under \textsc{TwoHopColoring}, 
which returns a coloring on the graph $\cG'$ that is created by connecting every node in $\cG$ to its immediate \emph{and} its two-hop neighbors. 
As such, for any unit $i$, 
no two units in its neighborhood $\cN(i)$ share the same color because any two units in $\cN(i)$ must be within each others' two-hop neighborhoods.

Second, 
$T_{\pre} D \geq |\cN(n)|$, 
i.e., there are at least as many columns in $B^{\pre}$ as there are rows.
To see why, 
observe that, 
under the proposed procedure,
$T_{\pre} = T' \bar{r} D = \lceil \frac{\textsc{NumColors}}{D-1} \rceil \bar{r} D \geq \lceil \frac{|\cN(n)|}{D-1} \rceil \bar{r} D$, 
where the inequality follows from the first fact, i.e., that no two units in $\cN(n)$ share the same color.

Third, 
by Lemma \ref{lem:same_color},
$B^{\pre}[i, :] = B^{\pre}[j, :]$ if and only if units $i$ and $j$ have been assigned the same color. 
However, by the first fact above,
this cannot occur when $i, j \in \cN(n)$.
As such, $B^{\pre}[\cN(n), : ]$ has $|\cN(n)|$ distinct rows. 
Since $B^{\pre}[\cN(n), : ]$ is a binary matrix \emph{and} $B^{\pre}[\cN(n), : ]$ has at least as many columns as rows,
these rows must be linearly independent. 
In other words,  $B^{\pre}[\cN(n), : ]$ has full row-rank, 
as we sought to prove.

\medskip

\paragraph{Second requirement of \textsc{TrainingTreatmentTest}.}
The second requirement holds by Step 4 of the experiment design in Section \ref{sec:exp_procedure}.
\ifarxiv
\end{proof}
\else
\Halmos
\endproof
\fi

\subsection{Proof of Lemma \ref{lem:treatment_schedule_complexity}}

\ifarxiv
\begin{proof}
\else
\proof{Proof.}
\fi
Recall that $T_{\pre}$ denotes the width of $A^{\pre}$.
Under the procedure in Section \ref{sec:exp_procedure},
the width of $A^{\pre}$ is given by $T' \bar{r} D$, 
where $T' = \lceil \frac{\textsc{NumColors}}{D - 1} \rceil$. 
Therefore, $T_{\pre} = \bar{r} D \lceil \frac{\textsc{NumColors}}{D - 1} \rceil$.
$T_{\pre} \leq \frac{\bar{r} D (d^2 + D)}{D - 1}$ follows from the facts that
(i) any graph $\cG''$ can be trivially colored using $\text{Degree}(\cG'') + 1$ colors and (ii) the degree of $\cG'$ is upper bounded by $d (d - 1)$ by the definition of $\cG'$ in Section \ref{sec:experimental_proc}
\ifarxiv
\end{proof}
\else
\Halmos
\endproof
\fi

\subsection{Tailoring experiment design to counterfactual treatments of interest}\label{app:tailor_exp_design}

Recall that, by Lemma \ref{lem:training_treatments_test_passes}, the experimental design procedure in Section \ref{sec:exp_procedure} produces a treatment schedule that passes \textsc{TrainingTreatmentTest} for \emph{any} $n$ and 
$\tilde{\ba}_{\cN(n)}$ of interest 
under Assumptions \ref{asm:model} and  \ref{asm:conditional_exo} and $\bar{r} = r$.
One can alternately tailor the treatment schedule to a specific $n$ and $\tilde{\ba}_{\cN(n)}$ of interest. 
One need only ensure that $\text{columnspace}(\tilde{B}^{\post}[\cN(n) , :]) \subseteq \text{columnspace}(B^{\pre}[\cN(n) , :])$ is satisfied, 
as required by \textsc{TrainingTreatmentTest}.
Such a modification is desirable because it might require fewer training samples. 

To make this modification, 
all that would change in the procedure in Section \ref{sec:experimental_proc} is the method \textsc{TwoHopColoring}. 
Specifically, 
one would form $\cG'$  by connecting each unit $i$ not to \emph{every} one of its immediate and two-hop neighbors, 
but only those units $j$ in the two-hop neighborhood for which $\tilde{a}_i \neq \tilde{a}_j$. 
That is, 
$(j, i) \in \cE'$ if $( (j, i) \in \cE ) \cup (\exists k \in [N] \setminus \{i, j\} : (j, k) , (k, i) \in \cE)$ and $\tilde{a}_i \neq \tilde{a}_j$.

\subsection{Proof of Proposition \ref{prop:reg_error}}

\ifarxiv
\begin{proof}
\else
\proof{Proof.}
\fi
Our goal is to apply Theorem \ref{thm:finite_sample_consistency}. 
However, the conditions and assumptions of Theorem \ref{thm:finite_sample_consistency} differ from those used in Proposition \ref{prop:reg_error}.
We proceed by showing that, under the assumptions in Proposition \ref{prop:reg_error}, 
we can recover the assumptions of Theorem \ref{thm:finite_sample_consistency} {and} obtain more precise estimates on the number of training samples $T_\pre$ and units $N$ needed for finite-sample consistency. 

We begin by characterizing the number of donors an ego-unit has under the conditions in Proposition  \ref{prop:reg_error}.
We then show that Assumptions \ref{asm:linear_span} and \ref{asm:subspace_inclusion} hold under the proposition conditions. 
\medskip

\paragraph{Number of donors.} 
There are three requirements for a donor, 
as given in Definition \ref{def:node_donors}.
The first requirement is that a donor has the same number of neighbors as $n$. 
Since $\cG$ is a $d$-regular graph, 
this requirement is automatically satisfied for all possible units. 

The second requirement for a unit $k$ to be a donor for unit $n$ is that there exists a permutation $\pi_k$ such that 
$A[\pi_k(\cN(k)),\cT_{\pre}] = A[\cN(n),\cT_{\pre}]$.
By the experiment design procedure in Section \ref{sec:exp_procedure},
this second requirement is satisfied as long as $n$ and $k$ are assigned the same colors. 
By Lemma \ref{lem:coloring_num_units}, 
there are at least $N - \Theta(\sqrt{N})$ ego-units for which there are at least $\sqrt{N}$ units that satisfy the second requirement. 
Let this set of ego-units be denoted by $E$.

{Therefore, at least $\sqrt{N}$ units satisfy the first and second requirements of a donor unit.
Suppose that these units are sub-sampled before checking whether they meet the third requirement of Definition \ref{def:node_donors}. 
Specifically, suppose that exactly $\sqrt{N}$ of them are randomly chosen and the rest discarded. (This subsampling method is what we refer to as the ``method of choosing donors'' in the proposition statement.)
Recall further that the third requirement for a unit $k$ to be a donor is that $\ba^\post_{\pi_k(\cN(k))} = \tilde{\ba}_{\cN(n)}$.}
By Lemma  \ref{lem:prediction_treatment_concentration} {and the subsampling condition},
the number of units that satisfy the third requirement of Definition \ref{def:node_donors} for any ego-unit in $n \in E$ (and therefore are considered ``donors'' for $n$) is:
\begin{align}
	{|\cI^{(n)}| =  \Theta \left( \frac{\sqrt{N}}{D^{d+1}}  \right) }, \label{eq:reg_graph_proof_1} 
\end{align}
with high probability. 
We can therefore replace $|\cI^{(n)}|$ in Theorem \ref{thm:finite_sample_consistency} with $\sqrt{N}/D^{d+1}$,
noting that this substitution holds with high probability for $N - \Theta(\sqrt{N})$ ego-units, 
as stated in Proposition \ref{prop:reg_error}.
\medskip

\paragraph{Assumption \ref{asm:linear_span}.}
Assumption \ref{asm:linear_span} is required in Theorem \ref{thm:finite_sample_consistency}. 
By Lemma \ref{lem:linear_span_gaussian} and Assumption \ref{asm:unif_lf}, 
Assumption \ref{asm:linear_span} holds if there are at least  $r|\cN(n)|$ donors. 
In a $d$-regular graph, 
this translates to needing at least $r(d+1)$ donors. 
Therefore, by \eqref{eq:reg_graph_proof_1},
we require that 
\begin{align*}
	\frac{\sqrt{N}}{D^{d+1}}
	&= \Omega( r(d + 1) )
	\\
	\implies N &= \Omega \left(
	r^2 d^2 D^{2d+2}
	\right) ,
\end{align*}
for
Assumption \ref{asm:linear_span} to hold, 
which matches the condition stated in Proposition \ref{prop:reg_error}. 
\medskip

\paragraph{Combining results.}
By Assumption \ref{asm:unif_lf}, $r = \bar{r}$, and Lemma \ref{lem:training_treatments_test_passes}, Assumption \ref{asm:subspace_inclusion} holds. 
Therefore, both Assumptions \ref{asm:linear_span} and \ref{asm:subspace_inclusion} hold under the conditions given in Proposition \ref{prop:reg_error}.
By Assumption \ref{asm:unif_lf}, 
Assumption \ref{asm:bounded} holds.
By Lemma \ref{lem:asm_8}, 
Assumption \ref{asm:rank_balanced_singular_vals} holds.
Furthermore, 
the condition in Proposition \ref{prop:reg_error} that there are at least $\frac{rD(d^2 + D)}{D - 1}$ comes from the fact that the experiment design in Section \ref{sec:exp_procedure} can always be carried out with at least $\frac{rD(d^2 + D)}{D - 1}$ training measurements. 
Combining these results with Theorem \ref{thm:finite_sample_consistency} gives
\begin{align*}
	&\left| 
	\hIPO{n} - \IPO{n}
	\right| 
	\\
	& \hspace{20pt}
	= O_P \left( 
	\log \left(  \frac{ T_\pre N }{D^{d+1}} \right)
	\left( 
	\frac{r_\pre^{3/4}}{(\xi''')^{3/2} T_\pre^{1/4} }
	+
	\frac{r_\pre^2}{(\xi''')^4 } 
	{\max} \left(
	\frac{1}{\sqrt{T_\pre}} , 
	\frac{ \sqrt{D^{d + 1}} }{ N^{1/4} } ,
	{\frac{ N^{1/4} }{ D^{(d+1)/2} T_\pre^{3/2}}}
	\right)
	\right) 
	\right) .
\end{align*}
Note that, by Lemma \ref{lem:asm_8}, 
$\xi' = (1 + 4rd^3)^{-1/2}$ and $\xi'' = (9 r (d + 1))^{-1}$.
Grouping terms and noting that $r_\pre \leq r (d+1)$ gives the result. 
\ifarxiv
\end{proof}
\else
\Halmos
\endproof
\fi

%% file: sections/app_simulations.tex
In this section, 
we provide full details behind the simulations produced in Section \ref{sec:experiments} and provide additional plots. 
\\

\noindent
\textbf{Setting}.
Let $\cG$ be a regular graph with degree $d$, and let the treatments be binary, i.e., $D = 2$.
In each of the experiments below, 
we will indicate the graph degree. 

At the start of each simulation, 
the latent factors $\bu_{k, n}$ and $\bw_{1, a}$ are drawn uniformly at random from $\Big[ -\frac{1}{\sqrt{r (d + 1)}} , \frac{1}{\sqrt{r (d + 1)}} \Big]^r$.
The latent factors $\bw_{\tau, a}$ are generated as random walk for $\tau > 1$, 
where each random step of the random walk is also drawn uniformly at random from $\Big[ -\frac{1}{\sqrt{r (d + 1)}} , \frac{1}{\sqrt{r (d + 1)}} \Big]^r$.

Our experiments use a simple donor-finding algorithm. 
In particular, instead of searching for donors over all possible permutations $\pi_j$, as defined in Definition \ref{def:node_donors}, we fix an ordering of units and restrict ourselves to the identity permutation $\pi_j(i) = i$.
In this way, the number of donors reported in our experiments is lower than the actual number of available donors. 
\\

\noindent
\textbf{Predictions}.
Figure \ref{fig:NSI}(a) shows an example of the estimates that NSI produces, 
where $\cG$ is a ring graph ($d = 2$) with $N = 1000$ units, $\epsilon^{(\bc_{\cN(i)})}_{\tau, i} \sim \cN(0, 0.1)$, $r = 2$, $T_{\pre} = 150$, and $T_\post = 50$.
Let the training treatments be assigned according to the experiment design in Section \ref{sec:experimental_proc}.

The prediction treatments ${\ba}^\post$ for $\tau \in \cT_{\post}$ are drawn uniformly at random from $[D]^N$. 
The plot is generated for a given  target treatment $ \tilde{\ba}$ of interest. 
As discussed in Section \ref{sec:setup}, we assume that the prediction and target treatments are constant across $\cT_\post$.
Consider the bottom plot and a specific unit $n$.
The solid line gives the ground truth potential outcomes for unit $n$ across measurements $t \in [200]$. 
The estimates produced by NSI are marked by asterisks $*$, with the 95 percent confidence interval in gray. 
The measurements to the left of the vertical line (i.e., in blue and green) correspond to the training set $\cT_{\pre}$ while those to the right (i.e., in red and orange) correspond to the prediction set $\cT_{\post}$. 
The top plot gives the spectrum $\{\hat{s}_\ell\}_{\ell=1}^{q}$ produced in Step 1 of Section \ref{sec:estimation_proc}, where the vertical line marks the singular value threshold $\kappa$ that is chosen in Step 1. 
In all of our experiments, $\kappa$ is chosen using a knee-point (otherwise known as elbow-point) method.
As shown in the bottom plot, the predictions closely match the ground-truth values. 
As shown on top, $6$ components are used to construct the estimates. 
Since the network-adjusted rank is $6 $ (the product of $r = 2$ and $|\cN(n)| = 3$), 
that NSI uses $6$ components explains why its estimates are fairly accurate. 

Further examples of the estimates NSI produces are given at the end of this section. 
\\

\noindent
\textbf{Consistency and asymptotic normality}.
Figure \ref{fig:NSI}(b) verifies that the NSI estimates are consistent and asymptotically normal. 
Specifically, we let $\cG$ be a ring graph (i.e., $d = 2$) with $N = 1000$ units, 
$\epsilon^{(\bc_{\cN(i)})}_{\tau, i} \sim \cN(0, 0.1)$, $r = 2$, $T_\pre = 150$, and $T_\post = 50$.
For each simulation, we randomly generate the latent factors in the same way as described above for Figure \ref{fig:NSI}(a).
We ran 500 simulations, 
then computed the NSI residuals $(\hIPO{n} - \IPO{n})$ for 50 units in $[N]$ and across all possible counterfactual treatments for each unit.
By all possible counterfactual treatments, we used NSI to estimate $\IPO{n}$ for $\tilde{\ba}_{\cN(n)} = (1, 0, 0)$, 
$\tilde{\ba}_{\cN(n)} = (0, 1, 0)$, $\tilde{\ba}_{\cN(n)} = (1, 1, 0)$, and so on. 
The rest of setup is identical to that used for Figure \ref{fig:NSI}(a). 

Figure \ref{fig:NSI}(b)  gives a histogram of the NSI residuals. 
A Gaussian distribution is fit to the residuals and given by the red line. 
This result verifies the consistency and asymptotic normality of NSI. 
\\

\noindent
\textbf{MSE trends}. 
Figure \ref{fig:NSI}(c) summarizes the performance of NSI across different parameters. 
The performance is given by the mean-squared error (MSE) across the prediction measurements $\cT_{\post}$, averaged across $50$ units. 
Each group of bars gives the MSE for regular graphs of degree $2$, $4$, $6$, and $8$, 
as indicated on the $x$-axis. 
Within each group of bars, 
the left (blue) bars are for $N = 1000$, $T_{\pre} = 100$, $T_{\post} = 50$; 
the middle (red) bars for $N = 1000$ and $T_{\pre} = T_{\post} = 50$; 
and the right (yellow) bars for $N = 500$ and $T_{\pre} = T_{\post} = 50$. 
Each bar is the average of $200$ simulations with $\epsilon^{(\bc_{\cN(i)})}_{\tau, i} \sim \cN(0, 0.1)$, and $r = 2$.
The training treatments $\ba^{\tau} = \ba^\pre$ for $\tau \in \cT_{\pre}$ are assigned randomly and remain constant across $\cT_\pre$. 
The prediction treatments are also generated randomly and remain constant across $\cT_\post$. 
In the experiments for Figure \ref{fig:NSI}(c), 
we compute the MSE for the synthetic control setting, 
that is, 
$\tilde{\ba} = \ba^\pre$ for all $\tau \in \cT_\post$. 
Note that studying the synthetic control setting does not bias the MSE, 
as the method we propose is agnostic to the counterfactual treatment of interest as long as \textsc{TrainingTreatmentTest} is passed. 
We use the synthetic control setting to simplify the computation, 
as \textsc{TrainingTreatmentTest} is always passed under synthetic control. 

As expected, 
the MSE typically increases with degree, fewer nodes, and less training time. 
\\

\noindent
\textbf{Comparing to other estimators}.
We also compare the NSI estimator to two others: the SI estimator \citep{agarwal2020synthetic} and a baseline estimator. 
The SI estimator is a method similar to NSI, but SI assumes that there is no spillover and therefore does not account for network interference.
The baseline estimator finds donor units that satisfy Definition \ref{def:node_donors}, then averages the donor units' observed outcomes. 
We compare the estimators for a ring graph.
We compare
the estimators for a ring graph under the same parameters as those used in Figure \ref{fig:NSI}(b) averaging
across $200$ simulations, $50$ units, and all possible counterfactual treatments.

The NSI numbers are given for $\kappa \geq d + 1 = 3$ (i.e., estimates for which the elbow points are lower than $3$ are removed). 
This heuristic is consistent with Theorems \ref{thm:finite_sample_consistency}-\ref{thm:asym_normality}, 
which hold when $\kappa = r_\pre$.
That is, the NSI estimates are consistent and asymptotically normal when the number of components is at least $r_\pre$. 
Since we do not know $r_\pre$ \emph{a priori}, 
we can lower bound it and discard NSI estimates that are produced using fewer components than the lower bound. 
From \eqref{eq:full_model}, 
$r_\pre \leq r|\cN(n)|$, 
which gives a lower bound $r_\pre \geq |\cN(n)|$ as long as $r \geq 1$.
Therefore, we can discard NSI estimates that are produced using fewer than $d + 1$ components in a $d$-regular graph. 
Similarly, the SI estimates that are produced using fewer than $1$ component are also discarded since $r$ is lower bounded by $1$. 
This is built on precisely the same intuition as that given for NSI's $d + 1$ lower bound;
the only difference is that the effective degree for SI is $0$ because SI ignores network interference. 
No donors are discarded for the baseline estimator. 

The MSEs and R-squared values for the NSI estimator, SI estimator, and baseline estimators are, respectively, \textbf{(0.1174, 0.8735)}, \textbf{(0.2310, 0.8149)}, and \textbf{(3.398, -2.957)}. 
Both the NSI and baseline estimators use donor sets that contain, on average, 41 units. The SI estimator uses donor sets with, on average, 166 units.  
As such, 
even though the SI estimator has more donors, 
the performance of NSI is better than that of SI, which is better than that of the baseline estimator. 
\\

\noindent 
\textbf{Additional simulations}. 
Below, we illustrate the results produced by NSI and SI. 
The setup is the same as that given in Figure \ref{fig:NSI}(a). 
For all the plots below, 
the unit of interest and simulation is fixed. 
The counterfactual treatment of interest varies across the rows. 
In each row, the left plot gives the results for NSI, 
and the right plot gives the results for SI.
Recall that SI and NSI differ in that NSI accounts for spillover effects while SI does not.  

We can make several observations from the two plots directly below. 
First, the number of donors is greater for SI than NSI (as can be seen by the range of the x-axis of the top plot on the left versus that of the top plot on the right).
Second, SI typically uses more components to construct its estimates as well (as can be seen by the number of components to the left of the vertical lines of the top plots). 
Both these trends hold true across the examples.  
Third, both NSI and SI perform well across the training set. 
However, SI performs poorly across the prediction set, indicating that it suffers in the presence of spillover. 
Even so, the confidence interval for SI is smaller than that for NSI, 
i.e., SI is overconfident in its estimates. 
Fourth, the number of components used by NSI (as marked the vertical line in the top-left plot) is 6, which matches the network-adjusted rank of 6 (the product of $r = 2$ and $|\cN(n)| = 3$ for a ring graph) and suggest that NSI would perform well. 
This is confirmed by the fact that the estimates are close to the ground truth.

\begin{figure}[H] 
	\label{fig:y ring_NSI_1}
	\centering
	\includegraphics[width=0.35\textwidth]{../images/NSI_ring_8.pdf}
	\qquad\qquad
	\includegraphics[width=0.35\textwidth]{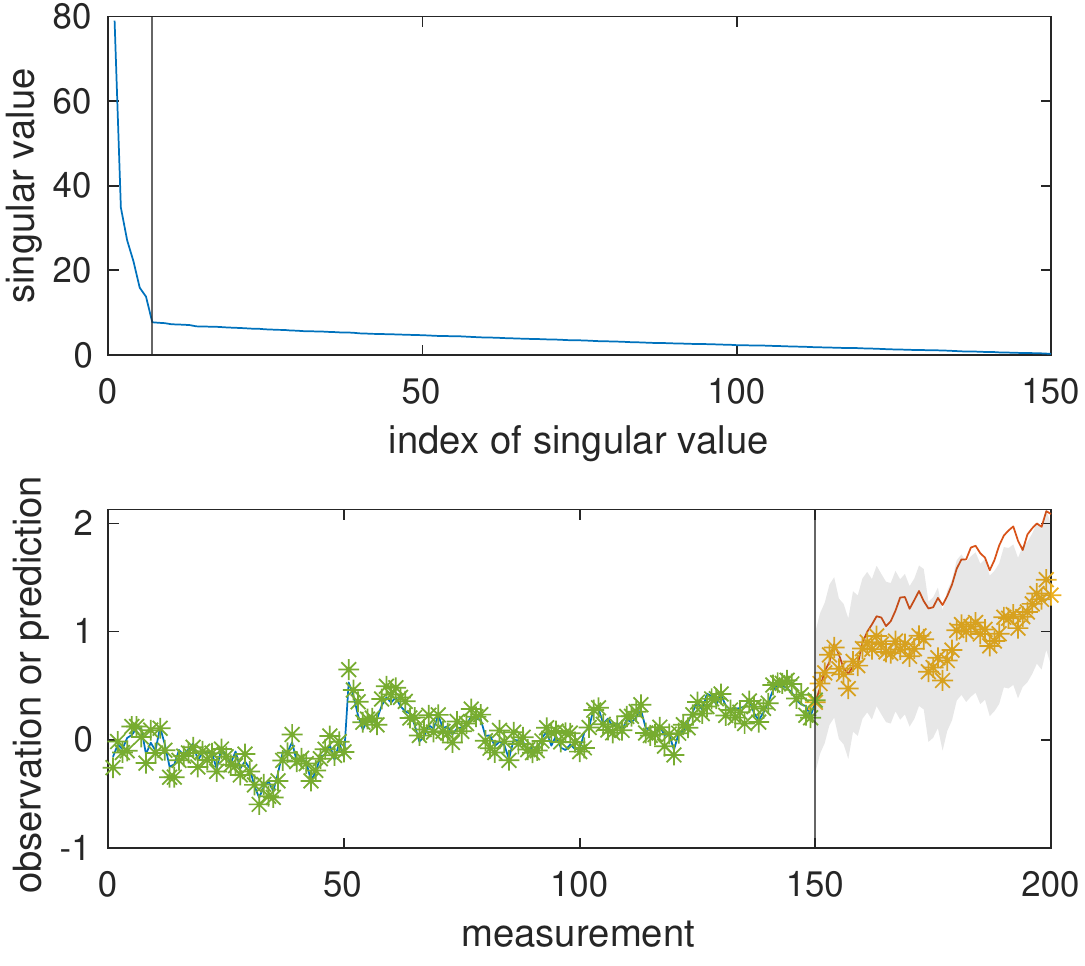}
\end{figure}

The next two pairs of plots show similar results. 
NSI performs well compared to SI, 
which is overconfident in its estimates. 
The number of components used by NSI is 6 in both cases, which suggests that it will produce good estimates. 

\begin{figure}[H] 
	\label{fig:y ring_NSI_2}
	\centering
	\includegraphics[width=0.35\textwidth]{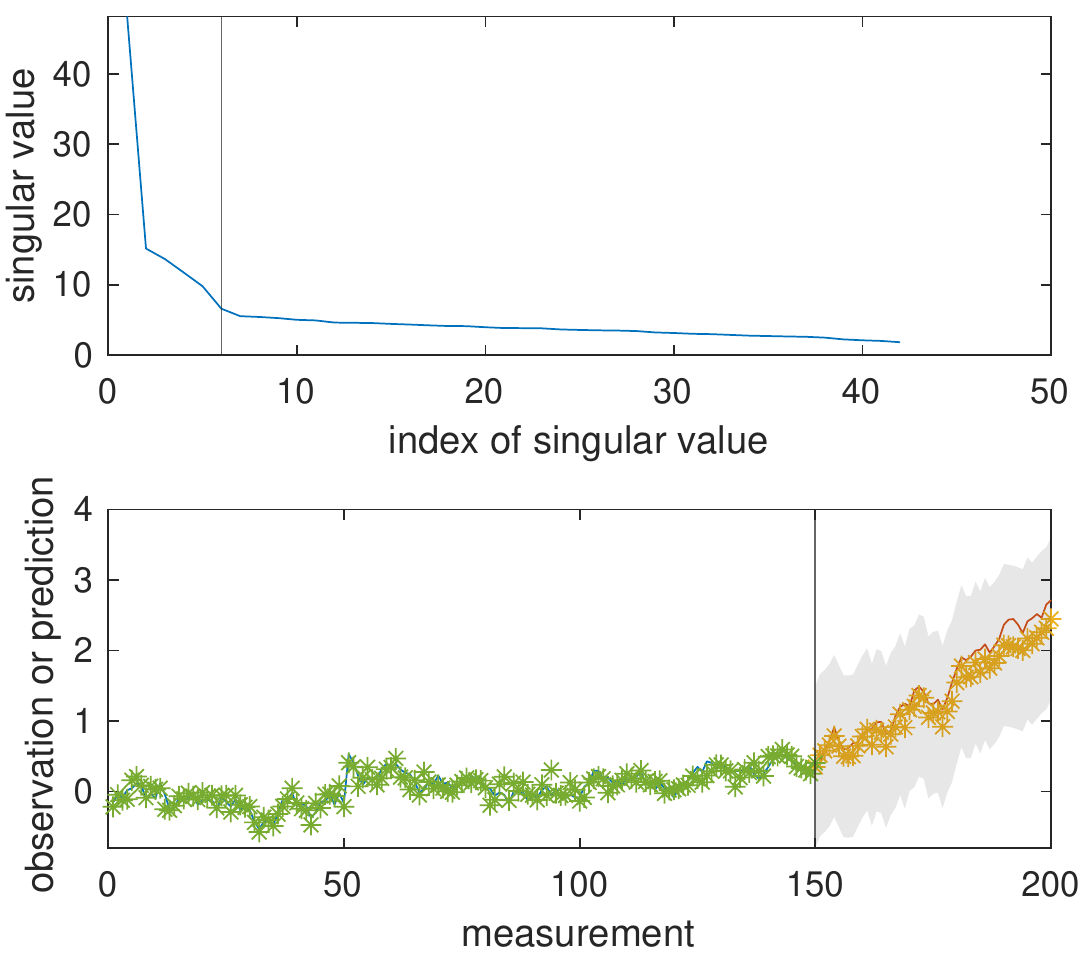}
	\qquad\qquad
	\includegraphics[width=0.35\textwidth]{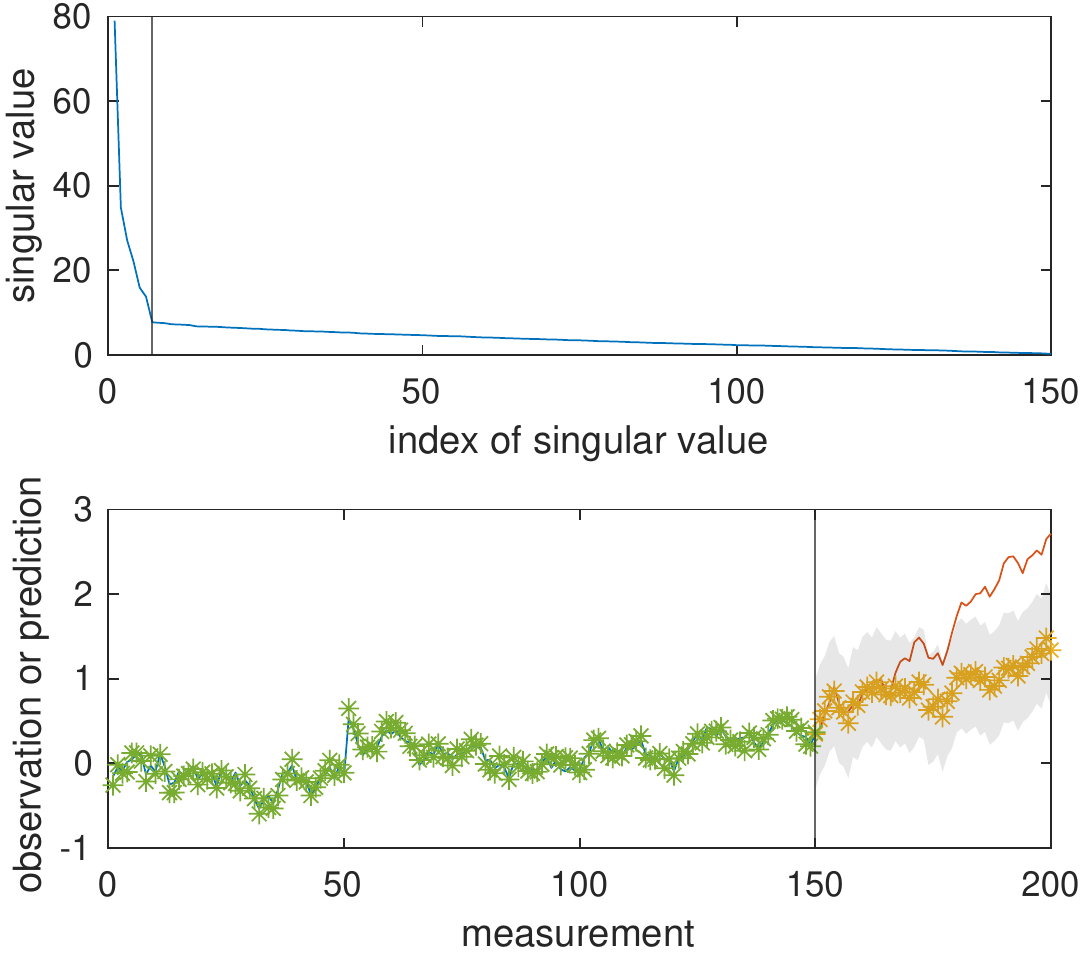}
	\\[18pt]
	\includegraphics[width=0.35\textwidth]{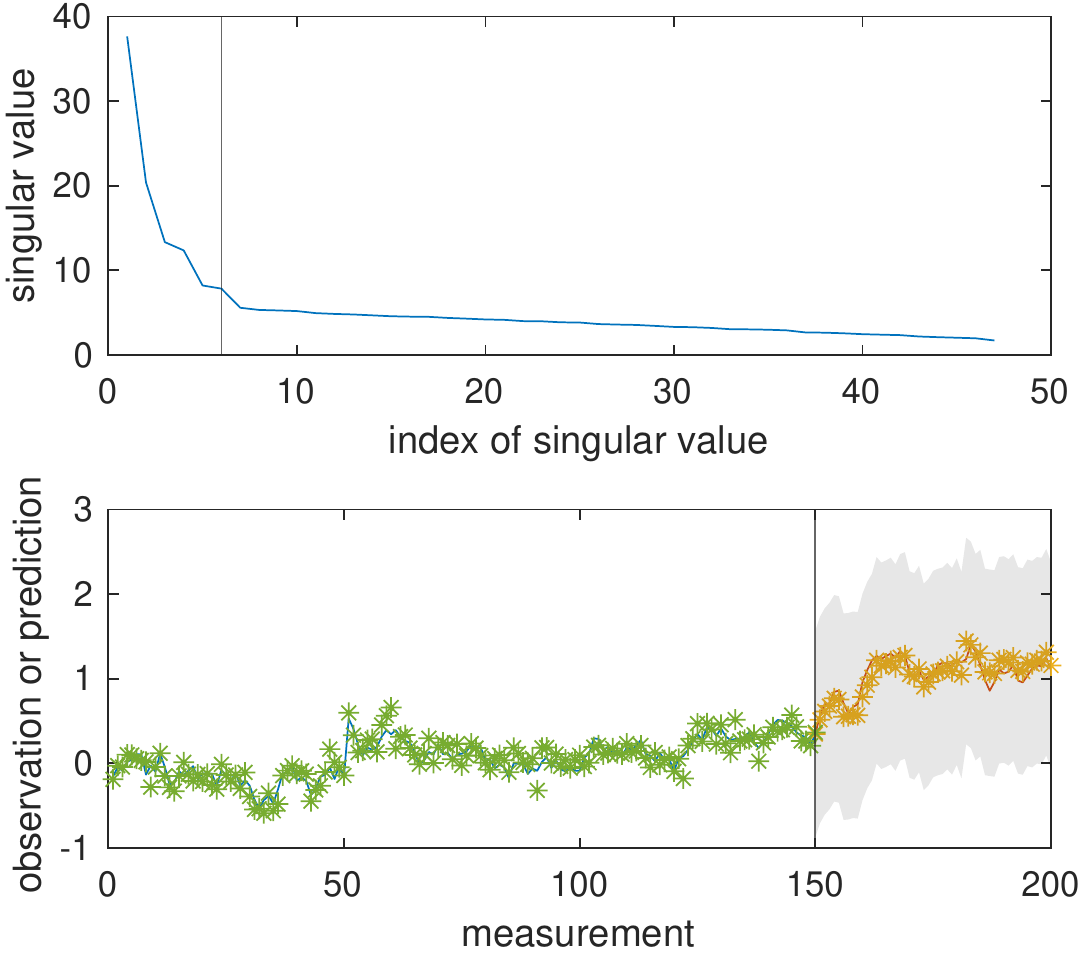}
	\qquad\qquad
	\includegraphics[width=0.35\textwidth]{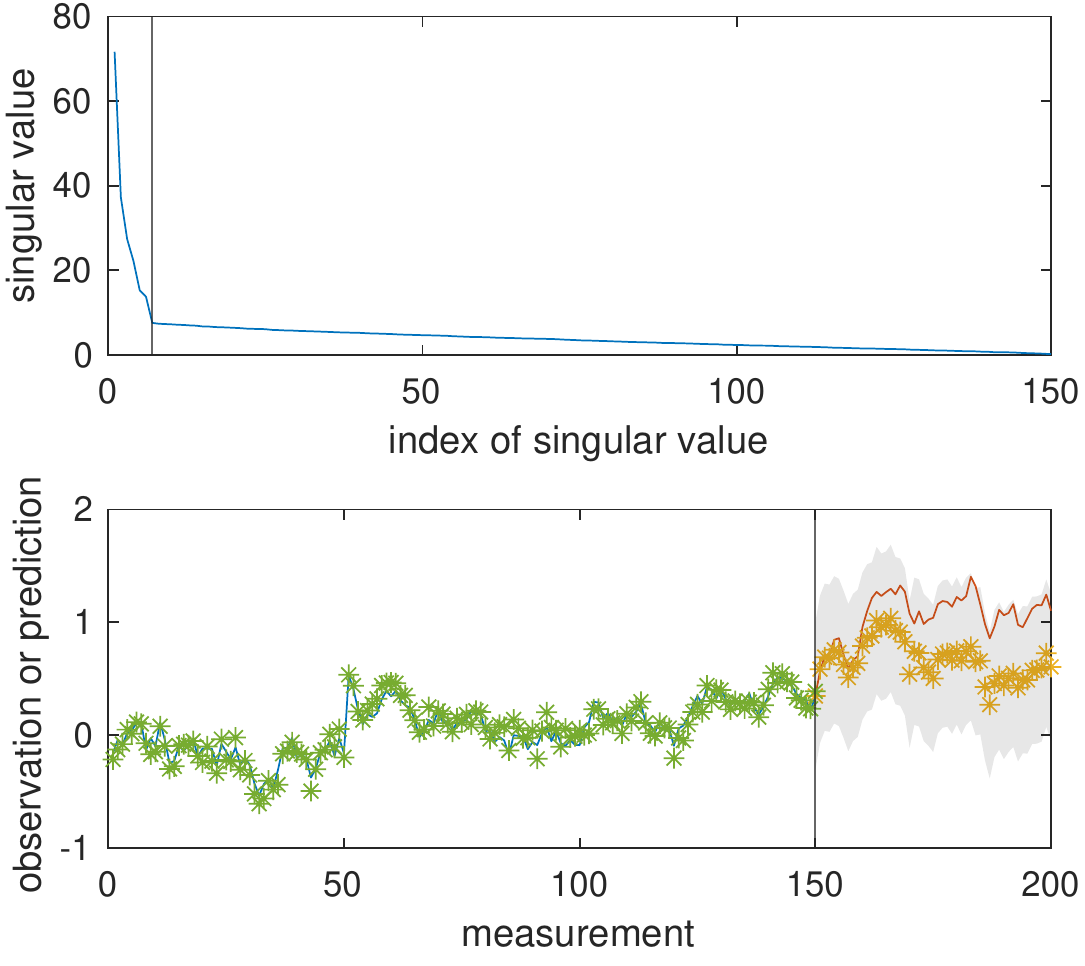}
\end{figure}

The following two plots illustrate an instance for which NSI performs poorly. 
Indeed, the number of components used by NSI is only 5.
As such, one would not expect NSI to do well. 
However, SI is still overconfident in its estimates (as there are ground truth values lie outside the gray area) whereas NSI's confidence interval covers the ground-truth potential outcomes.

\begin{figure}[H] 
	\label{fig:y ring_NSI_4}
	\centering
	\includegraphics[width=0.35\textwidth]{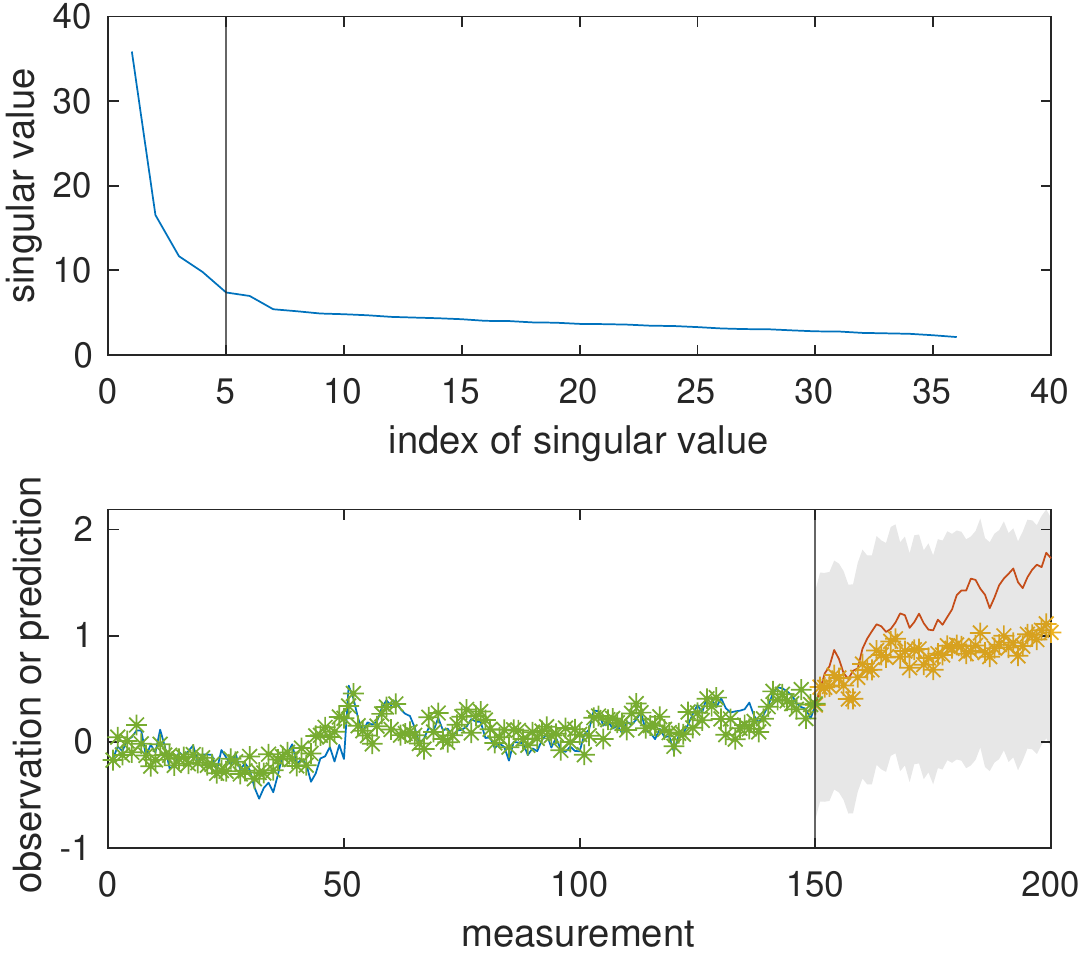}
	\qquad\qquad
	\includegraphics[width=0.35\textwidth]{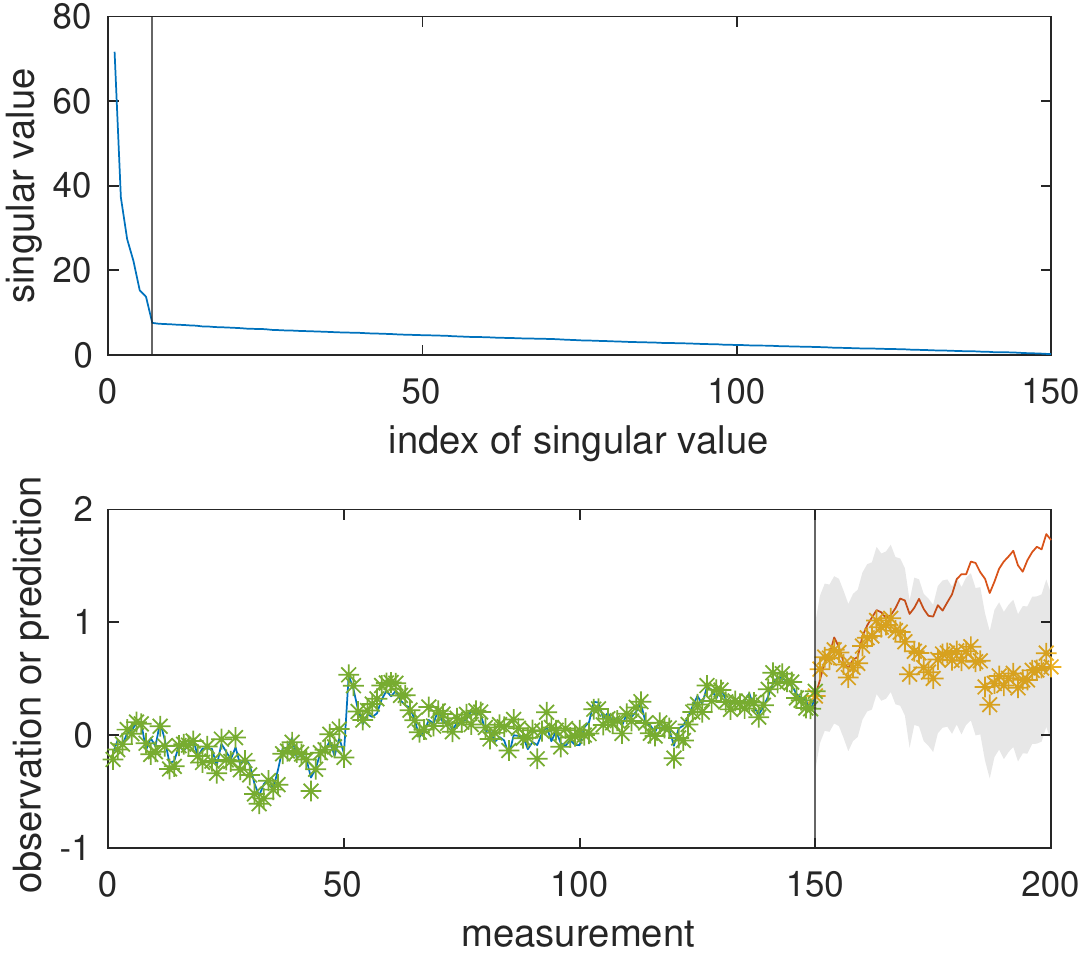}
\end{figure}

It is worth noting that in many cases  (including the one directly above), 
there seems to be leftover spectral energy beyond the $\kappa$ chosen by the knee point method. 
As such, the NSI estimates could be improved by letting $\kappa = 6$. 
The fact that $\kappa = 5$ is due to the automated knee point method that we utilize, 
and it motivates the incorporation of human oversight---that, when $\kappa$ is too small and there is leftover spectral energy, $\kappa$ can be increased. 
Generally speaking, one can increase $\kappa$ until the training MSE is small, then apply that $\kappa$ to the predictions.

%% file: main_arxiv.bbl
\begin{thebibliography}{}

\bibitem[Abadie, 2021]{abadie2021using}
Abadie, A. (2021).
\newblock {Using Synthetic Controls: Feasibility, Data Requirements, and
  Methodological Aspects}.
\newblock {\em Journal of Economic Literature}, 59(2):391--425.

\bibitem[Agarwal et~al., 2023a]{agarwal2023synthetic}
Agarwal, A., Agarwal, A., and Vijaykumar, S. (2023a).
\newblock {Synthetic Combinations: A Causal Inference Framework for
  Combinatorial Interventions}.
\newblock {\em arXiv preprint arXiv:2303.14226}.

\bibitem[Agarwal et~al., 2021a]{causalmatrixcompletion}
Agarwal, A., Dahleh, M., Shah, D., and Shen, D. (2021a).
\newblock {Causal Matrix Completion}.
\newblock {\em arXiv preprint arXiv:2109.15154}.

\bibitem[Agarwal et~al., 2020a]{PCR_2}
Agarwal, A., Shah, D., and Shen, D. (2020a).
\newblock {On Principal Component Regression in a High-dimensional
  Error-in-variables Setting}.
\newblock {\em arXiv preprint arXiv:2010.14449}.

\bibitem[Agarwal et~al., 2020b]{agarwal2020synthetic}
Agarwal, A., Shah, D., and Shen, D. (2020b).
\newblock {Synthetic Interventions}.
\newblock {\em arXiv preprint arXiv:2006.07691v4}.

\bibitem[Agarwal et~al., 2023b]{agarwal2023SI}
Agarwal, A., Shah, D., and Shen, D. (2023b).
\newblock {Synthetic A/B Testing Using Synthetic Interventions}.
\newblock {\em arXiv preprint arXiv:2006.07691}.

\bibitem[Agarwal et~al., 2021b]{PCR_1}
Agarwal, A., Shah, D., Shen, D., and Song, D. (2021b).
\newblock {On Robustness of Principal Component Regression}.
\newblock {\em Journal of the American Statistical Association},
  116(536):1731--1745.

\bibitem[Arkhangelsky et~al., 2019]{arkhangelsky2019synthetic}
Arkhangelsky, D., Athey, S., Hirshberg, D.~A., Imbens, G.~W., and Wager, S.
  (2019).
\newblock {Synthetic Difference in Differences}.
\newblock Technical report, National Bureau of Economic Research.

\bibitem[Aronow, 2012]{Aronow12}
Aronow, P.~M. (2012).
\newblock {A General Method for Detecting Interference Between Units in
  Randomized Experiments}.
\newblock {\em Sociological Methods \& Research}, 41(1):3--16.

\bibitem[Aronow and Samii, 2017]{aronow2017estimating}
Aronow, P.~M. and Samii, C. (2017).
\newblock {Estimating Average Causal Effects Under General Interference, with
  Application to a Social Network Experiment}.

\bibitem[Aronow et~al., 2017]{AronowSamii17}
Aronow, P.~M., Samii, C., et~al. (2017).
\newblock {Estimating Average Causal Effects Under General Interference, with
  Application to a Social Network Experiment}.
\newblock {\em The Annals of Applied Statistics}, 11(4):1912--1947.

\bibitem[Athey et~al., 2018]{AtheyEcklesImbens17}
Athey, S., Eckles, D., and Imbens, G.~W. (2018).
\newblock {Exact P-values for Network Interference}.
\newblock {\em Journal of the American Statistical Association},
  113(521):230--240.

\bibitem[Auerbach and Tabord-Meehan, 2021]{auerbach2021local}
Auerbach, E. and Tabord-Meehan, M. (2021).
\newblock {The Local Approach to Causal Inference Under Network Interference}.
\newblock Technical report.

\bibitem[Bajari et~al., 2021]{bajari2021multiple}
Bajari, P., Burdick, B., Imbens, G.~W., Masoero, L., McQueen, J., Richardson,
  T., and Rosen, I.~M. (2021).
\newblock {Multiple Randomization Designs}.
\newblock {\em arXiv preprint arXiv:2112.13495}.

\bibitem[Bargagli-Stoffi et~al., 2020]{bargagli2020heterogeneous}
Bargagli-Stoffi, F.~J., Tort{\`u}, C., and Forastiere, L. (2020).
\newblock {Heterogeneous Treatment and Spillover Effects Under Clustered
  Network Interference}.
\newblock Technical report.

\bibitem[Basse and Airoldi, 2018a]{BasseAiroldi17}
Basse, G.~W. and Airoldi, E.~M. (2018a).
\newblock {Limitations of Design-based Causal Inference and A/B Testing Under
  Arbitrary and Network Interference}.
\newblock {\em Sociological Methodology}, 48(1):136--151.

\bibitem[Basse and Airoldi, 2018b]{BasseAiroldi15}
Basse, G.~W. and Airoldi, E.~M. (2018b).
\newblock {Model-assisted Design of Experiments in the Presence of
  Network-correlated Outcomes}.
\newblock {\em Biometrika}, 105(4):849--858.

\bibitem[Belloni et~al., 2022]{belloni2022neighborhood}
Belloni, A., Fang, F., and Volfovsky, A. (2022).
\newblock {Neighborhood Adaptive Estimators for Causal Inference Under Network
  Interference}.
\newblock {\em arXiv preprint arXiv:2212.03683}.

\bibitem[Bertrand et~al., 2004]{bertrand2004much}
Bertrand, M., Duflo, E., and Mullainathan, S. (2004).
\newblock {How Much Should We Trust Differences-in-differences Estimates?}
\newblock {\em The Quarterly journal of economics}, 119(1):249--275.

\bibitem[Bhattacharya et~al., 2020]{pmlr-v115-bhattacharya20a}
Bhattacharya, R., Malinsky, D., and Shpitser, I. (2020).
\newblock {Causal Inference Under Interference And Network Uncertainty}.
\newblock In Adams, R.~P. and Gogate, V., editors, {\em {Proceedings of The
  35th Uncertainty in Artificial Intelligence Conference}}, volume 115 of {\em
  Proceedings of Machine Learning Research}, pages 1028--1038. PMLR.

\bibitem[Bowers et~al., 2013]{BowersFredricksonPanagopoulos12}
Bowers, J., Fredrickson, M.~M., and Panagopoulos, C. (2013).
\newblock {Reasoning about Interference Between Units: A General Framework}.
\newblock {\em Political Analysis}, 21(1):97--124.

\bibitem[Cai et~al., 2015]{cai2015social}
Cai, J., De~Janvry, A., and Sadoulet, E. (2015).
\newblock {Social Networks and the Decision to Insure}.
\newblock {\em American Economic Journal: Applied Economics}, 7(2):81--108.

\bibitem[Chatterjee, 2015]{chatterjee2015matrix}
Chatterjee, S. (2015).
\newblock {Matrix Estimation by Universal Singular Value Thresholding}.
\newblock {\em The Annals of Statistics}, 43(1):177--214.

\bibitem[Chin, 2019]{chin2019regression}
Chin, A. (2019).
\newblock {Regression Adjustments for Estimating the Global Treatment Effect in
  Experiments with Interference}.
\newblock {\em Journal of Causal Inference}, 7(2).

\bibitem[Cortez et~al., 2022a]{cortez2022exploiting}
Cortez, M., Eichhorn, M., and Yu, C.~L. (2022a).
\newblock {Exploiting Neighborhood Interference with Low Order Interactions
  Under Unit Randomized Design}.
\newblock {\em arXiv preprint arXiv:2208.05553}.

\bibitem[Cortez et~al., 2022b]{cortez2022graph}
Cortez, M., Eichhorn, M., and Yu, C.~L. (2022b).
\newblock {Staggered Rollout Designs Enable Causal Inference Under Interference
  Without Network Knowledge}.
\newblock {\em arXiv preprint arXiv:2205.14552}.

\bibitem[De~Paula, 2017]{de2017econometrics}
De~Paula, A. (2017).
\newblock {Econometrics of Network Models}.
\newblock In {\em {Advances in economics and econometrics: Theory and
  applications, eleventh world congress}}, pages 268--323. Cambridge University
  Press Cambridge.

\bibitem[De~Paula et~al., 2018]{de2018recovering}
De~Paula, A., Rasul, I., and Souza, P. (2018).
\newblock {Recovering Social Networks from Panel Data: Identification,
  Simulations and an Application}.

\bibitem[De~Paula et~al., 2019]{de2019identifying}
De~Paula, {\'A}., Rasul, I., and Souza, P. (2019).
\newblock {Identifying Network Ties from Panel Data: Theory and an Application
  to Tax Competition}.
\newblock {\em arXiv preprint arXiv:1910.07452}.

\bibitem[DiTraglia et~al., 2020]{DiTraglia2020}
DiTraglia, F.~J., Garcia-Jimeno, C., O'Keeffe-O'Donovan, R., and
  Sanchez-Becerra, A. (2020).
\newblock {Identifying Causal Effects in Experiments with Spillovers and
  Non-compliance}.
\newblock {\em arXiv preprint arXiv:2011.07051}.

\bibitem[Eckles et~al., 2017]{EcklesKarrerUgander17}
Eckles, D., Karrer, B., and Ugander, J. (2017).
\newblock {Design and Analysis of Experiments in Networks: Reducing Bias from
  Interference}.
\newblock {\em Journal of Causal Inference}, 5(1).

\bibitem[Forastiere et~al., 2021]{forastiere2021identification}
Forastiere, L., Airoldi, E.~M., and Mealli, F. (2021).
\newblock {Identification and Estimation of Treatment and Interference Effects
  in Observational Studies on Networks}.
\newblock {\em Journal of the American Statistical Association},
  116(534):901--918.

\bibitem[Gui et~al., 2015]{GuiXuBhasinHan15}
Gui, H., Xu, Y., Bhasin, A., and Han, J. (2015).
\newblock {Network A/B Testing: From Sampling to Estimation}.
\newblock In {\em {Proceedings of the 24th International Conference on World
  Wide Web}}, pages 399--409. International World Wide Web Conferences Steering
  Committee.

\bibitem[Jagadeesan et~al., 2020]{JagadeesanPillaiVolfovsky17}
Jagadeesan, R., Pillai, N.~S., and Volfovsky, A. (2020).
\newblock {Designs for Estimating the Treatment Effect in Networks with
  Interference}.
\newblock {\em The Annals of Statistics}, 48(2):679--712.

\bibitem[Jensen and Toft, 2011]{jensen2011graph}
Jensen, T.~R. and Toft, B. (2011).
\newblock {\em {Graph Coloring Problems}}.
\newblock John Wiley \& Sons.

\bibitem[Johari et~al., 2022]{johari2022experimental}
Johari, R., Li, H., Liskovich, I., and Weintraub, G.~Y. (2022).
\newblock {Experimental Design in Two-sided Platforms: An Analysis of Bias}.
\newblock {\em Management Science}.

\bibitem[Karwa and Airoldi, 2018]{karwa2018systematic}
Karwa, V. and Airoldi, E.~M. (2018).
\newblock {A Systematic Investigation of Classical Causal Inference Strategies
  Under Mis-specification Due to Network Interference}.
\newblock Technical report.

\bibitem[Leung, 2019]{leung2019causal}
Leung, M.~P. (2019).
\newblock {Causal Inference Under Approximate Neighborhood Interference}.
\newblock Technical report.

\bibitem[Li et~al., 2021]{li2021causal}
Li, W., Sussman, D.~L., and Kolaczyk, E.~D. (2021).
\newblock {Causal Inference Under Network Interference with Noise}.
\newblock Technical report.

\bibitem[Liu et~al., 2016]{liu2016inverse}
Liu, L., Hudgens, M.~G., and Becker-Dreps, S. (2016).
\newblock {On Inverse Probability-weighted Estimators in the Presence of
  Interference}.
\newblock {\em Biometrika}, 103(4):829--842.

\bibitem[Ma and Tresp, 2021]{ma2021causal}
Ma, Y. and Tresp, V. (2021).
\newblock {Causal Inference Under Networked Interference and Intervention
  Policy Enhancement}.
\newblock In {\em {International Conference on Artificial Intelligence and
  Statistics}}, pages 3700--3708. PMLR.

\bibitem[Manski, 2013]{Manski13}
Manski, C.~F. (2013).
\newblock {Identification of Treatment Response with Social Interactions}.
\newblock {\em The Econometrics Journal}, 16(1).

\bibitem[Matou{\v{s}}ek, 2008]{matouvsek2008variants}
Matou{\v{s}}ek, J. (2008).
\newblock {On Variants of the Johnson--lindenstrauss Lemma}.
\newblock {\em Random Structures \& Algorithms}, 33(2):142--156.

\bibitem[Ogburn et~al., 2017]{ogburn2017causal}
Ogburn, E.~L., Sofrygin, O., Diaz, I., and Van~der Laan, M.~J. (2017).
\newblock {Causal Inference for Social Network Data}.
\newblock Technical report.

\bibitem[Perez-Heydrich et~al., 2014]{perez2014assessing}
Perez-Heydrich, C., Hudgens, M.~G., Halloran, M.~E., Clemens, J.~D., Ali, M.,
  and Emch, M.~E. (2014).
\newblock {Assessing Effects of Cholera Vaccination in the Presence of
  Interference}.
\newblock {\em Biometrics}, 70(3):731--741.

\bibitem[Pouget-Abadie et~al.,
  2017]{PougetAbadieSaveskiSaintJacquesDuanXuGhoshAiroldi17}
Pouget-Abadie, J., Saveski, M., Saint-Jacques, G., Duan, W., Xu, Y., Ghosh, S.,
  and Airoldi, E.~M. (2017).
\newblock {Testing for Arbitrary Interference on Experimentation Platforms}.
\newblock Technical report.

\bibitem[Satopaa et~al., 2011]{satopaa2011finding}
Satopaa, V., Albrecht, J., Irwin, D., and Raghavan, B. (2011).
\newblock {Finding a ``kneedle''' in a Haystack: Detecting Knee Points in
  System Behavior}.
\newblock In {\em {2011 31st international conference on distributed computing
  systems workshops}}, pages 166--171. IEEE.

\bibitem[Saveski et~al., 2017]{saveski2017detecting}
Saveski, M., Pouget-Abadie, J., Saint-Jacques, G., Duan, W., Ghosh, S., Xu, Y.,
  and Airoldi, E.~M. (2017).
\newblock {Detecting Network Effects: Randomizing Over Randomized Experiments}.
\newblock In {\em {Proceedings of the 23rd ACM SIGKDD International Conference
  on Knowledge Discovery and Data Mining}}, pages 1027--1035. ACM.

\bibitem[S{\"a}vje et~al., 2021]{SavjeAronowHudgens17}
S{\"a}vje, F., Aronow, P.~M., and Hudgens, M.~G. (2021).
\newblock {Average Treatment Effects in the Presence of Unknown Interference}.
\newblock {\em The Annals of Statistics}, 49(2):673--701.

\bibitem[Shah et~al., 2020]{shah2020sample}
Shah, D., Song, D., Xu, Z., and Yang, Y. (2020).
\newblock {Sample Efficient Reinforcement Learning Via Low-rank Matrix
  Estimation}.
\newblock {\em arXiv preprint arXiv:2006.06135}.

\bibitem[Sussman and Airoldi, 2017a]{SussmanAiroldi17}
Sussman, D.~L. and Airoldi, E.~M. (2017a).
\newblock {Elements of Estimation Theory for Causal Effects in the Presence of
  Network Interference}.
\newblock Technical report.

\bibitem[Sussman and Airoldi, 2017b]{sussman2017elements}
Sussman, D.~L. and Airoldi, E.~M. (2017b).
\newblock {Elements of Estimation Theory for Causal Effects in the Presence of
  Network Interference}.
\newblock {\em arXiv preprint arXiv:1702.03578}.

\bibitem[Tchetgen and VanderWeele, 2012]{TchetgenVanderWeele12}
Tchetgen, E. J.~T. and VanderWeele, T.~J. (2012).
\newblock {On Causal Inference in the Presence of Interference}.
\newblock {\em Statistical Methods in Medical Research}, 21(1):55--75.
\newblock PMID: 21068053.

\bibitem[Toulis and Kao, 2013]{ToulisKao13}
Toulis, P. and Kao, E. (2013).
\newblock {Estimation of Causal Peer Influence Effects}.
\newblock In {\em {International Conference on Machine Learning}}, pages
  1489--1497.

\bibitem[Ugander et~al., 2013]{UganderKarrerBackstromKleinberg13}
Ugander, J., Karrer, B., Backstrom, L., and Kleinberg, J. (2013).
\newblock {Graph Cluster Randomization: Network Exposure to Multiple
  Universes}.
\newblock In {\em {Proceedings of the 19th ACM SIGKDD international conference
  on Knowledge discovery and data mining}}, pages 329--337. ACM.

\bibitem[Vazquez-Bare, 2022]{vazquez-bare2022}
Vazquez-Bare, G. (2022).
\newblock {Identification and Estimation of Spillover Effects in Randomized
  Experiments}.
\newblock {\em Journal of Econometrics}.

\bibitem[Verbitsky-Savitz and Raudenbush, 2012]{verbitsky2012causal}
Verbitsky-Savitz, N. and Raudenbush, S.~W. (2012).
\newblock {Causal Inference Under Interference in Spatial Settings: a Case
  Study Evaluating Community Policing Program in Chicago}.
\newblock {\em Epidemiologic Methods}, 1(1):107--130.

\bibitem[Viviano, 2020]{viviano2020experimental}
Viviano, D. (2020).
\newblock {Experimental Design Under Network Interference}.
\newblock Technical report.

\bibitem[Yu et~al., 2022]{yu2022graph}
Yu, C.~L., Airoldi, E.~M., Borgs, C., and Chayes, J.~T. (2022).
\newblock {Estimating Total Treatment Effect in Randomized Experiments with
  Unknown Network Structure}.
\newblock {\em arXiv preprint arXiv:2205.12803}.

\bibitem[Zack et~al., 1977]{zack1977automatic}
Zack, G.~W., Rogers, W.~E., and Latt, S.~A. (1977).
\newblock {Automatic Measurement of Sister Chromatid Exchange Frequency.}
\newblock {\em Journal of Histochemistry \& Cytochemistry}, 25(7):741--753.

\end{thebibliography}
